\documentclass[preprint,11pt,authoryear]{elsarticle}

\usepackage{graphics}
\usepackage{graphicx}
\graphicspath{{Figures/}}
\usepackage{epsfig}
\usepackage{subfigure}
\usepackage{amssymb}
\usepackage{amsthm}
\usepackage{bm}
\usepackage{amsmath}
\usepackage{color}
\usepackage{geometry}
\usepackage{multirow}
\usepackage{setspace}
\allowdisplaybreaks
\usepackage{enumerate}
\usepackage{booktabs}
\usepackage{indentfirst}
\usepackage{booktabs} 
\usepackage{hyperref}  
\usepackage{float}  
\hypersetup{
	colorlinks=true,       
}

\biboptions{}

\journal{International Journal of Solids and Structures}

\begin{document}

\begin{frontmatter}



\title{Voltage-controlled non-axisymmetric vibrations \\of soft electro-active tubes with strain-stiffening effect \footnote{\textbf{Dedicated to Yibin Fu, in esteem and friendship}}}


\author[1]{Fangzhou Zhu}
\author[1]{Bin Wu\corref{cor1}}
\ead{bin.wu@zju.edu.cn}
\author[1,2]{Michel Destrade}
\author[1]{Huiming Wang}
\author[1]{Ronghao Bao}
\author[1,3,4]{Weiqiu Chen}

\cortext[cor1]{Corresponding author.}
\address[1]{Key Laboratory of Soft Machines and Smart Devices of Zhejiang Province and \\Department of Engineering Mechanics, Zhejiang University, 310027 Hangzhou, P.R. China;\\[6pt]}
\address[2]{School of Mathematical and Statistical Sciences, University of Galway, University Road, Galway, Ireland; \\[6pt]}
\address[3]{Huanjiang Laboratory, 311816 Zhuji, Zhejiang, P.R. China;\\[6pt]}
\address[4]{Soft Matter Research Center, Zhejiang University, 310027 Hangzhou, P.R. China.}


\begin{abstract}
	
Material properties of soft electro-active (SEA) structures are significantly sensitive to external electro-mechanical biasing fields (such as pre-stretch and electric stimuli), which generate remarkable knock-on effects on their dynamic characteristics. In this work, we analyze the electrostatically tunable non-axisymmetric vibrations of an incompressible SEA cylindrical tube under the combination of a radially applied electric voltage and an axial pre-stretch. Following the theory of nonlinear electro-elasticity and the associated linearized theory for superimposed perturbations, we derive the nonlinear static response of the SEA tube to the inhomogeneous biasing fields for the Gent ideal dielectric model. Using the State Space Method, we efficiently obtain the frequency equations for voltage-controlled small-amplitude three-dimensional non-axisymmetric vibrations, covering a wide range of behaviors, from the purely radial breathing mode to torsional modes, axisymmetric longitudinal modes, and prismatic diffuse modes. We also perform an exhaustive numerical analysis to validate the proposed approach compared with the conventional displacement method, as well as to elucidate the influences of the applied voltage, axial pre-stretch, and strain-stiffening effect on the nonlinear static response and vibration behaviors of the SEA tube. The present study clearly indicates that manipulating electro-mechanical biasing fields is a feasible way to tune the small-amplitude vibration characteristics of an SEA tube. The results should  benefit experimental work on, and design of, voltage-controlled resonant devices made of SEA tubes.
	
\end{abstract}

\end{frontmatter}




\section{Introduction}

Among the ever-increasing range of smart materials being currently developed, soft electro-active (SEA) materials such as dielectric elastomers can convert or transduce electrical energy to or from mechanical energy \citep{pelrine1998electrostriction}. SEA structures demonstrate superior electro-mechanical coupling properties, as they reduce remarkably in thickness and expand in area when exposed to applied electro-mechanical biasing fields, in contrast to typical piezoelectric materials which are too brittle to undergo large deformations. 
Other excellent features, such as reversible large deformation, rapid response, low weight and low cost, high elastic energy density, and high conversion efficiency, have attracted wide academic and industrial interest and led to various broad practical applications in soft robotics, biomedical devices, flexible electronics, tunable resonators as well as active waveguides, phononic crystals and metamaterials \citep{carpi2011dielectric, anderson2012multi,zhao2014harnessing, lu2020mechanics, zhu2010resonant, zhao2016application, wang2020tunable, chen2022voltage, zhao2022vibrations}.

Strong nonlinearity and electro-mechanical coupling make it quite difficult to establish a general continuum mechanics framework. Pioneering works on nonlinear theory of electro-elasticity were conducted by \cite{toupin1956elastic, toupin1963dynamical} more than half a century ago for static and dynamic analyses of finitely deformed elastic dielectrics. Since the 1980s, there have been numerous reformulations of a general nonlinear continuum theory for electro-magneto-mechanical couplings \citep{maugin2013continuum, eringen2012electrodynamics}, paralleled with the development of various smart materials and structures  with wide-ranging applications. Furthermore, the emergence of SEA materials in recent decades has encouraged new interpretations, advancements, and applications of nonlinear electro-elasticity theory \citep{mcmeeking2005electrostatic,dorfmann2006nonlinear,suo2008nonlinear,liu2013energy,dorfmann2014nonlinear}. 
A nonlinear continuum framework, accounting for the nonlinear interaction between mechanical and electromagnetic fields, as documented in the monograph by \cite{dorfmann2014nonlinear}, has now successfully been applied to the analysis of electro-active and magneto-active materials undergoing significant deformations \citep{rudykh2013stability,xie2016bifurcation, jandron2018numerical, fu2018localized, psarra2019wrinkling, su2019finite, su2020pattern, wu2021wrinkling}.

The linearized incremental theory based on the nonlinear electro-elasticity theory \citep{baumhauer1973nonlinear,tiersten1981electroelastic, maugin2013continuum, eringen2012electrodynamics, baesu2003incremental, dorfmann2010electroelastic} is commonly employed to investigate how biasing fields (induced by, for example, prestretch, internal pressure, and electric stimuli) affect the superimposed small-amplitude dynamic properties of SEA structures. \cite{dorfmann2010electroelastic,dorfmann2014nonlinear} developed a compact version of the linearized incremental theory in both the Lagrangian and updated Lagrangian descriptions to examine the small-amplitude motions superimposed on finite biasing fields, paying particular attention to SEA materials. 
We recommend the detailed review by \cite{Wu2016theory} for a comprehensive comparison of different versions of nonlinear electro-elasticity theories and relevant linearized incremental theories, showing ultimately that ostensibly different theories in the literature on this topic are actually equivalent with no substantive differences.

\textcolor{black}{The analysis of electro-mechanical instabilities has significant theoretical and practical implications. Practically speaking, SEA materials and structures could experience multiple failure mechanisms, such as pull-in or snap-through instability \citep{zhao2007method, su2018wrinkles}, electric breakdown \citep{zurlo2017catastrophic}, macroscopic and microscopic buckling instability \citep{bertoldi2011instabilities,rudykh2014multiscale,goshkoderia2017electromechanical}, localized necking of SEA membranes \citep{fu2018localized}, bending instability of SEA slabs or bilayers \citep{su2019finite,su2020pattern}, barrelling axisymmetric instability of SEA tubes \citep{melnikov2018bifurcation}, prismatic instability of SEA tubes \citep{bortot2018prismatic}, bulging instability of SEA tubes or balloons \citep{lu2015electro,wang2017anomalous}, and pear-shaped bifurcation from SEA spherical balloons \citep{xie2016bifurcation}. In particular, \cite{bertoldi2011instabilities} analytically identified four different instability criteria for multilayered soft dielectrics: (i) loss of positive definiteness of the tangent electro-elastic constitutive operator, (ii) existence of diffuse modes of bifurcation or microscopic instability modes, (iii) loss of strong ellipticity of the homogenized continuum or macroscopic instability modes, and (iv) electric breakdown. In another work, \cite{rudykh2014multiscale} extensively investigated the multiscale instabilities in layered dielectric elastomers and explained the crucial effect of microstructures on the onset of instabilities: (i) macroscopic instabilities predominate when the stiffer phase’s volume fraction is moderate, (ii) interfacial instabilities start to show up at low stiffer phase volume fractions, and (iii) instabilities of a finite scale, comparable to the microstructure size, appear at high volume fractions of the stiffer phase.}


Investigations on the small-amplitude dynamic behaviors of smart systems made of SEA materials subject to biasing fields induced by pre-stretch, internal pressure and electric stimuli, also have theoretical as well as practical significance, see the recent review article by \cite{zhao2022vibrations} on vibrations and waves in soft dielectric structures. 
\textcolor{black}{Based on Dorfmann and Ogden's linearized incremental theory of nonlinear electro-elasticity \citep{dorfmann2010electroelastic,dorfmann2014nonlinear}, considerable efforts have been devoted to studying small-amplitude elastic waves propagating in finitely deformed SEA structures, such as
bulk waves in compressible dielectrics \citep{galich2016manipulating}, 
surface waves in a deformed SEA half-space \citep{dorfmann2010electroelastic}, Rayleigh-Lamb waves in a deformed ideal dielectric plate \citep{shmuel2012rayleigh,ziser2017experimental,broderick2020electro}, torsional, axisymmetric, non-axisymmetric and circumferential waves in a pre-stretched SEA tube subject to an axial or a radial electrical biasing field \citep{shmuel2013axisymmetric, shmuel2015manipulating, su2016propagation, wu2017guided, wu2020nonlinear, dorfmann2020waves},
bulk waves in extremely deformed soft auxetic materials \citep{galich2015influence}, shear or longitudinal wave propagation and tunable band gaps in periodic dielectric laminates \citep{galich2017shear, chen2020effects}, and electrostatically tunable band gaps in finitely deformed SEA fiber composites \citep{shmuel2013electrostatically}. 
In particular, to overcome the difficulty in solving the inhomogeneous biasing fields generated by the application of a radial electric voltage in an SEA tube, \cite{wu2017guided} proposed the State Space Method (SSM), which combines the state-space formalism with the approximate laminate technique, to effectively investigate the elastic waves propagating in SEA tubes.}


Electro-mechanical biasing fields lead to changes in the effective material characteristics and geometry, and also allow for active tuning of the vibration behaviors to acquire desired operating performance. 
Thus, tunable SEA resonators have a wide range of potential applications, including the design of tunable SEA loudspeakers for sound generation, the use of adaptive acoustic absorbers for noise reduction, and the development of active and adaptive vibration isolators and dampers that take advantage of viscoelasticity and stiffness tunability. For example, \cite{dubois2008voltage} examined experimentally and theoretically the voltage controllability of the resonance frequency for SEA polymer membranes, and found that tuning the voltage might reduce the resonance frequency by up to 77\% from its initial value. \cite{zhu2010resonant} analytically and experimentally found that the natural frequencies of a circular dielectric membrane can be tuned by varying the in-plane pre-stretch, out-of-plane pressure and voltage. 
\cite{sugimoto2013lightweight} proposed a lightweight push-pull acoustic transducer using dielectric films for sound generation in advanced audio systems, and their experiments showed that push-pull driving can effectively suppress harmonic distortion. 
Furthermore, \cite{hosoya2015hemispherical} constructed, examined, and evaluated a hemispherical breathing mode loudspeaker driven by a dielectric actuator to determine its repeatability, sound pressure, vibration mode profiles, and acoustic radiation patterns. 
To absorb sound energy, \cite{lu2015electronically} developed an electronically tunable duct silencer, which is formed with dielectric membranes and back cavity, and uses external control signals. 
With the Space State Method, \cite{zhu2020electrostatically} studied axisymmetric torsional and longitudinal vibrations in an SEA tube subject to inhomogeneous biasing fields induced by the combined action of axial pre-stretch and radial voltage, and established electrostatically tunable axisymmetric vibration characteristics. By applying alternating voltages with opposite phases to a dielectric actuator, \cite{zhang2015tunable} proposed a vibration damper to reduce vibration. \cite{sarban2011tubular} effectively achieved active vibration isolation after fabricating a core-free rolled tubular SEA actuator and studying its dynamic properties. 
As a biomedical application, \cite{son2012large} suggested coupling a SEA tube sensor to an artery segment to give it structural support while also keeping track of its local condition data. 
\cite{zhu2010nonlinear}  demonstrated theoretically the tunability of the natural frequency of breathing modes in a dielectric balloon by varying the pressure or voltage.  
\cite{mao2019electrostatically} used the SSM to investigate the feasibility of tuning the three-dimensional (3D) and small-amplitude torsional and spheroidal vibrations in a dielectric spherical balloon by means of varying internal pressure and radial electric voltage.
\textcolor{black}{Recently, the SSM was utilized by \cite{cao2024axisymmetric} to explore the influences of electro-mechanical biasing fields and fluid added mass effect on the linearized axisymmetric vibration of multilayered SEA circular plates in contact with fluid.}

\begin{figure}[h!] 
	\centering  
	\subfigure{
		\includegraphics[width=0.8\textwidth]{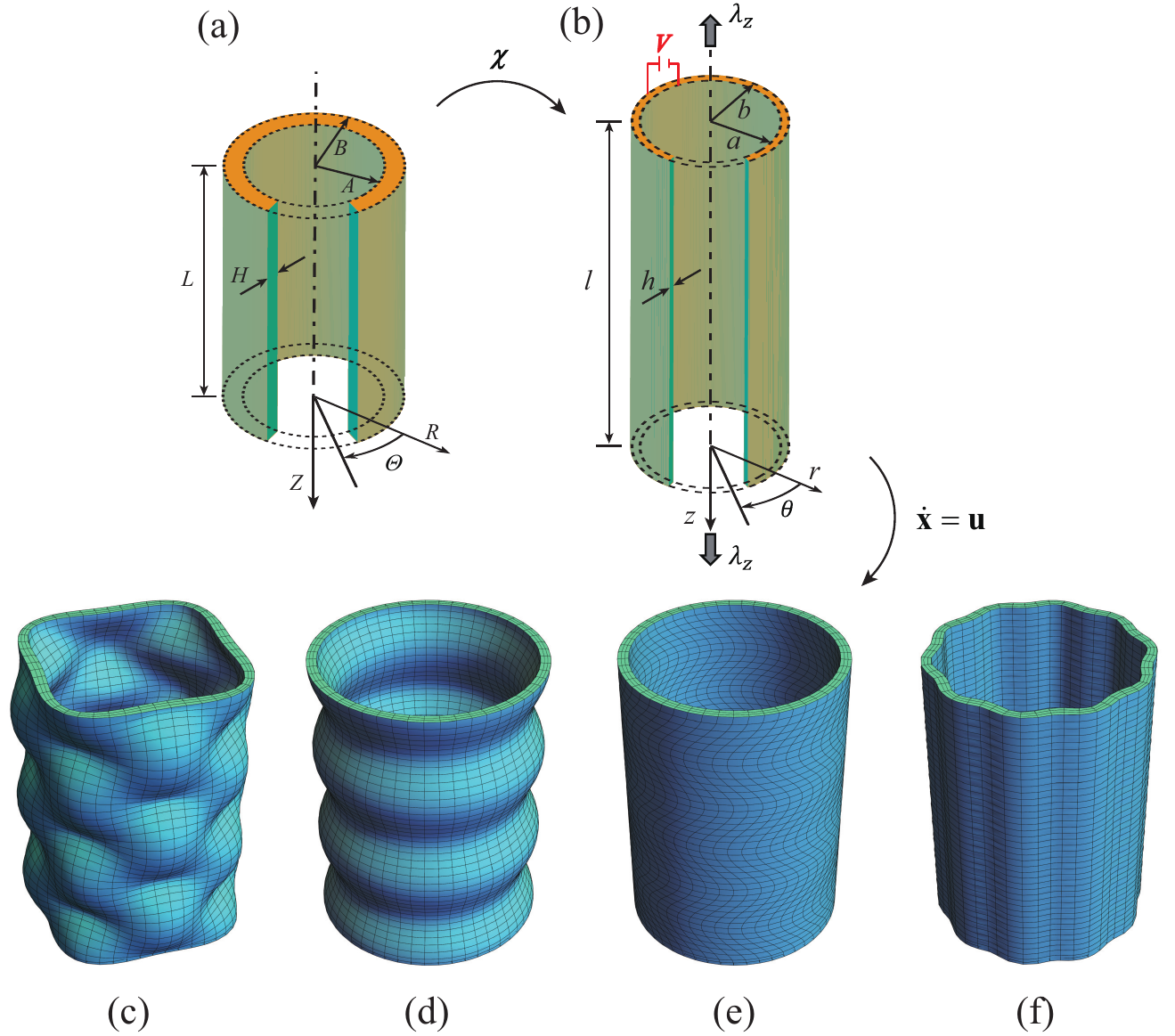}}
	\caption{Schematic diagram of an SEA tube with cylindrical coordinates and geometric sizes: (a) undeformed configuration and (b) deformed configuration subject to inhomogeneous biasing fields generated by the combined action of radial electric voltage $(V)$ and axial pre-stretch $(\lambda_{z})$. Incremental motion fields created by (c) non-axisymmetric vibrations, (d) longitudinal vibrations (L vibrations), (e) torsional vibrations (T vibrations), and (f) prismatic vibrations.} 
	\label{Physical_description}
\end{figure}

The first objective of this paper is to investigate the strain-stiffening effect on axisymmetric torsional and longitudinal vibrations (hereafter abbreviated as T vibrations and L vibrations) of an SEA tube as a continuation of our previous work \citep{zhu2020electrostatically} wherein the strain-stiffening effect was not taken into consideration. The second purpose of this paper is to clarify how the inhomogeneous biasing fields induced by the radial electric voltage and axial pre-stretch (see Figs.~\ref{Physical_description}(a) and \ref{Physical_description}(b)) and the strain-stiffening effect impact the superimposed non-axisymmetric small-amplitude vibrations in the SEA tube.

The structure of this paper is as follows. Section \ref{section2} derives the basic formulations governing the axisymmetric static deformations and induced inhomogeneous biasing fields of an SEA tube characterized by the Gent ideal dielectric model. In Section \ref{section3}, we employ the state-space formalism combined with an approximate laminate technique to obtain the frequency equations of the non-axisymmetric vibrations and prismatic vibrations of the deformed SEA tube with generalized rigidly supported boundary condition. Numerical calculations are presented in Section \ref{section4} to examine the nonlinear static response to the applied radial voltage of the SEA tube, to validate the excellent convergence rate and accuracy of the SSM, and to demonstrate the influences of the applied voltage, axial pre-stretch, and strain-stiffening effect on the axisymmetric and non-axisymmetric vibration characteristics. Finally, we give a conclusive summary in Section \ref{section5} and some relevant mathematical derivations are provided in \ref{AppendixA}. 


\section{Equations of nonlinear electro-elasticity} \label{section2}

The basic equations governing the finite electro-elastic deformations of an incompressible soft electro-elastic body are presented first in Sec.~\ref{2_1}, as established by \cite{dorfmann2006nonlinear,dorfmann2014nonlinear}. Then we specialize the basic equations of nonlinear electro-elasticity to the finite static axisymmetric deformations of an SEA tube subject to a radial electric field and an axial pre-stretch in Sec.~\ref{2_2}.

\subsection{Finite electro-elasticity theory} \label{2_1}

Here we consider a soft deformable continuous electro-elastic body undergoing a static finite  deformation. The \textit{underformed stress-free reference configuration} at time $t_0$ is denoted by ${\mathcal{B}}_r$ with its boundary and outward unit normal vector denoted as ${\partial\mathcal{B}}_r$ and $\mathbf N$, respectively.
An arbitrary material point $X$ in the stress-free reference configuration is labelled by a position vector $\mathbf{X}$. At time $t$, stimulated by external electro-mechanical loads, the electro-elastic body ${\mathcal{B}}_r$ deforms to the \textit{deformed or current} configuration ${\mathcal{B}}_t$. Naturally, the material point $X$ occupies a new position $\mathbf{x}$, and the deformation is described by the mapping $\mathbf{x}=\bm{\chi} \left( \mathbf{X}, t\right) $ where $\bm{\chi}$ is a continuous and twice differentiable vector function. The boundary and outward unit normal of the current configuration are denoted as ${\partial\mathcal{B}}_t$ and ${\mathbf{n}}_t$, respectively, and $\mathbf{F}=\partial \mathbf{x} / \partial \mathbf{X}= \rm Grad\hskip 1pt \bm{\chi}$ is the deformation gradient tensor, where `$\rm Grad$' is the gradient operator with respect to $\mathbf{X}$ in the reference configuration $\mathcal{B}_r$. The quantity $J=\rm{det}\hskip 1pt \mathbf{F}>0$ measures local volume changes, and is equal to one identically for an incompressible material.

The equation of motion, Gauss's law, and Faraday's law can be expressed as follows, respectively, within the `quasi-electrostatic approximation' and in the absence of mechanical body forces as well as free charges and electric currents:
\begin{equation}\label{governing_equations}
	\rm{div}\hskip 1pt \bm{\tau} = \rho \partial^2 \mathbf{x}/\partial \textit{t}^2, \qquad \rm{div}\hskip 1pt \mathbf{D} = 0, \qquad \rm{curl}\hskip 1pt \mathbf{E} = \mathbf{0},
\end{equation}
where `$\rm{div}$' and `$\rm{curl}$' are the divergence and curl operators with respect to $\mathbf{x}$ in the current configuration $\mathcal{B}_t$, $\rho$ is the mass density (which is unchanged during the motion due to the material incompressibility), and $\bm{\tau}$, $\mathbf{D}$ and $\mathbf{E}$ are the total Cauchy stress tensor, Eulerian electric displacement vector and electric field vector in $\mathcal{B}_t$, respectively.

Following nonlinear electro-elasticity theory \citep{dorfmann2014nonlinear}, it is convenient to present the nonlinear constitutive relations for incompressible SEA materials in terms of a \textit{total energy function} or \textit{modified free energy function}, $\Omega(\mathbf{F},\bm{\mathcal{D}})$ per unit reference volume in $\mathcal{B}_r$ as
\begin{equation} \label{nonlinear_constitutive_relations_for_incompressibility}
	\mathbf{T}=\dfrac{\partial \Omega}{\partial \mathbf{F}}-p \mathbf{F}^{-1}, 
	\qquad 
	\bm{\mathbf{\cal E}}=\dfrac{\partial \Omega}{\partial \bm{\mathcal{D}}},
\end{equation}
where $\mathbf{T}=\mathbf{F}^{-1} \bm{\tau}$ is the total nominal stress tensor, $\bm{\mathcal{D}}=\mathbf{F}^{-1}\mathbf{D}$ and $\bm{\mathbf{\cal E}}=\mathbf{F}^\mathsf{T} \mathbf{E}$ are the nominal electric displacement vector and nominal electric field vector, respectively, which are the Lagrangian counterparts of $\mathbf{D}$ and $\mathbf{E}$, and $p$ is a Lagrange multiplier due to the incompressibility constraint. For an incompressible isotropic SEA material ($I_3 \equiv \mathrm{det} \, \mathbf{C}= J^2 = 1$), the energy density function $\Omega(\mathbf{F},\bm{\mathcal{D}})$ depends on five scalar quantities, for example the following five invariants:
\begin{equation} \label{five_invariants}
	I_1 = \mathrm{tr} \mathbf{C}, \quad I_2 = \tfrac{1}{2}[(\mathrm{tr}\mathbf{C})^2-\mathrm{tr}(\mathbf{C}^2)], \quad I_4= \bm{\mathcal{D}} \cdot \bm{\mathcal{D}}, \quad I_5 = \bm{\mathcal{D}} \cdot (\mathbf{C} \bm{\mathcal{D}}), \quad I_6 = \bm{\mathcal{D}} \cdot (\mathbf{C}^2 \bm{\mathcal{D}}),
\end{equation}
where $\mathbf{C} = \mathbf{F}^{\mathsf{T}} \mathbf{F}$ is the right Cauchy-Green deformation tensor, with the superscript $^\mathsf{T}$ indicating the transpose operator. 

Therefore, combination of Eqs.~(\ref{nonlinear_constitutive_relations_for_incompressibility}) and (\ref{five_invariants}) provides the total Cauchy stress tensor $\bm{\tau}$ and the Eulerian electric field vector $\mathbf{E}$ as
\begin{equation} \label{Constitutive_equations}
	\begin{array}{l}
		\bm{\tau} = 2 \Omega_1 \mathbf{b} + 2 \Omega_2 (I_1 \mathbf{b} - \mathbf{b}^2) - p \mathbf{I} + 2 \Omega_5 \mathbf{D} \otimes \mathbf{D} + 2 \Omega_6 (\mathbf{D} \otimes \mathbf{b} \mathbf{D} + \mathbf{b} \mathbf{D} \otimes \mathbf{D}), \\\\
		\mathbf{E} = 2 (\Omega_4 \mathbf{b}^{-1} \mathbf{D} + \Omega_5 \mathbf{D} + \Omega_6 \mathbf{b} \mathbf{D}),
	\end{array}
\end{equation}
where $\mathbf{I}$ is the identity tensor in ${\mathcal{B}}_t$, $\mathbf{b} = \mathbf{F} \mathbf{F}^{\mathsf{T}}$ is the left Cauchy-Green deformation tensor, and the shorthand notation $\Omega_m =\partial \Omega/ \partial I_m \;(m= 1, 2, 4, 5, 6)$ is adopted throughout this paper.

There is no electric field in the surrounding vacuum when an electric voltage is applied to the surfaces of the SEA body covered with flexible electrodes. Thus, the mechanical and electric boundary conditions to be satisfied on the boundary $\partial \mathcal{B}_t$ are written in Eulerian form as
\begin{equation}
	\bm{\tau} \mathbf{n}_t = \mathbf{t}_a, \quad \mathbf{E} \times \mathbf{n}_t =\mathbf{0}, \quad \mathbf{D} \cdot \mathbf{n}_t = -\sigma_f, 
\end{equation}
where $\mathbf{t}_a$ is the applied mechanical traction vector per unit area of $\partial \mathcal{B}_t$ and $\sigma_f$ is the free surface charge density on $\partial \mathcal{B}_t$.

\subsection{Finite axisymmetric deformations of an SEA tube} \label{2_2}

The problem of nonlinear axisymmetric deformations of an SEA tube subject to a radial electric field, internal/external pressures, and an axial pre-stretch has previously been discussed by \cite{zhu2010large, shmuel2013axisymmetric, zhou2014electromechanical, shmuel2015manipulating,  melnikov2016finite, su2016propagation, wu2017guided, bortot2018prismatic}. In this section, we briefly outline for completeness the formulations governing the nonlinear axisymmetric deformations for arbitrary energy function when the SEA tube is subject to a radial voltage combined with an axial pre-stretch (see Fig.~\ref{Physical_description}(b)). 
We then specialize the results to the Gent ideal dielectric model and obtain explicit expressions for both the nonlinear static response and the radially inhomogeneous biasing fields.

The schematic diagrams of an SEA tube with flexible electrodes before and after activation are illustrated in Figs.~\ref{Physical_description}(a) and \ref{Physical_description}(b), respectively. For convenience, we use the cylindrical coordinate systems $\left( R, \mit{\Theta}, Z \right) $ and $\left( r, \theta, z \right) $ to describe the undeformed and deformed configurations, respectively. In the undeformed configuration, the inner and outer radii and the thickness of the tube are specified as $A$, $B$, and $H=B-A$, respectively, with the tube length denoted by $L$. 
To realize electro-mechanically tunable vibration characteristics of the SEA tube, a radial electric voltage $V$ is applied to the electrodes of the tube, which is also simultaneously subject to a uniform axial pre-stretch $\lambda_{z}$. As a result, the inner and outer radii, the thickness and the length of the deformed tube are $a$, $b$, $h$, and $l$, respectively. The inner-to-outer radius ratios in the undeformed and deformed configurations are defined as $\eta = A/B$ and $\overline{\eta} = a/b$, respectively. 

For an incompressible, initially isotropic tube, cylindrically axisymmetric deformations are described by
\begin{equation} \label{deformaion_equations}
	r = \sqrt{\lambda^{-1}_{z} \left(R^2 - A^2 \right) + a^2 } ,
	\quad
	\theta = \mit{\Theta} ,
	\quad
	z = \lambda_{z}Z,
\end{equation}
where $\lambda_{z}=l/L$. Thus, the deformation gradient tensor can be represented as
\begin{equation}
	\mathbf{F} = \left[ \begin{array}{ccc}
		\dfrac{\partial r}{\partial R}  &  \dfrac{\partial r}{R \partial \mit{\Theta}}  &  \dfrac{\partial r}{\partial Z} \\\\
		\dfrac{r \partial \Theta}{\partial R}  &  \dfrac{r \partial \theta}{R \partial  \mit{\Theta}}  &  \dfrac{r \partial \theta}{\partial Z} \\\\
		\dfrac{\partial z}{\partial R}  &  \dfrac{\partial z}{R \partial \mit{\Theta}}  &  \dfrac{\partial z}{\partial Z}
	\end{array}
	\right] 
	=
		\text{diag}\left[ \lambda^{-1}_{\theta} \lambda^{-1}_{z} , \lambda_{\theta}, \lambda_{z} \right],
\end{equation}
where `$\text{diag}$' denotes the diagonal matrix, $\lambda_{\theta}=r/R$ is the circumferential principal stretch, and $\lambda_{r} = \lambda^{-1}_{\theta} \lambda^{-1}_{z} $ is the radial principal stretch. Thus, the left and right Cauchy-Green tensors can be written as $\mathbf{b} = \mathbf{C} = \text{diag}\left[ 	\lambda^{-2}_{\theta} \lambda^{-2}_{z}, \lambda^{2}_{\theta}, \lambda^{2}_{z}     \right] $ in their respective eigenvector bases.

For cylindrically axisymmetric deformations and an applied radial electric field, the biasing Eulerian electric displacement vector $\mathbf{D}$ only has a radial component $D_{r}$ and the non-zero component of its Lagrangian counterpart, $\bm{\mathcal{D}}=\mathbf{F}^{-1}\mathbf{D}$, is ${\mathcal{D}}_{r} = \lambda_{\theta} \lambda_{z} D_{r}$. As a result, the five independent scalar invariants $I_{m}$ in Eq.~(\ref{five_invariants}) can be written now in the form
\begin{equation} \label{Five_independent_scalar_invariants}
	\begin{array}{l}
		I_{1} = \lambda^{-2}_{\theta} \lambda^{-2}_{z} + \lambda^{2}_{\theta} + \lambda^{2}_{z}, \quad 
		I_{2} = \lambda^{2}_{\theta} \lambda^{2}_{z} + \lambda^{-2}_{\theta} + \lambda^{-2}_{z}, \\\\
		I_{4} = \lambda^{2}_{\theta} \lambda^{2}_{z} D^2_{r},
		\quad
		I_{5} = \lambda^{-2}_{\theta} \lambda^{-2}_{z} I_{4},
		\quad
		I_{6} = \lambda^{-4}_{\theta} \lambda^{-4}_{z} I_{4},
	\end{array}
\end{equation}
which, when substituted into the initial constitutive relations (\ref{Constitutive_equations}), yields the non-zero components of the total stress tensor $\bm{\tau}$ and the Eulerian electric field vector $\mathbf{E}$ as
\begin{equation}
	\begin{array}{l} \label{Eulerian-electric_field}
		\tau_{rr} = 2 \lambda^{-2}_{\theta} \lambda^{-2}_{z} \left[  \Omega_{1} + \Omega_{2} \left(  \lambda^{2}_{\theta} + \lambda^{2} _{z}\right)  \right] + 2 \left(  \Omega_{5} + 2 \Omega_{6} \lambda^{-2}_{\theta} \lambda^{-2}_{z} \right) D^{2}_{r} - p,  
		\\\\
		\tau_{\theta \theta} = 2 \lambda^{2}_{\theta} \left[ \Omega_{1} + \Omega_{2} \left( \lambda^{-2}_{\theta} \lambda^{-2}_{z} + \lambda^{2}_{z} \right)  \right] -p, 
		\quad
		\tau_{zz} = 2 \lambda^{2}_{z} \left[ \Omega_{1} + \Omega_{2}\left(  \lambda^{-2}_{\theta} \lambda^{-2}_{z} + \lambda^{2}_{\theta} \right)  \right] - p, \\\\
		E_{r} = 2 \left(  \Omega_{4} \lambda^{2}_{\theta} \lambda^{2}_{z} + \Omega_{5} + \Omega_{6} \lambda^{-2}_{\theta} \lambda^{-2}_{z} \right) D_{r}. 
	\end{array}
\end{equation}
From Eq.~(\ref{Five_independent_scalar_invariants}), we see that the five invariants can be written in terms of three independent variables only, for instance: $\lambda_{\theta}$, $\lambda_{z}$ and $I_{4}$. 
Then we define a new reduced energy density function $\Omega^{*}(\lambda_{\theta}, \lambda_{z}, I_{4})=\Omega(I_1, I_2, I_4, I_5, I_6)$ and obtain from Eqs.~(\ref{Five_independent_scalar_invariants}) and (\ref{Eulerian-electric_field}) the following relations:
\begin{equation}\label{relations_from_Omega_star}
	\lambda_{\theta} \Omega^{*}_{\lambda_{\theta}} = \tau_{\theta \theta} - \tau_{rr}, \quad \lambda_{z} \Omega^{*}_{\lambda_{z}} = \tau_{zz} - \tau_{rr}, \quad E_{r} = 2 \lambda^{2}_{\theta} \lambda^{2}_{z} \Omega^{*}_{4} D_{r},
\end{equation}
where $\Omega^{*}_{\lambda_{\theta}}=\partial \Omega^{*}/\partial \lambda_{\theta}$,  $\Omega^{*}_{\lambda_{z}}=\partial \Omega^{*}/\partial \lambda_{z}$, and  $\Omega^{*}_4=\partial \Omega^{*}/\partial {I_4}$.

For axisymmetric deformations invariant along the axis, all the initial physical quantities depend only on $r$. In this case, Faraday's law (\ref{governing_equations})$_{3}$ is then fulfilled automatically. With the help of Eq.~(\ref{relations_from_Omega_star})$_{1}$, the equation of motion (\ref{governing_equations})$_{1}$ and  Gauss's law (\ref{governing_equations})$_{2}$ simplify to
\begin{equation}\label{blackuced_eom_and_G}
	 \dfrac{\partial \tau_{rr}}{\partial r} = \dfrac{\tau_{\theta \theta} - \tau_{rr}}{r} = \dfrac{\lambda_{\theta} \Omega^{*}_{\lambda_{\theta}}}{r}, \quad 
	 \dfrac{\partial D_{r}}{\partial r} + \dfrac{D_{r}}{r} =\dfrac{1}{r}\dfrac{\partial (r D_{r})}{\partial r}= 0,
\end{equation}
respectively. 

The inner and outer electrode-coated surfaces of the deformed tube carry equal free charges with opposite sign (i.e. $Q(a)+Q(b)=0$). The electric field vanishes in the surrounding vacuum based on Gauss's theorem and neglecting edge effects. 
Then integrating Eq.~(\ref{blackuced_eom_and_G})$_{2}$, we can obtain the solution of the radial electric displacement as
\begin{equation} \label{Component_radial_displacement}
D_{r} = \dfrac{Q(a)}{2 \pi r \lambda_{z} L} = -\dfrac{Q(b)}{2 \pi r \lambda_{z} L}.
\end{equation}
The electric field vector is curl-free, and hence an electrostatic potential $\varphi$ is introduced to write $\mathbf E = - \text{Grad} \varphi$. Inserting Eq.~(\ref{Component_radial_displacement}) into Eq.~(\ref{relations_from_Omega_star})$_{3}$ and integrating the resulting equation from $a$ to $b$ results in the electric potential difference $V=\varphi(a)-\varphi(b)$ between the inner and outer surfaces as
\begin{equation} \label{Voltage}
V = \varphi \left( a \right) - \varphi \left( b \right) = \lambda_{z} \dfrac{Q(a)}{\pi L} \int_{a}^{b} \lambda_{\theta}^{2} \Omega_{4}^{*} \dfrac{{\rm d} r}{r}.
\end{equation}

Integrating Eq.~(\ref{blackuced_eom_and_G})$_{1}$ from $a$ to $r$ and using the change of variable $\mathrm{d}r/r =
\mathrm{d}\lambda_{\theta}/[\lambda_{\theta} (1-\lambda_{\theta}^2 \lambda_z)]$ (which is derived from Eq.~(\ref{deformaion_equations}) with the definition $\lambda_{\theta}=r/R$), we obtain
\begin{equation} \label{IO_Force}
\tau_{rr}(r) - \tau_{rr}(a) =\int_{\lambda_{\theta}}^{\lambda_{a}} \dfrac{\Omega^{*}_{\lambda_{\theta}}} {\lambda_{\theta}^{2}\lambda_{z} - 1} \mathrm{d} \lambda_{\theta},
\quad \Rightarrow \quad
\tau_{rr}(b) - \tau_{rr}(a) =\int_{\lambda_{b}}^{\lambda_{a}} \dfrac{\Omega^{*}_{\lambda_{\theta}}} {\lambda_{\theta}^{2}\lambda_{z} - 1} \mathrm{d} \lambda_{\theta},
\end{equation}
where $\lambda_{a}=a/A$ and $\lambda_{b}=b/B$ are circumferential stretches of the inner and outer surfaces of the SEA tube, respectively. Here, we assume that both the inner and outer surfaces of the tube are traction-free, i.e., $\tau_{rr}\left( a \right) = \tau_{rr}\left( b \right) = 0$. Thus, Eq.~(\ref{IO_Force})$_2$ can be written as
\begin{equation} \label{I_and_O_Force}
\int_{\lambda_{b}}^{\lambda_{a}} \dfrac{\Omega^{*}_{\lambda_{\theta}}}{\lambda_{\theta}^{2}\lambda_{z} - 1} \mathrm{d} \lambda_{\theta} = 0,
\end{equation}
which establishes a general expression of nonlinear axisymmetric response of the circumferential stretch $\lambda_{a}$ to the electrical variable $V$ or $Q$ (which is included in $\Omega^*$) for an arbitrary energy function. Similarly, we can obtain from Eq.~(\ref{IO_Force})$_1$ the radial normal stress as
\begin{equation} \label{sigma_rr}
\tau_{rr} \left( r \right)  = \int_{\lambda_{\theta}}^{\lambda_{a}} \dfrac{\Omega^{*}_{\lambda_{\theta}}} {\lambda_{\theta}^{2}\lambda_{z} - 1} \mathrm{d} \lambda_{\theta}.
\end{equation}

The formulations obtained above are completely universal, valid for any isotropic SEA tube. We now specialize the preceding results to the \textit{Gent ideal dielectric model}, which is characterized by the following (reduced) total energy density function:
\begin{equation} \label{Gent_model}
	\begin{array}{l}
	\Omega(I_1, I_5) = -\dfrac{\mu {G}}{2}\ln \left[ 1 - \dfrac{I_{1} - 3}{G} \right] + \dfrac{I_{5}}{2 \varepsilon},
	\\\\
	\Omega^{*}(\lambda_{\theta}, \lambda_{z}, I_{4}) = -\dfrac{\mu {G}}{2} \ln \left[ 1- \dfrac{\lambda_{\theta}^{-2} \lambda_{z}^{-2} + \lambda_{\theta}^{2} + \lambda_{z}^{2} - 3   }{G} \right] + \dfrac{1}{2 \varepsilon} \lambda_{\theta}^{-2} \lambda_{z}^{-2} I_{4},
	\end{array}
\end{equation}
where $\mu$ is the shear modulus of the SEA tube in the absence of biasing fields, $\varepsilon$ is the permittivity of an ideal dielectric material, and the dimensionless Gent constant $G$ reflects the limiting chain extensibility of rubber networks \citep{gent1996new}, accounting for the strain-stiffening effect. Moreover, the Gent ideal dielectric model (\ref{Gent_model}) reduces to the neo-Hookean model in the limit of $G \rightarrow \infty$. 

Substituting Eq.~(\ref{Gent_model})$_2$ into Eq.~(\ref{Voltage}) and integrating, we obtain the explicit expression between the \textit{dimensionless electric voltage} $\overline{V} = V \sqrt{\varepsilon/\mu}/H $ and \textit{dimensionless surface free charge} $\overline{Q} = Q(a)/ (2\pi HL \sqrt{\mu \varepsilon}) $ as
\begin{equation} \label{voltage_charge}
	\overline{V} = - \dfrac{\overline{Q}}{\lambda_{z}} \ln \overline{\eta}.
\end{equation}
Substitution of Eq.~(\ref{Gent_model})$_2$ into Eq.~(\ref{I_and_O_Force}) gives
\begin{equation} \label{Expansion_Force}
	\int_{\lambda_{b}}^{\lambda_{a}}  
	\dfrac{G}{G-\lambda_{\theta}^{-2}\lambda_{z}^{-2} - \lambda_{\theta}^2 - \lambda_{z}^{2} + 3} \dfrac{\lambda_{\theta}\left(1 - \lambda_{\theta}^{-4} \lambda_{z}^{-2} \right) }{\lambda_{\theta}^{2} \lambda_{z} - 1} \mathrm{d} \lambda_{\theta} 
	-
	\int_{\lambda_{b}}^{\lambda_{a}} \dfrac{D_{r}^{2}}{\mu \varepsilon} \dfrac{1}{\lambda_{\theta} (\lambda_{\theta}^{2} \lambda_{z} - 1)} \mathrm{d} \lambda_{\theta}  
	=0.
\end{equation}
Making use of the relation $\mathrm{d}r/r =\mathrm{d}\lambda_{\theta}/[\lambda_{\theta} (1-\lambda_{\theta}^2 \lambda_z)]$ and Eq.~(\ref{Component_radial_displacement}), we can obtain the second integration part in Eq.~(\ref{Expansion_Force}) as
\begin{equation} \label{2nd_integration}
	\int_{\lambda_{b}}^{\lambda_{a}} \dfrac{D_{r}^{2}}{\mu \varepsilon} \dfrac{1}{\lambda_{\theta} (\lambda_{\theta}^{2} \lambda_{z} - 1)} \mathrm{d} \lambda_{\theta} 
	=
	- \dfrac{1}{2 \mu \varepsilon} \left[ \dfrac{Q(a)}{2 \pi \lambda_{z} L} \right]^{2} \left( \dfrac{1}{b^2} - \dfrac{1}{a^2} \right).  
\end{equation}
The first integration part in Eq.~(\ref{Expansion_Force}) can be derived as follows:
\begin{equation}  \label{1st_integration}
	\begin{array}{lll} 
		\int_{\lambda_{b}}^{\lambda_{a}}  
		\dfrac{G}{G-\lambda_{\theta}^{-2}\lambda_{z}^{-2} - \lambda_{\theta}^2 - \lambda_{z}^{2} + 3} \dfrac{\lambda_{\theta}\left(1 - \lambda_{\theta}^{-4} \lambda_{z}^{-2} \right) }{\lambda_{\theta}^{2} \lambda_{z} - 1} \mathrm{d} \lambda_{\theta} 
		\\\\ = 
		G \lambda_{z}^{-2} \int_{\lambda_{b}}^{\lambda_{a}} \dfrac{\lambda_{\theta}^{2} \lambda_{z} + 1}{\lambda_{\theta} \left[ \left( G - \lambda_{z}^2 + 3 \right) \lambda_{\theta}^2 - \lambda_{z}^{-2} - \lambda_{\theta}^4  \right] } \mathrm{d} \lambda_{\theta} 
		\\\\ = 
		-G \lambda_{z}^{-2} \int_{\lambda_{b}}^{\lambda_{a}} \dfrac{\lambda_{\theta}^{2} \lambda_{z} + 1}{\lambda_{\theta} \left( \lambda_{\theta}^2 - \lambda_{01} \right) \left( \lambda_{\theta}^2 - \lambda_{02} \right) } \mathrm{d} \lambda_{\theta}, 
	\end{array}
\end{equation}
where $\lambda_{0j}\left( j = 1 ,2 \right) $ are the two roots of the following quadratic polynomial equation of $\lambda_{\theta}^2$ :
\begin{equation}
	\left( G - \lambda_{z}^2 + 3 \right) \lambda_{\theta}^2 - \lambda_{z}^{-2} - \lambda_{\theta}^4 = 0.
\end{equation}
Therefore, integrating Eq.~(\ref{1st_integration}) yields
\begin{equation} \label{23}
	-G \lambda_{z}^{-2} \int_{\lambda_{b}}^{\lambda_{a}} \dfrac{\lambda_{\theta}^{2} \lambda_{z} + 1}{\lambda_{\theta} \left( \lambda_{\theta}^2 - \lambda_{01} \right) \left( \lambda_{\theta}^2 - \lambda_{02} \right) } \mathrm{d} \lambda_{\theta}
	=
	-G \lambda_{z}^{-2} \left[ F\left( \lambda_{a} \right) - F\left( \lambda_{b} \right) \right], 
\end{equation}
where
\begin{equation}
	F\left( \lambda_{\theta} \right) = \dfrac{1}{\lambda_{01} \lambda_{02}} \ln \lambda_{\theta} + \dfrac{1 + \lambda_{01} \lambda_{z}}{2 \lambda_{01} \left(\lambda_{01} - \lambda_{02} \right) } \ln \left(\lambda_{\theta}^2 - \lambda_{01} \right) - \dfrac{1 + \lambda_{02} \lambda_{z}}{2 \lambda_{02} \left( \lambda_{01} - \lambda_{02} \right) } \ln \left(\lambda_{\theta}^2 - \lambda_{02} \right). 
\end{equation}
Substituting Eqs.~(\ref{2nd_integration}), (\ref{1st_integration}) and (\ref{23}) into Eq.~(\ref{Expansion_Force}), we obtain the nonlinear axisymmetric deformation relation between $\lambda_{a}$ and $\overline{Q}$ as
\begin{equation} \label{stretch_charge}
	\overline{Q}^2 = 2 G \dfrac{\lambda_{a}^2}{1 - \overline{\eta}^2} \left[ F\left( \lambda_{b} \right) - F\left( \lambda_{a} \right) \right] \left(\dfrac{\eta}{1 - \eta}\right)^2. 
\end{equation}
By means of Eqs.~(\ref{voltage_charge}) and (\ref{stretch_charge}), we finally get the nonlinear axisymmetric response between $\lambda_{a}$ and $\overline{V}$ for the Gent ideal dielectric model as
\begin{equation} \label{nonlinear_response_Gent}
	\overline{V} = - \sqrt{2 G \lambda_{z}^{-2} \dfrac{\lambda_{a}^2}{1 - \overline{\eta}^2} \left[ F\left( \lambda_{b} \right) - F\left( \lambda_{a} \right) \right] } \dfrac{\eta}{1 - \eta} \ln \overline{\eta}.
\end{equation}

Moreover, the radially inhomogeneous biasing fields (including the circumferential stretch $\lambda_{\theta}$,
the radial electric displacement $D_r$, the radial normal stress $\tau_{rr}$, and the Lagrange multiplier $p$) needed to calculate the natural frequency of electrostatically tunable non-axisymmetric vibrations can be derived, using Eqs.~(\ref{deformaion_equations})$_1$, (\ref{Component_radial_displacement}), (\ref{sigma_rr}), and (\ref{Eulerian-electric_field})$_1$, as
\begin{equation} \label{biasing_fields}
	\begin{array}{l}
		\lambda_{\theta}=\dfrac{r}{R} = \dfrac{\xi}{\sqrt{\lambda_{z} \left[ \xi^2 - {\lambda_{a}^2 \eta^2}/{\left( 1 - \eta \right)^2 } \right] + {\eta^2}/{\left( 1 - \eta \right)^2} }},
		\quad
		\overline{D}_{r} = - \dfrac{\overline{V}}{\xi \ln \overline{\eta}},
		\\\\
		\overline{\tau}_{rr} = - G \lambda_{z}^{-2} \left[ F\left( \lambda_{a} \right) - F\left( \lambda_\theta \right) \right] + \dfrac{\overline{Q}^2}{2 \lambda_{z}^2} \left[ \dfrac{1}{\xi^2} - \dfrac{\left( 1 - \eta \right)^2 }{\lambda_{a}^2 \eta^2} \right],
		\\\\
		\overline{p} = \lambda_{\theta}^{-2} \lambda_{z}^{-2} \dfrac{G}{G - I_{1} + 3} + \overline{D}_{r}^2 - \overline{\tau}_{rr},
	\end{array}
\end{equation}
where $\xi = r/H$ is the dimensionless radial coordinate in the deformed configuration, $\overline{D}_{r} = D_{r} / \sqrt{\mu \varepsilon}$ is the \textit{dimensionless radial electric displacement}, and $\overline{\tau}_{rr}=\tau_{rr}/\mu$ and $\overline{p} = p / \mu $ are the dimensionless radial normal stress and dimensionless Lagrange multiplier, respectively. 

For the case of the neo-Hookean ideal dielectric model $\Omega = \mu (I_{1} -3)/2 + I_{5}/(2 \varepsilon)$ and after taking the limit of $G \rightarrow \infty$, Eqs.~(\ref{nonlinear_response_Gent}) and (\ref{biasing_fields}) reduce to Eqs.~(14) and (15) in the paper of \cite{zhu2020electrostatically}.

\section{Incremental fields and non-axisymmetric vibration analysis}\label{section3}

To analyze the non-axisymmetric vibrations of a finitely deformed SEA tube, we employ the linearized incremental theory of electro-elasticity \citep{dorfmann2010electroelastic, dorfmann2014nonlinear}, the main parts of which are first summarized for sake of completeness. Then, the incremental governing equations in cylindrical coordinates $\left( r, \theta, z \right)$ are recast into the state-space formalism, which we use, combined with the approximate laminate technique, to obtain the frequency equations.

\subsection{Linearized incremental theory} \label{Section3.1}

We superimpose a time-dependent infinitesimal incremental motion $\mathbf{\dot{x}}(\mathbf{X}, t)$ and an infinitesimal incremental electric displacement $\bm{\mathcal{\dot{D}}}_0$ upon a finitely deformed configuration $\mathcal{B}_0$ (with the boundary $\partial \mathcal{B}_0$ and outward unit $\mathbf{n}$). Here and henceforth, a superposed dot indicates the increment in the quantity concerned. According to the incremental theory of electro-elasticity \citep{dorfmann2010electroelastic, dorfmann2014nonlinear}, we can write the linearized incremental incompressibility condition, incremental governing equations and incremental constitutive relations for incompressible SEA materials in \textit{updated Lagrangian} form as
\begin{equation} \label{Material_incompressibility}
\mathrm{div}\hskip 1pt \mathbf{u} = \mathrm{tr}\hskip 1pt \mathbf{H} = 0,
\end{equation}
\begin{equation} \label{Lag_form_cons}
	\mathrm{div} \hskip 1pt\mathbf{\dot{T}}_0 = \rho \partial^2 \mathbf{u}/\partial \textit{t}^2, 
	\quad 
	\mathrm{div}\hskip 1pt \bm{\mathcal{\dot{D}}}_0 = 0, 
	\quad 
	\mathrm{curl} \hskip 1pt\bm{\mathbf{\dot{\cal E}}}_0=\mathbf{0},
\end{equation}
respectively, where
\begin{equation} \label{incremental_constitutive_relations}
\mathbf{\dot{T}}_0 = \bm{\mathcal{A}}_0 \mathbf{H} + \bm{\mathbf{\Gamma}}_0 \bm{\mathcal{\dot{D}}}_0 + p \mathbf{H}- \dot{p}\mathbf{I}, \quad \bm{\mathcal{\dot{E}}}_0 = \bm{\Gamma}^{\mathsf{T}}_0 \mathbf{H} + \bm{\mathcal{M}}_0 \bm{\mathcal{\dot{D}}}_0.
\end{equation}
Here, $\mathbf{u}(\mathbf{x},t)= \mathbf{\dot{x}}(\mathbf{X},t)$ is the incremental mechanical displacement vector, $\mathbf{H}= \mathrm{grad} \hskip 1pt \mathbf{u}$ is the incremental displacement gradient tensor, $\dot{p}$ is the incremental Lagrange multiplier, and $\mathbf{\dot{T}}_0$, $\bm{\mathcal{\dot{D}}}_0$, and $\bm{\mathbf{\dot{\cal E}}}_0$ are the \textit{push-forward} counterparts of the increments of the total nominal stress and of the Lagrangian electric displacement and electric field, respectively. The subscript `$0$' is used to indicate the resultant push-forward variables. 

In Eq.~(\ref{incremental_constitutive_relations}), $\bm{\mathcal{A}}_0$, $\bm{\mathbf{\Gamma}}_0$ and $\bm{\mathcal{M}}_0$ are fourth-, third- and second-order tensors, respectively, which are referred to as the \textit{instantaneous} electro-elastic moduli tensors. Their component forms satisfy
\begin{equation}
	\begin{array}{l}
		{\mathcal{A}}_{0piqj} =  F_{p \alpha} F_{q \beta} \mathcal{A}_{\alpha i \beta j} = {\mathcal{A}}_{0qjpi}, 
		\quad 
		{\Gamma}_{0piq} = F_{p \alpha} F^{-1}_{\beta q} \Gamma_{\alpha i \beta} = {\Gamma}_{0 i p q},
		\\\\ 
		{\mathcal{M}}_{0ij} = F^{-1}_{\alpha i} F^{-1}_{\beta j} \mathcal{M}_{\alpha \beta} = {M}_{0ji}, 
	\end{array}
\end{equation}
where $\bm{\mathcal{A}}$, $\mathbf{\Gamma}$ and $\bm{\mathcal{M}}$ are the \textit{referential} electro-elastic moduli tensors related to the total energy function $\bm{\Omega}(\mathbf{F}, \bm{\mathcal{D}})$, with components
\begin{equation}
	\begin{array}{l}
		{\mathcal{A}}_{\alpha i \beta j} = \dfrac{\partial^2 \Omega}{\partial F_{i \alpha} \partial F_{j \beta}}, 
		\quad 
		{{\Gamma}}_{\alpha i \beta} = \dfrac{\partial^2 \Omega}{\partial F_{i \alpha} \partial \mathcal{D}_{\beta}}, 
		\quad 
		{\mathcal{M}}_{\alpha \beta} = \dfrac{\partial^2 \Omega}{\partial \mathcal{D}_{\alpha} \partial \mathcal{D}_{\beta}}.
	\end{array}
\end{equation}

With no external fields in the surrounding vacuum, the updated Lagrangian forms of the incremental mechanical and electric boundary conditions satisfied on $\partial \mathcal{B}_0$ take the following simple forms,
\begin{equation}\label{incremental boundary conditions}
	\dot{\mathbf{T}}^{\mathsf{T}}_{0} \mathbf{n} = \dot{\mathbf{t}}^{A}_{0}, \quad \dot{\bm{\mathcal{E}}}_0 \times \mathbf{n} = \mathbf{0}, \quad \dot{\bm{\mathcal{D}}}_0 \cdot \mathbf{n} = -\dot{\sigma}_{\mathrm{F}0},
\end{equation}
where $\dot{\mathbf{t}}^{A}_{0}$ is the updated Lagrangian incremental traction vector per unit area of $\partial \bm{\mathcal{B}}_0$ and $\dot{\sigma}_{\mathrm{F}0}$ is the incremental surface charge density on $\partial \bm{\mathcal{B}_0}$.

\subsection{Incremental equations and state-space formalism in cylindrical coordinates}

In this subsection, the incremental governing equations (\ref{Material_incompressibility})-(\ref{incremental_constitutive_relations}) are first specialized to the cylindrical coordinates $\left(r, \theta, z \right) $. The basic incremental governing equations of the deformed SEA tube are the incremental incompressibility constraint,
\begin{equation} \label{incremental_incompressible}
\dfrac{\partial u_{r}}{\partial r} + \dfrac{1}{r} \left( \dfrac{\partial u_{\theta}}{\partial \theta} + u_{r} \right) + \dfrac{\partial u_{z}}{\partial z} = 0,
\end{equation}
together with the incremental equations of motion and incremental Gauss's law,
\begin{equation} \label{incremental_Kinematics_cons}
\begin{array}{l}
\dfrac{\partial \dot{T}_{0rr}}{\partial r} + \dfrac{1}{r} \dfrac{\partial \dot{T}_{0 \theta r}}{\partial \theta} + \dfrac{\dot{T}_{0rr} - \dot{T}_{0 \theta \theta}}{r} + \dfrac{\partial \dot{T}_{0zr}}{\partial z} = \rho \dfrac{\partial^2 u_{r}}{\partial t^2},
\\\\
\dfrac{\partial \dot{T}_{0r \theta}}{\partial r} + \dfrac{1}{r} \dfrac{\partial \dot{T}_{0 \theta \theta}}{\partial \theta} + \dfrac{\dot{T}_{0 \theta r} + \dot{T}_{0 r \theta}}{r} + \dfrac{\partial \dot{T}_{0z \theta}}{\partial z} = \rho \dfrac{\partial^2 u_{\theta}}{\partial t^2},
\\\\
\dfrac{\partial \dot{T}_{0rz}}{\partial r} + \dfrac{1}{r} \dfrac{\partial \dot{T}_{0 \theta z}}{\partial \theta} + \dfrac{\dot{T}_{0rz}}{r} + \dfrac{\partial \dot{T}_{0zz}}{\partial z} = \rho \dfrac{\partial^2 u_{z}}{\partial t^2},
\end{array} 
\end{equation}
\begin{equation} \label{incremental_electro_cons}
\dfrac{\partial \mathcal{\dot{D}}_{0r}}{\partial r} + \dfrac{1}{r} \left( \dfrac{\partial \mathcal{\dot{D}}_{0 \theta}}{\partial \theta} + {\mathcal{\dot{D}}}_{0r} \right) + \dfrac{\partial \mathcal{\dot{D}}_{0z}}{\partial z} = 0,
\end{equation}
and the incremental constitutive equations,
\begin{equation} \label{expansion_incre_cons}
\begin{array}{l}
\dot{T}_{0rr} = c_{11} \dfrac{\partial u_{r}}{\partial r} + c_{12} \dfrac{1}{r}\left( \dfrac{\partial u_{\theta}}{\partial \theta} + u_{r} \right) + c_{13} \dfrac{\partial u_{z}}{\partial z} + e_{11} \dfrac{\partial \dot{\varphi}}{\partial r} - \dot{p},
\\\\
\dot{T}_{0 \theta \theta} = c_{12} \dfrac{\partial u_{r}}{\partial r} + c_{22} \dfrac{1}{r}\left( \dfrac{\partial u_{\theta}}{\partial \theta} + u_{r} \right) + c_{23} \dfrac{\partial u_{z}}{\partial z} + e_{12} \dfrac{\partial \dot{\varphi}}{\partial r} - \dot{p},
\\\\
\dot{T}_{0zz} = c_{13} \dfrac{\partial u_{r}}{\partial r} + c_{23} \dfrac{1}{r}\left( \dfrac{\partial u_{\theta}}{\partial \theta} + u_{r} \right) + c_{33} \dfrac{\partial u_{z}}{\partial z} + e_{13} \dfrac{\partial \dot{\varphi}}{\partial r} - \dot{p},
\\\\
\dot{T}_{0rz} = c_{58} \dfrac{\partial u_{r}}{\partial z} + c_{55} \dfrac{\partial u_{z}}{\partial r} + e_{35} \dfrac{\partial \dot{\varphi}}{\partial z}, 
\qquad
\dot{T}_{0zr} = c_{88} \dfrac{\partial u_{r}}{\partial z} + c_{58} \dfrac{\partial u_{z}}{\partial r} + e_{35} \dfrac{\partial \dot{\varphi}}{\partial z},
\\\\
\dot{T}_{0 \theta z} = c_{44} \dfrac{1}{r} \dfrac{\partial u_{z}}{\partial \theta} + c_{47} \dfrac{\partial u_{\theta}}{\partial z},
\qquad
\dot{T}_{0 z \theta} = c_{77} \dfrac{\partial u_{\theta}}{\partial z} + c_{47} \dfrac{1}{r} \dfrac{\partial u_{z}}{\partial \theta},
\\\\
\dot{T}_{0 r \theta} = c_{66} \dfrac{\partial u_{\theta}}{\partial r} + c_{69} \dfrac{1}{r} \left( \dfrac{\partial u_{r}}{\partial \theta} - u_{\theta} \right) + e_{26} \dfrac{1}{r} \dfrac{\partial \dot{\varphi}}{\partial \theta},
\\\\
\dot{T}_{0 \theta r} = c_{99} \dfrac{1}{r} \left( \dfrac{\partial u_{r}}{\partial \theta} - u_{\theta} \right) + c_{69} \dfrac{\partial u_{\theta}}{\partial r} + e_{26} \dfrac{1}{r} \dfrac{\partial \dot{\varphi}}{\partial \theta},
\\\\
{\mathcal{\dot{D}}}_{0r} = e_{11} \dfrac{\partial u_{r}}{\partial r} + e_{12} \dfrac{1}{r} \left( \dfrac{\partial u_{\theta}}{\partial \theta} + u_{r} \right) + e_{13} \dfrac{\partial u_{z}}{\partial z} - \varepsilon_{11} \dfrac{\partial \dot{\varphi}}{\partial r},
\\\\
{\mathcal{\dot{D}}}_{0 \theta} = e_{26} \left[ \dfrac{1}{r}\left( \dfrac{\partial u_{r}}{\partial \theta} - u_{\theta} \right) + \dfrac{\partial u_{\theta}}{\partial r}  \right] - \varepsilon_{22} \dfrac{1}{r} \dfrac{\partial \dot{\varphi}}{\partial \theta},
\\\\
{\mathcal{\dot{D}}}_{0 z} = e_{35} \left( \dfrac{\partial u_{z}}{\partial r} + \dfrac{\partial u_{r}}{\partial z} \right) - \varepsilon_{33} \dfrac{\partial \dot{\varphi}}{\partial z}. 
\end{array}
\end{equation}
where $c_{ij}$, $e_{ij}$ and $\varepsilon_{ij}$ are the effective material parameters associated with the instantaneous electro-elastic moduli $\bm{\mathcal{A}}_0$, $\bm{\mathbf{\Gamma}}_0$ and $\bm{\mathcal{M}}_0$ (their explicit expressions are provided in Eq.~(41) in the paper by \cite{wu2017guided}). In the process of deriving Eq.~(\ref{expansion_incre_cons}), the incremental displacement gradient tensor $\mathbf{H}$ was specialized to the cylindrical coordinates. In addition, an incremental electric potential $\dot{\varphi}$, defined by $\bm{\mathbf{\dot{\cal E}}}_0=-\text{grad} \dot{\varphi}$, was introduced so that the incremental Faraday law (\ref{Lag_form_cons})$_3$ is satisfied automatically.
%
%

Note that the non-zero components of the instantaneous electro-elastic moduli tensors ${\mathcal{A}}_{0 p i q j}$, ${{\Gamma}}_{0 p i q}$ and ${\mathcal{M}}_{0 i j}$ were derived by \cite{wu2017guided} for the nonlinear axisymmetric deformations of incompressible isotropic SEA tubes subject to a combination of radial electric displacement and axial pre-stretch (see their \emph{Appendix B} for specific expressions). It is worth emphasizing that instantaneous physical properties of the SEA tubes are significantly sensitive to applied electro-mechanical biasing fields, which generate remarkable knock-on influences on the dynamic characteristics of superimposed small-amplitude motions, as demonstrated below.

As discussed in Sec.~\ref{2_2}, the physical quantities for cylindrically axisymmetric deformations in the SEA tube subject to the radial electric voltage are radially inhomogeneous, leading to the $r$-dependence of the instantaneous electro-elastic moduli. Thus, the resultant incremental displacement equations are, in general, a set of coupled partial differential equations with variable coefficients, which are likely intractable to solve analytically and difficult to solve numerically. 
Here we employ the state-space method (SSM) \citep{wu2017guided, zhu2020electrostatically}, combining the incremental state-space formalism with the approximate laminate technique, to obtain the frequency equations.
Following a standard derivation procedure (omitted here for simplicity), we transform the original incremental governing equations (\ref{incremental_incompressible})-(\ref{expansion_incre_cons}) into the following \emph{incremental state equation}:
\begin{equation} \label{state_equation}
	\dfrac{\partial \mathbf{S}}{\partial r} = \mathbf{M}\left(\partial_\theta,\partial_z, \partial_t; \; \mathcal{A}_{0 p i q j}, {\Gamma}_{0 p i q}, \mathcal{M}_{0 i j}, \rho; \; r \right) \mathbf{S}, 
\end{equation}
which is a first-order system of differential equations with respect to $r$, where $\partial_\gamma=\partial/\partial_\gamma \; (\gamma=\theta, z, t)$, $\mathbf{S} = \left[ u_{r}, u_{\theta}, u_{z}, \dot{\varphi}, \dot{T}_{0rr}, \dot{T}_{0r \theta}, \dot{T}_{0rz}, {\mathcal{\dot{D}}}_{0r}\right]^{\mathsf{T}}$ is the incremental state vector, with $ u_{r}$, $u_{\theta}$, $u_{z}$, $\dot{\varphi}$, $\dot{T}_{0rr}$, $\dot{T}_{0r \theta}$, $\dot{T}_{0rz}$ and $ {\mathcal{\dot{D}}}_{0r}$ being the state variables, and $\mathbf{M}$ is an 8 $\times$ 8 system matrix, which depends on the  radial coordinate, the instantaneous electro-elastic moduli, and partial derivatives with respect to $\theta$, $z$ and $t$. The specific expressions for the elements of the system matrix $\mathbf{M}$ can be found in the \emph{Appendix C} in the paper of \cite{wu2017guided}; they are not reproduced here for brevity. 
We emphasize that the state equation (\ref{state_equation}) is applicable for any form of energy function. 

\subsection{Approximate laminate technique}

We assume that generalized rigidly supported conditions \citep{ding2001three} are applied to the two ends of the deformed SEA tube, so that the end cross-sections are in smooth contact with rigid plattens at both ends. In addition, the electric inductions in the surrounding vacuum near the tube ends can be neglected, so that the incremental electric displacement at both ends is zero. Therefore, the incremental mechanical and
electric boundary conditions are
\begin{equation} \label{incremental_BC}
	u_{z} = \dot{T}_{0zr} = \dot{T}_{0 z \theta}= \mathcal{\dot{D}}_{0 z} = 0, \quad \left( z = 0, l \right). 
\end{equation}

To satisfy the incremental boundary conditions (\ref{incremental_BC}), we assume the following formal solutions for the harmonic non-axisymmetric free vibrations of the deformed SEA tube:
\begin{equation} \label{Assumed_solution}
	\mathbf{S} = \left[ 
	\begin{array}{c}
		u_{r} 
		\\
		u_{\theta}  
		\\
		u_{z}
		\\
		\dot{\varphi}
		\\ 
		\dot{T}_{0rr}
		\\
		\dot{T}_{0r \theta}
		\\ 
		\dot{T}_{0rz}
		\\
		\mathcal{\dot{D}}_{0r}
	\end{array} \right]
               = \left[ 
     \begin{array}{c}
     	H U_{r} \left( \xi \right) \cos \left( m \theta \right) \cos \left( n \pi \zeta \right)  
     	\\
     	H U_{\theta} \left( \xi \right) \sin \left( m \theta \right) \cos \left( n \pi \zeta \right)  
     	\\
     	H U_{z} \left( \xi \right) \cos \left( m \theta \right) \sin \left( n \pi \zeta \right)
     	\\
     	H \sqrt{\mu / \varepsilon} \mit\Phi \left( \xi \right) \cos \left( m \theta \right) \cos \left( n \pi \zeta \right)  
     	\\ 
     	\mu \mit{\Sigma}_{0rr} \left( \xi \right) \cos \left( m \theta \right) \cos \left( n \pi \zeta \right) 
     	\\
     	\mu \mit\Sigma_{0r \theta} \left( \xi \right) \sin \left( m \theta \right) \cos \left( n \pi \zeta \right)
     	\\ 
     	\mu \mit\Sigma_{0rz} \left( \xi \right) \cos \left( m \theta \right) \sin \left( n \pi \zeta \right) 
     	\\
     	\sqrt{\mu \varepsilon} \mit\Delta_{0r} \left( \xi \right) \cos \left( m \theta \right) \cos \left( n \pi \zeta \right) 
     \end{array} \right] e^{\rm i \omega t},
\end{equation}
where $\xi = r/H$ and $\zeta = z/l$ are the dimensionless radial and axial coordinates, respectively; $m$ and $n$ denote the circumferential mode number and axial mode number, respectively; $\rm i$ = $\sqrt{-1}$ is the imaginary unit, and $\omega$ is the circular frequency of free vibrations. Note that all the unknown functions ($U_{r} \left( \xi \right), U_{\theta} \left( \xi \right)$, etc.) in Eq.~(\ref{Assumed_solution}) are dimensionless. Then, inserting Eq.~(\ref{Assumed_solution}) into the state equation (\ref{state_equation}) and non-dimensionalizing related variables, we obtain the dimensionless form of the incremental state equation as
\begin{equation} \label{dimensionless_state_equation}
	\dfrac{\rm d \overline{\mathbf{S}} \left( \xi \right)}{\rm d \xi} 
	= 
	\overline{\mathbf{M}} \left( \xi \right) \overline{\mathbf{S}} \left( \xi \right),
\end{equation}
where $\overline{\mathbf{S}} = \left[ U_{r}, U_{\theta}, U_{z}, \mit\Phi, \mit\Sigma_{0rr}, \mit\Sigma_{0r \theta}, \mit\Sigma_{0rz}, \mit\Delta_{0r}  \right]^\mathsf{T} $ is the dimensionless incremental state vector and $\overline{\mathbf{M}}$ is the 8 $\times$ 8 dimensionless system matrix, with its four partitioned $4 \times 4$ sub-matrices $\mathbf{\overline{M}}_{ij} \, (i,j=1,2)$ given by
\begin{equation*}
\begin{array}{l}
\mathbf{\overline{M}}_{11} 
= \left[ \begin{array}{cccc}
- \dfrac{1}{\xi}   &  - \dfrac{m}{\xi}  &  - \alpha  &  0 
\\\\
\beta_{1} \dfrac{m}{\xi}  & \dfrac{\beta_{1}}{\xi}  &  0  &  \beta_{2} \dfrac{m}{\xi} 
\\\\
\beta_{3} \alpha  &  0  &  0  & \beta_{4} \alpha 
\\\\
\dfrac{\overline{q}_{1}}{\xi}  &  \overline{q}_{1} \dfrac{m}{\xi}  & \overline{q}_{2} \alpha  &  0
\end{array} \right], 
\quad
\mathbf{\overline{M}}_{12} = \left[ \begin{array}{cccc}
0  &  0  &  0  &  0 
\\\\
0  &  \dfrac{1}{\overline{c}_{66}}  &  0  &  0 
\\\\
0  &  0  &  \dfrac{1}{\overline{c}_{55}}  &  0 
\\\\
0  &  0  &  0  &  - \dfrac{1}{\overline{\varepsilon}_{11}} 
\end{array} \right],
\end{array}
\end{equation*}
\begin{small}  
	\begin{equation} \label{elements_M}
	\begin{array}{l}
	\mathbf{\overline{M}}_{21} =
	\\\\ 
	\left[ \begin{array}{cccc}
	\overline{q}_{7} \dfrac{m^2}{\xi^2} + \dfrac{\overline{q}_{3}}{\xi^2} + \overline{q}_{8} \alpha^2 - \varpi^2  &  \dfrac{m \left( \overline{q}_{3} + \overline{q}_{7} \right) }{\xi^2}  &  \overline{q}_{4} \dfrac{\alpha}{\xi}  &\overline{q}_{9} \dfrac{m^2}{\xi^2} + \overline{q}_{10} \alpha^2   
	\\\\
	\left(\overline{q}_{3} + \overline{q}_{7} \right) \dfrac{m}{\xi^2}  & \overline{q}_{3} \dfrac{m^2}{\xi^2} + \dfrac{\overline{q}_{7}}{\xi^2} + \overline{c}_{77} \alpha^2 - \varpi^2  &  \left( \overline{q}_{4} + \overline{c}_{47} \right) \dfrac{m}{\xi} \alpha  &  \overline{q}_{9} \dfrac{m}{\xi^2} 
	\\\\
	\overline{q}_{5} \dfrac{\alpha}{\xi}  &  \left( \overline{c}_{47} + \overline{q}_{5} \right) \dfrac{m}{\xi} \alpha  &  \overline{q}_{6} \alpha^2 + \overline{c}_{44} \dfrac{m^2}{\xi^2} - \varpi^2  & 0 
	\\\\
	\overline{q}_{9} \dfrac{m^2}{\xi^2} + \overline{q}_{10} \alpha^2  &  \overline{q}_{9} \dfrac{m}{\xi^2}  & 0 &  -  \overline{q}_{11} \dfrac{m^2}{\xi^2} - \overline{q}_{12} \alpha^2
	\end{array} \right],
	\end{array}
	\end{equation}
\end{small}
\begin{equation*}
\mathbf{\overline{M}}_{22} = \left[ \begin{array}{cccc}
0  &  - \beta_{1} \dfrac{m}{\xi}  &  -\beta_{3} \alpha  &  - \dfrac{\overline{q}_{1}}{\xi}
\\\\
\dfrac{m}{\xi}  &  - \dfrac{\beta_{1} + 1}{\xi}  & 0 &  -\overline{q}_{1} \dfrac{m}{\xi}
\\\\
\alpha  &  0  &  -\dfrac{ 1 }{\xi}  &  - \overline{q}_{2} \alpha
\\\\
0  &  -\beta_{2} \dfrac{m}{\xi}  &  - \beta_{4} \alpha  &  -\dfrac{1}{\xi} 
\end{array} \right],
\end{equation*}
in which the dimensionless quantities are defined as
\begin{equation} \label{di_quantity}
\begin{array}{l}
\alpha = n \pi H/l = n \pi H/(\lambda_z L), \quad \overline{c}_{ij} = c_{ij}/\mu, \quad \overline{e}_{ij} = e_{ij}/\sqrt{\mu \varepsilon}, \quad \overline{\varepsilon}_{ij} = \varepsilon_{ij}/\varepsilon,
\\\\ 
\beta_{1} = \overline{c}_{69}/\overline{c}_{66}, 
\quad 
\beta_{2} = \overline{e}_{26}/\overline{c}_{66}, 
\quad 
\beta_{3} = \overline{c}_{58}/\overline{c}_{55},
\quad
\beta_{4} = \overline{e}_{35}/\overline{c}_{55}, 
\quad
\overline{q}_{{j}} = q_{{j}} \sqrt{\varepsilon/ \mu} \enspace (j=1,2), 
\\\\
\overline{q}_{{j}} = q_{{j}}/\mu \enspace (j=3 \sim 8),
\quad
\overline{q}_{{j}} = q_{{j}}/ \sqrt{\mu \varepsilon} \enspace (j=9, 10),
\quad
\overline{q}_{{j}} = q_{{j}}/\varepsilon \enspace (j=11, 12),
\end{array}
\end{equation}
and $\varpi = \omega H/\sqrt{\mu/\rho}$ is the dimensionless circular frequency.

Note that different combinations of mode numbers $m$ and $n$ result in different types of vibrations. It is obvious from Eqs.~(\ref{dimensionless_state_equation}) and (\ref{elements_M}) that the eight unknown incremental state variables $u_{r}, u_{\theta}, u_{z},$ $\dot{\varphi}, \dot{T}_{0rr}, \dot{T}_{0r \theta}, \dot{T}_{0rz}, {\mathcal{\dot{D}}}_{0r}$ are fully coupled for non-axisymmetric vibrations with $m\neq0$ and $n\neq0$ (see Fig.~\ref{Physical_description}(c)). 

For the incremental axisymmetric vibrations with $m=0$ and $n\neq0$, the incremental fields are independent of the $\theta$ coordinate (i.e.~$\partial/\partial \theta=0$) and the state equation (\ref{dimensionless_state_equation}) reduces to two uncoupled classes of incremental axisymmetric vibrations: the \emph{axisymmetric longitudinal vibrations (L vibrations)} where the mechanical displacement components $u_{r}$ and $u_{z}$ coupled with the incremental electrical quantities are non-zero (see Fig.~\ref{Physical_description}(d)); and the \emph{purely torsional vibrations (T vibrations)} with the sole displacement component $u_{\theta}$ (see Fig.~\ref{Physical_description}(e)). 
The \emph{cylindrical breathing mode} with $m=n=0$, characterized by the sole radial displacement component $u_r$,  is a special mode of the L vibrations. The state equations governing the incremental axisymmetric vibrations of deformed SEA tubes were obtained by \cite{zhu2020electrostatically} (see their Eqs.~(30), (31), (41) and (42)) and are omitted here for brevity.

Additionally, we consider incremental fields independent of the $z$ coordinate (i.e.~$m\neq0$ and $n=0$), so that $\partial/\partial z=0$ holds. To satisfy the incremental boundary conditions (\ref{incremental_BC}), the harmonic solutions for the incremental vibrations independent of $z$ are assumed as
\begin{equation} \label{Assumed_solution_Pris}
\mathbf{S}_{\mathrm{P}} = \left[ 
\begin{array}{c}
u_{r} 
\\
u_{\theta}
\\
\dot{\varphi}
\\ 
\dot{T}_{0rr}
\\
\dot{T}_{0r \theta}
\\
\mathcal{\dot{D}}_{0r}
\end{array} \right]
= \left[ 
\begin{array}{c}
H U_{r} \left( \xi \right) \cos \left( m \theta \right)  
\\
H U_{\theta} \left( \xi \right) \sin \left( m \theta \right) 
\\
H \sqrt{\mu / \varepsilon} \mit\Phi \left( \xi \right) \cos \left( m \theta \right)  
\\ 
\mu \mit\Sigma_{0rr} \left( \xi \right) \cos \left( m \theta \right)
\\
\mu \mit\Sigma_{0r \theta} \left( \xi \right) \sin \left( m \theta \right)
\\
\sqrt{\mu \varepsilon} \mit\Delta_{0r} \left( \xi \right) \cos \left( m \theta \right)
\end{array} \right] e^{\rm i \omega t},
\end{equation}
and hence, the incremental state equation (\ref{dimensionless_state_equation}) specializes to
\begin{equation} \label{PAS_state_eq}
	\dfrac{\mathrm{ d \overline{\mathbf{S}}_{P}}\left( \xi \right)}{\rm d \xi} 
	= 
	\mathrm{\overline{\mathbf{M}}_{P} }\left( \xi \right) \mathrm{\overline{\mathbf{S}}_{P}} \left( \xi \right),
\end{equation}
where $\mathrm{\overline{\mathbf{S}}_{P}}= \left[ U_{r}, U_{\theta}, \mit\Phi, \mit\Sigma_{0rr}, \mit\Sigma_{0r \theta}, \mit\Delta_{0r}  \right]^\mathsf{T} $ is the dimensionless incremental state vector, and the $6 \times 6$ dimensionless system matrix $\mathrm{\overline{\mathbf{M}}_{P}}$ is obtained from Eq.~(\ref{elements_M}) as
\begin{equation} \label{PAS_matrix}
		\mathrm{\overline{\mathbf{M}}_{P}}
		= 
		\left[ \begin{array}{cccccc}
			- \dfrac{1}{\xi}   &  - \dfrac{m}{\xi}   &  0   &  0  &  0  &  0   
			\\\\
			\beta_{1} \dfrac{m}{\xi}  & \dfrac{\beta_{1}}{\xi}  &  \beta_{2} \dfrac{m}{\xi}  &  0  &  \dfrac{1}{\overline{c}_{66}}  &  0
			\\\\
			\dfrac{	\overline{q}_{1}}{\xi} & \overline{q}_{1} \dfrac{m}{\xi } & 0  
			&  0  &  0  &  -\dfrac{1}{\overline{\varepsilon}_{11}} 
			\\\\
			\overline{q}_{7} \dfrac{m^2}{\xi^2} + \overline{q}_{3} \dfrac{1}{\xi^2} - \varpi^2 &  \left( \overline{q}_{3} + \overline{q}_{7} \right) \dfrac{m}{\xi^2}   &  \overline{q}_{9} \dfrac{m^2}{\xi^2}  &  0  &  -\beta_{1} \dfrac{m}{\xi}  &  -\dfrac{\overline{q}_{1}}{\xi}
			\\\\
			\left( \overline{q}_{3} + \overline{q}_{7} \right) \dfrac{m}{\xi^2}   & \overline{q}_{3} \dfrac{m^2}{\xi^2} +  \dfrac{\overline{q}_{7}}{\xi^2} - \varpi^2   &  \overline{q}_{9} \dfrac{m}{\xi^2}  &     	\dfrac{m}{\xi}  &  -\dfrac{\beta_{1}+1}{\xi}  &  - \overline{q}_{1} \dfrac{m}{\xi} 
			\\\\
			\overline{q}_{9} \dfrac{m^2}{\xi^2}    &  \overline{q}_{9} \dfrac{m}{\xi^2}    &  -\overline{q}_{11} \dfrac{m^2}{\xi^2}  &  0  &  - \beta_{2} \dfrac{m}{\xi}  &  -\dfrac{1}{\xi}
		\end{array} \right].   
\end{equation}

Therefore, the incremental vibrations independent of $z$ ($m\neq0$ and $n=0$) are described by Eqs.~(\ref{Assumed_solution_Pris})-(\ref{PAS_matrix}), which are identified as the \emph{prismatic vibrations}, where the non-zero mechanical displacement components $u_{r}$ and $u_{\theta}$ are coupled with the incremental electrical quantities and the SEA tube remains prismatic while its cross-section loses its circular shape (see Fig.~\ref{Physical_description}(f)). Furthermore, prismatic vibrations do not depend on the length-to-thickness ratio $(L/H)$. Note that the nomenclature of \emph{prismatic vibrations} corresponds to the elastic counterpart of prismatic diffuse modes in the realm of instability \citep{haughton1979bifurcation2, bortot2018prismatic}. 
When $m=0$, Eqs.~(\ref{Assumed_solution_Pris})-(\ref{PAS_matrix}) governing the prismatic vibrations reduce to those of the cylindric breathing mode. 

Now we adopt the approximate laminate technique \citep{fan1992exact, chen2002free} and divide the tube into a laminate with $N$ equal thin sublayers. The thickness of each sublayer is $h/N$, which is sufficiently small for the system matrices $\overline{\mathbf{M}}$ and $\mathrm{\overline{\mathbf{M}}_{P}}$ within each sublayer to be approximately regarded as constant. Without loss of generality, the dimensionless deformed radial coordinate itself and the effective material parameters take the values at the mid-plane of each sublayer. 

Here, we use $r_{j0}=a+(j-1)h/N$, $r_{j1}= a+jh/N$, and  $r_{jm}= a+(2j-1)h/(2N)$ to denote the deformed radial coordinates of the inner, outer and middle surfaces of the $j$-th sublayer. Their dimensionless forms of the deformed radial coordinate are
\begin{equation}
\begin{array}{l}
		\xi_{j0}=\dfrac{r_{j0}}{H} = \dfrac{\lambda_{a} \eta}{1-\eta} + \left( j-1 \right) \dfrac{\lambda_{b}(1-\overline{\eta})}{N(1-\eta)}, 
		\quad
		\xi_{j1}=\dfrac{r_{j1}}{H} = \dfrac{\lambda_{a} \eta}{1-\eta} + 
		j \dfrac{\lambda_{b}(1-\overline{\eta})}{N(1-\eta)},
		\\\\
		\xi_{jm}= \dfrac{\lambda_{a} \eta}{1-\eta} + \left( 2j-1 \right) \dfrac{\lambda_{b}(1-\overline{\eta})}{2N(1-\eta)}.
\end{array}
\end{equation}
respectively. Applied to each sublayer, the incremental state equation (\ref{dimensionless_state_equation}) becomes
\begin{equation} \label{State_equation_for_each_layer}
	\dfrac{\rm d \overline{\mathbf{S}}\left( \xi \right)}{\rm d \xi} 
	= 
	\overline{\mathbf{M}}_{j} \left( \xi_{jm} \right) \overline{\mathbf{S}} \left( \xi \right), \quad (j=1, 2, 3, \cdots N)
\end{equation}
where $\mathbf{\overline{M}}_{j}\left( \xi_{jm} \right) $ is the approximated constant system matrix within the $j$-th thin sublayer, obtained by setting $\xi = \xi_{jm}$. Consequently, the formal solution to Eq.~(\ref{State_equation_for_each_layer}) in the $j$-th sublayer is
\begin{equation}
	\begin{array}{l}
			\mathbf{\overline{S}}\left( \xi \right) 
		=
		\mathrm{exp}\left[ \left( \xi - \xi_{j0} \right) \mathbf{\overline{M}}_{j} \left( \xi_{jm} \right)   \right]  \mathbf{\overline{S}}\left( \xi_{j0} \right),
		\\\\ 
	    \left(\xi_{j0} \leq \xi \leq \xi_{j1}; \quad j=1,2,3, \cdots N \right), 
	\end{array}
\end{equation}
which yields the transfer relation between the incremental state vectors at the inner and outer surfaces of the $j$-th sublayer as
\begin{equation} \label{transfer_relation}
	\mathbf{\overline{S}} \left( \xi_{j1} \right) 
	=
	\mathrm{exp} \left[ \dfrac{\lambda_{b}(1-\overline{\eta})}{N(1-\eta)} \mathbf{\overline{M}}_{j}\left( \xi_{jm} \right)  \right] \mathbf{\overline{S}}\left( \xi_{j0} \right).   
\end{equation}

Making use of the continuity conditions at the fictitious interfaces between equally divided sublayers that require the eight state variables be continuous, we can derive from Eq.~(\ref{transfer_relation}) the following transfer relation connecting the incremental state vectors $\overline{\mathbf{S}}^{ou}$ and $\overline{\mathbf{S}}^{in}$ at the outer and inner surfaces of the deformed SEA tube:
\begin{equation} \label{transfer_relation_INOU}
	\overline{\mathbf{S}}^{ou}
	=
	\mathbf{Z} \hskip 1pt \overline{\mathbf{S}}^{in}
	=
	\Pi^{1}_{j=N}  \mathrm{exp}\left[ \dfrac{\lambda_{b}(1-\overline{\eta})}{N(1-\eta)} \mathbf{\overline{M}}_{j}\left( \xi_{jm} \right) \right] \overline{\mathbf{S}}^{in},
\end{equation}
where $\mathbf{Z}$ is the global transfer matrix. Similar derivations are applicable to the case of the prismatic vibrations described by $\mathrm{\overline{\mathbf{S}}_{P}}$ and $\mathrm{\overline{\mathbf{M}}_{P}}$.

\subsection{Frequency equations of non-axisymmetric vibrations}
To proceed, we assume that the inner and outer surfaces of the SEA tube are traction-free and that the applied electric voltage between these two surfaces is kept fixed. Thus, the incremental mechanical and electric boundary conditions (\ref{incremental boundary conditions}) are:
\begin{equation} \label{new_incremental_boundary_conditions}
	\dot{\varphi}^{in} = \dot{\varphi}^{ou} =0,
	\quad
	\dot{T}_{0rr}^{in} = \dot{T}_{0r \theta}^{in} = \dot{T}_{0rz}^{in} = \dot{T}_{0rr}^{ou} = \dot{T}_{0r \theta}^{ou} = \dot{T}_{0rz}^{ou} = 0.
\end{equation}

Combining Eq.~(\ref{new_incremental_boundary_conditions}) with Eq.~(\ref{transfer_relation_INOU}) results in a set of independent linear algebraic equations:
\begin{equation}  \label{coefficient_Matrix}
	\left[ 
	\begin{array}{cccc}
		Z_{41} & Z_{42} & Z_{43} & Z_{48}
		\\
		Z_{51} & Z_{52} & Z_{53} & Z_{58}  
		\\
		Z_{61} & Z_{62} & Z_{63} & Z_{68}
		\\
		Z_{71} & Z_{72} & Z_{73} & Z_{78}
	\end{array} \right]
    \left[
	\begin{array}{c}
	    U_{r}^{\mathrm{in}} 
	    \\
    	U_{\theta}^{\mathrm{in}}
     	\\
    	U_{z}^{\mathrm{in}} 
    	\\
    	\mit\Delta_{0r}^{\mathrm{in}} 
   \end{array} \right]
      = 
    \left[
   \begin{array}{c}
	   0
       \\
       0
       \\
       0 
       \\
       0 
   \end{array} \right],
\end{equation}
where $Z_{ij}$ denote the elements of the global transfer matrix $\mathbf{Z}$. For non-trivial solutions to exist, the determinant of the coefficient matrix in Eq.~(\ref{coefficient_Matrix}) must vanish. Thus, we have
\begin{equation} \label{frequency_eq_of_anti}
		\left| 
	\begin{array}{cccc}
		Z_{41} & Z_{42} & Z_{43} & Z_{48}
       \\
        Z_{51} & Z_{52} & Z_{53} & Z_{58}  
       \\
       Z_{61} & Z_{62} & Z_{63} & Z_{68}
       \\
       Z_{71} & Z_{72} & Z_{73} & Z_{78}
	\end{array} \right| 
         = 0,
\end{equation}
which is the characteristic frequency equation for small-amplitude non-axisymmetric vibrations of the
activated SEA tube subject to radially inhomogeneous biasing fields for mode numbers $m \geq 1$ and $n \geq 1$.

Following the same derivation procedure, we can also acquire the characteristic frequency equation of the incremental prismatic vibrations ($m \neq 0, n=0$) as:
\begin{equation}
	\left| 
	\begin{array}{ccc}
		Z_{\mathrm{p}_{31}} & Z_{\mathrm{p}_{32}} & Z_{\mathrm{p}_{36}} 
		\\
		Z_{\mathrm{p}_{41}} & Z_{\mathrm{p}_{42}} & Z_{\mathrm{p}_{46}}
		\\
		Z_{\mathrm{p}_{51}} & Z_{\mathrm{p}_{52}} & Z_{\mathrm{p}_{56}}    
	\end{array} \right| 
	= 0,
\end{equation}
where $Z_{\mathrm{p}_{ij}}$ are the elements of the global transfer matrix $\mathrm{\mathbf{Z}_{p}}$ for the prismatic vibrations.

For the incremental axisymmetric vibrations including the L vibrations, the T vibrations and the breathing vibrations, we refer to the frequency equations (40) and (43) in the paper by \cite{zhu2020electrostatically}.

Note that the frequency equations derived above are applicable for any form of the energy function of the incompressible isotropic SEA tubes. 
To calculate numerically the natural frequency of electrostatically tunable non-axisymmetric vibrations, we specialize the analysis to the Gent ideal dielectric model (\ref{Gent_model}). Then the dimensionless effective material parameters appearing in Eqs.~(\ref{elements_M}) and (\ref{di_quantity}) are 
\begin{equation} \label{effective}
	\begin{array}{l}
	    \overline{c}_{44} = \lambda_{\theta}^2 G_{1},
	    \quad
		\overline{c}_{47} = \overline{p},
	    \quad
        \overline{c}_{77} = \lambda_{z}^2 G_{1},
	    \quad
	    \overline{c}_{55}=\overline{c}_{66} = \lambda_{\theta}^{-2} \lambda_{z}^{-2} G_{1},	
	    \quad
	    \overline{\varepsilon}_{11} = 1,
     	\\\\
		\beta_{1} = \beta_{3} = \lambda_{\theta}^2 \lambda_{z}^2 G_{1}^{-1}\left( \overline{p} - \overline{D}_{r}^2 \right),
		\quad
		\beta_{2} = \beta_{4} = -\lambda_{\theta}^2 \lambda_{z}^2 G_{1}^{-1} \overline{D}_{r}, 
		\quad
		\overline{q}_{1} = \overline{q}_{2} = 2 \overline{D}_{r},
		\\\\
		\overline{q}_{3} = G_{1} \left( \lambda_{\theta}^2 + \lambda_{\theta}^{-2} \lambda_{z}^{-2} \right)  + 2G_{2} \left( \lambda_{\theta}^4 - 2\lambda_{z}^{-2} + \lambda_{\theta}^{-4} \lambda_{z}^{4}\right)  + \overline{D}_{r}^2 + 2 \overline{p},
		\\\\
		\overline{q}_{4} = \overline{q}_{5} = G_{1}\lambda_{\theta}^{-2} \lambda_{z}^{-2} + 2 G_{2} \left( \lambda_{\theta}^2 \lambda_{z}^2 - \lambda_{z}^{-2} - \lambda_{\theta}^{-2} + \lambda_{\theta}^{-4} \lambda_{z}^{-4} \right) + \overline{D}_{r}^2 + \overline{p},
		\\\\
		\overline{q}_{6} = G_{1} \left( \lambda_{z}^2 + \lambda_{\theta}^{-2} \lambda_{z}^{-2} \right) + 2 G_{2} \left( \lambda_{z}^{4} - 2 \lambda_{\theta}^{-2} + \lambda_{\theta}^{-4} \lambda_{z}^{-4} \right) + \overline{D}_{r}^2 + 2 \overline{p},
		\\\\
		\overline{q}_{7} = \lambda_{\theta}^2 G_{1} - \lambda_{\theta}^2 \lambda_{z}^2 G_{1}^{-1} \left( \overline{D}_{r}^4 + \overline{p}^2 - 2 \overline{p} \overline{D}_{r}^2 \right) - \overline{D}_{r}^2,
		\quad
		\overline{q}_{8} = \overline{q}_{7} + \left(\lambda_{z}^2 - \lambda_{\theta}^2 \right) G_{1}, 
		\\\\
		\overline{q}_{9} = \overline{q}_{10} = - \left( 1 - \beta_{1} \right) \overline{D}_{r} ,
		\quad
		\overline{q}_{11} = \overline{q}_{12} = \lambda_{\theta}^2 \lambda_{z}^2 G_{1}^{-1} \overline{D}_{r}^2 + 1,
	\end{array}
\end{equation}
where $G_{1}$ = $G/\left( G - I_{1} +3 \right) $ and $G_{2}$ = $G/\left( G - I_{1} +3 \right)^2 $. 
In the limit of $G \rightarrow \infty$, we have $G_{1} \rightarrow 1$ and $G_{2} \rightarrow 0$, and the effective material parameters listed in Eq.~(\ref{effective}) reduce to their equivalent forms for the neo-Hookean ideal dielectric model (see Eq. (44) in \cite{zhu2020electrostatically}).

\section{Numerical results and discussions} \label{section4}

In this section, we investigate the influences of the electro-mechanical biasing fields (i.e., the combined action of radial voltage $\overline{V}$ and axial pre-stretch $\lambda_{z}$) and strain-stiffening effect on the nonlinear response and non-axisymmetric vibration characteristics of the SEA tube characterized by the Gent ideal dielectric model (\ref{Gent_model}).

\subsection{Axisymmetric nonlinear static response} \label{section4.1}

We first investigate the axisymmetric nonlinear static response under the combined action of electric voltage and axial pre-stretch. 

\begin{figure}[H] 
	\centering  
	\subfigure{
		\label{NL_res.sub.1}
		\includegraphics[width=0.45\textwidth]{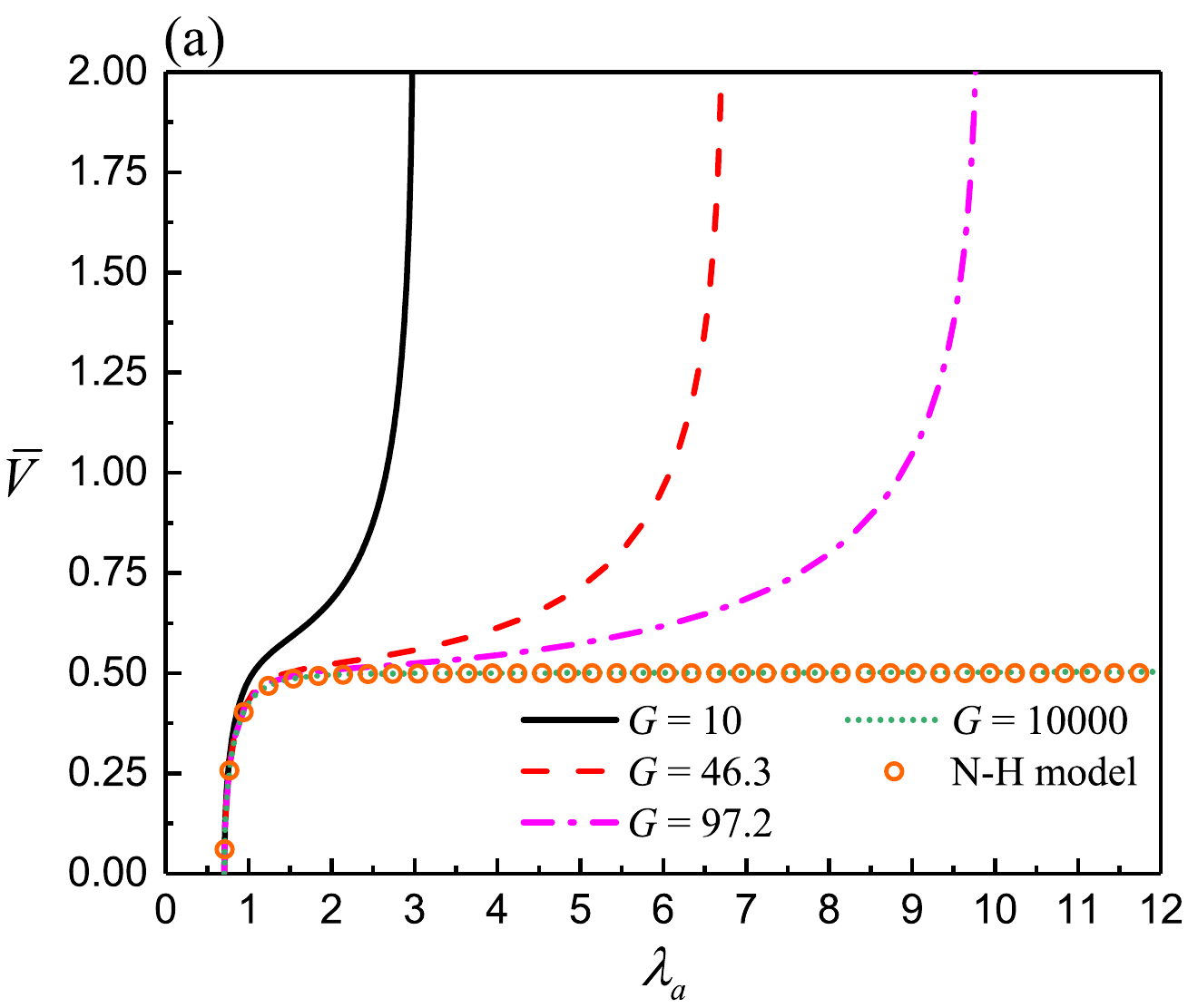}}
	\subfigure{
		\label{NL_res.sub.2}
		\includegraphics[width=0.45\textwidth]{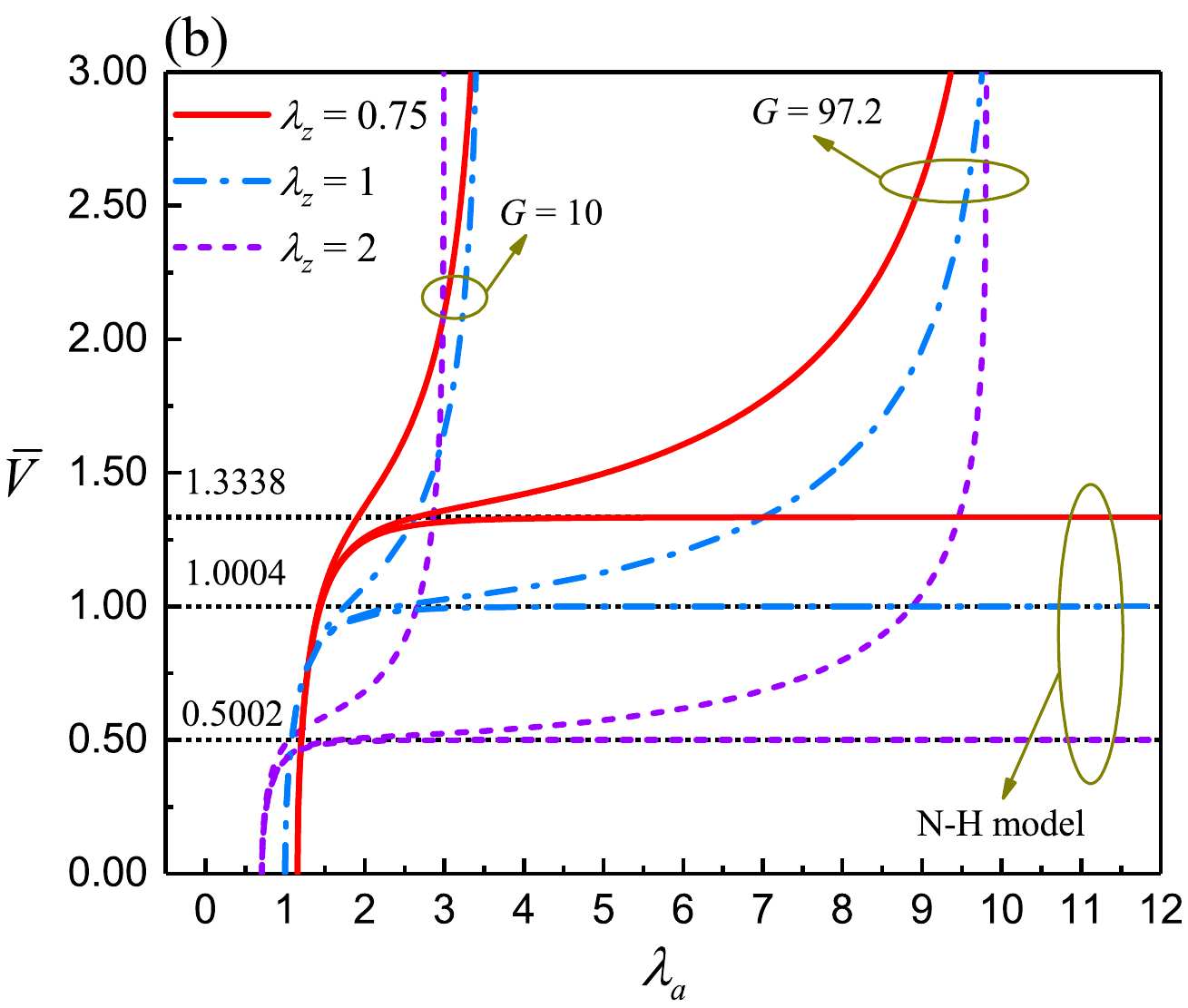}}
	\caption{Nonlinear response curves of the dimensionless electric voltage $\overline{V}$ versus the circumferential stretch $\lambda_{a}$ at the inner surface of a thin and slender SEA tube ($\eta=0.9$ and $L/H=10$) for (a) different values of the Gent constant $G$ with a fixed axial pre-stretch $\lambda_{z}$= 2, and (b) different values of the Gent constant $G$ and axial pre-stretch $\lambda_{z}$. } 
	\label{Nonlinear_response}
\end{figure}

Based on the nonlinear response equations described in Section.~\ref{2_2} for the neo-Hookean and Gent ideal dielectric models, we plot in Fig.~\ref{Nonlinear_response} the nonlinear static response variation curves (i.e. the dimensionless electric voltage $\overline{V}$ versus the circumferential stretch $\lambda_{a}$) of the SEA tube under different electro-mechanical biasing fields. In Fig.~\ref{Nonlinear_response}, the Gent constant $G$ models the strain-stiffening effect of the SEA tube near a limiting stretch and the `N-H model' legend denotes the neo-Hookean ideal dielectric model. It should be emphasized that the electro-mechanical limit-point instability (i.e., the $\overline{V}-\lambda_{a}$ reaches a maximum) occurs in neo-Hookean SEA tubes \citep{shmuel2013axisymmetric, shmuel2015manipulating, wu2017guided, zhu2020electrostatically} but not in  Gent SEA tubes because of the strain-stiffening effect \citep{zhou2014electromechanical, bortot2018prismatic}. Fig.~\ref{NL_res.sub.1} displays the nonlinear response curves for different values of $G$ with a fixed pre-stretch $\lambda_{z} = 2$, and the variation curve for the neo-Hookean model is also included. It shows that $\lambda_{a}$ increases monotonically as the radial electric voltage $\overline{V}$ increases, which means physically that the tube reduces in thickness and expands in the radial direction with increasing $\overline{V}$. When the applied voltage keeps growing, the tube reaches its limiting stretch $\lambda_{\text{lim}}$ due to the effect of strain-stiffening for a Gent SEA tube. The smaller $G$ is, the smaller the limiting stretch $\lambda_{\text{lim}}$ is, and the stronger the strain-stiffening effect is, while for large $G$ values (e.g. $G=10^4$), the curve tends to that of the neo-Hookean model.

In Fig.~\ref{NL_res.sub.2}, we additionally take the influence of axial pre-stretch $\lambda_{z}$ into account. We mark out the electro-mechanical instability voltages $\overline{V}_{\mathrm{EMI}}$ (i.e., $\overline{V}_{\mathrm{EMI}}=0.5002$ for $\lambda_{z}=2$, $\overline{V}_{\mathrm{EMI}}=1.0004$ for $\lambda_{z}=1$, and $\overline{V}_{\mathrm{EMI}}=1.3338$ for $\lambda_{z}=0.75$) for the neo-Hookean model under three different pre-stretches for reference. For a fixed $G$ value, when subject to larger axial pre-stretch, lower radial voltage is required to obtain equal circumferential stretch $\lambda_{a}$, which means physically that the SEA tube is easier to deform for a larger axial pre-stretch.

\subsection{Validation of the State Space Method (SSM)} \label{Convergence_and_validation}

The effectiveness of the SSM is first verified in terms of its convergence and accuracy for the 3D free vibration analysis of the Gent SEA tube subject to biasing fields in this section.

\begin{table}[H] 
	\centering
	\setlength{\abovecaptionskip}{5pt}
	\setlength{\belowcaptionskip}{5pt}
	\caption{The first three natural frequencies $\varpi$ of the non-axisymmetric vibration with $m=n=1$ of a \emph{thin and slender} SEA tube ($\eta=0.9$ and $L/H=10$) based on the SSM with different numbers of discretized layers (NoL) ($\lambda_z = 2$, $\overline{V} = 0.2$ and $G=97.2$). }
	{\footnotesize
		\setlength{\tabcolsep}{3mm}{
			\begin{tabular}{ccccccccc}
				\toprule  
				NoL & 20 & 40 & 60 & 80 & 100 & 120 & 140 & 160  \\
				\midrule  
				$\mathrm{1^{st}}$ & 0.30495 & 0.30495 & 0.30495 & 0.30495 & 0.30495 & 0.30495 & 0.30495 & 0.30495 \\
				$\mathrm{2^{nd}}$ & 0.35255 & 0.35255 & 0.35255 & 0.35255 & 0.35255 & 0.35255 & 0.35255 & 0.35255 \\
				$\mathrm{3^{rd}}$ & 0.44378 & 0.44378 & 0.44378 & 0.44378 & 0.44378 & 0.44378 & 0.44378 & 0.44378  \\
				\bottomrule 
	\end{tabular}}} \label{tab1}
\end{table}
\begin{table}[H] 
	\centering
	\setlength{\abovecaptionskip}{5pt}
	\setlength{\belowcaptionskip}{5pt}
	\caption{The first three natural frequencies $\varpi$ of the non-axisymmetric vibration with $m=n=1$ of a \emph{thick and short} SEA tube ($\eta=0.2$ and $L/H=2.5$) based on the SSM with different numbers of discretized layers (NoL) ($\lambda_z = 2$, $\overline{V} = 0.2$ and $G=97.2$). }
	{\footnotesize
		\setlength{\tabcolsep}{3mm}{
			\begin{tabular}{ccccccccc}
				\toprule  
				NoL & 20 & 40 & 60 & 80 & 100 & 120 & 140 & 160  \\
				\midrule  
				$\mathrm{1^{st}}$ & 1.20181 & 1.20180 & 1.20180 & 1.20180 & 1.20180 & 1.20180 & 1.20180 & 1.20180  \\
				$\mathrm{2^{nd}}$ & 1.97680 & 1.97698 & 1.97702 & 1.97703 & 1.97704 & 1.97704 & 1.97704 & 1.97704  \\
				$\mathrm{3^{rd}}$ & 2.85347 & 2.85284 & 2.85273 & 2.85269 & 2.85267 & 2.85266 & 2.85265 & 2.85265  \\
				\bottomrule 
	\end{tabular}}} \label{tab2} 
\end{table}

For the convergence analysis, we list in Tables \ref{tab1} and \ref{tab2} the first three natural frequencies $\varpi$ of the non-axisymmetric vibration calculated by the SSM with different numbers of discretized layers (NoL) for a \emph{thin and slender} SEA tube and a \emph{thick and short} SEA tube of the Gent ideal dielectric model, respectively. Moreover, the circumferential mode number $m$ and axial mode number $n$ are all equal to one. Clearly, we can find the natural frequencies with arbitrary precision using the current SSM because the results clearly indicate an excellent convergence rate with increasing NoL. The calculated frequencies based on the SSM are close to those of the original SEA tube that is subject to radially inhomogeneous biasing fields. Thus, 120 discretized layers will be chosen hereafter, which is assumed to have high accuracy.

The radially inhomogeneous biasing fields are induced by the applied radial electric voltage. However, when there is no applied radial electric voltage, the deformation of the pre-stretched SEA tube is homogeneous, which makes it feasible to obtain the exact frequency solutions to the non-axisymmetric vibrations through the conventional displacement method. The detailed derivations for the analytical frequency equations of the non-axisymmetric vibrations were provided in \emph{Appendix C} of our previous work \citep{zhu2020electrostatically} for an arbitrary energy function. 
\ref{AppendixA} in this paper additionally gives the effective material parameters and analytical frequency equations of three kinds of special vibrations (including the breathing mode, T vibrations and prismatic vibrations) in a pre-stretched hyperelastic tube characterized by the Gent model. Therefore, the accuracy of SSM can be verified by making a comparison to the results obtained from the exact solutions. 

\begin{figure}[h!]
	\centering  
	\subfigure{
		\label{Validation.sub.1}
		\includegraphics[width=0.48\textwidth]{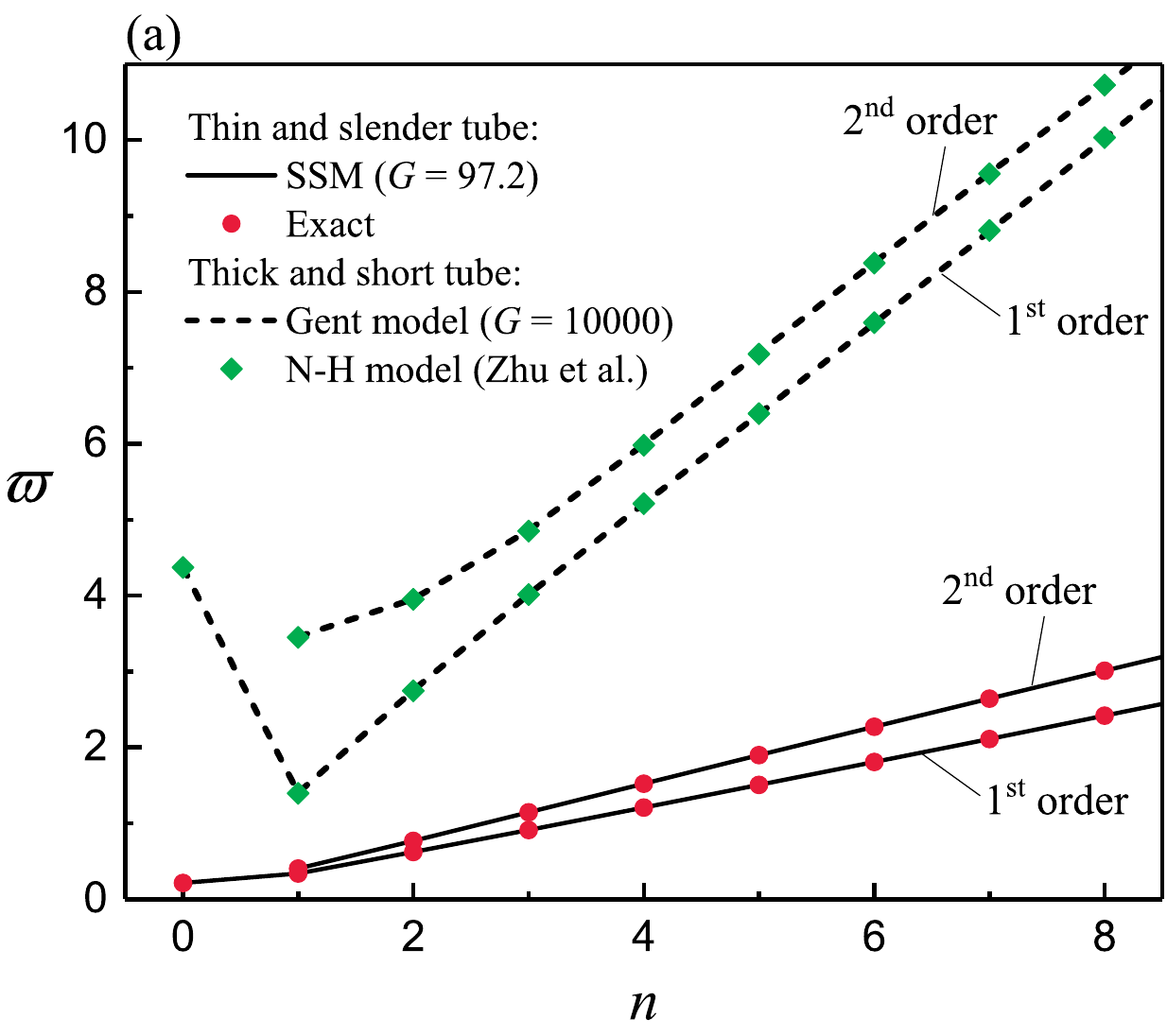}}
	\subfigure{
		\label{Validation.sub.2}
		\includegraphics[width=0.48\textwidth]{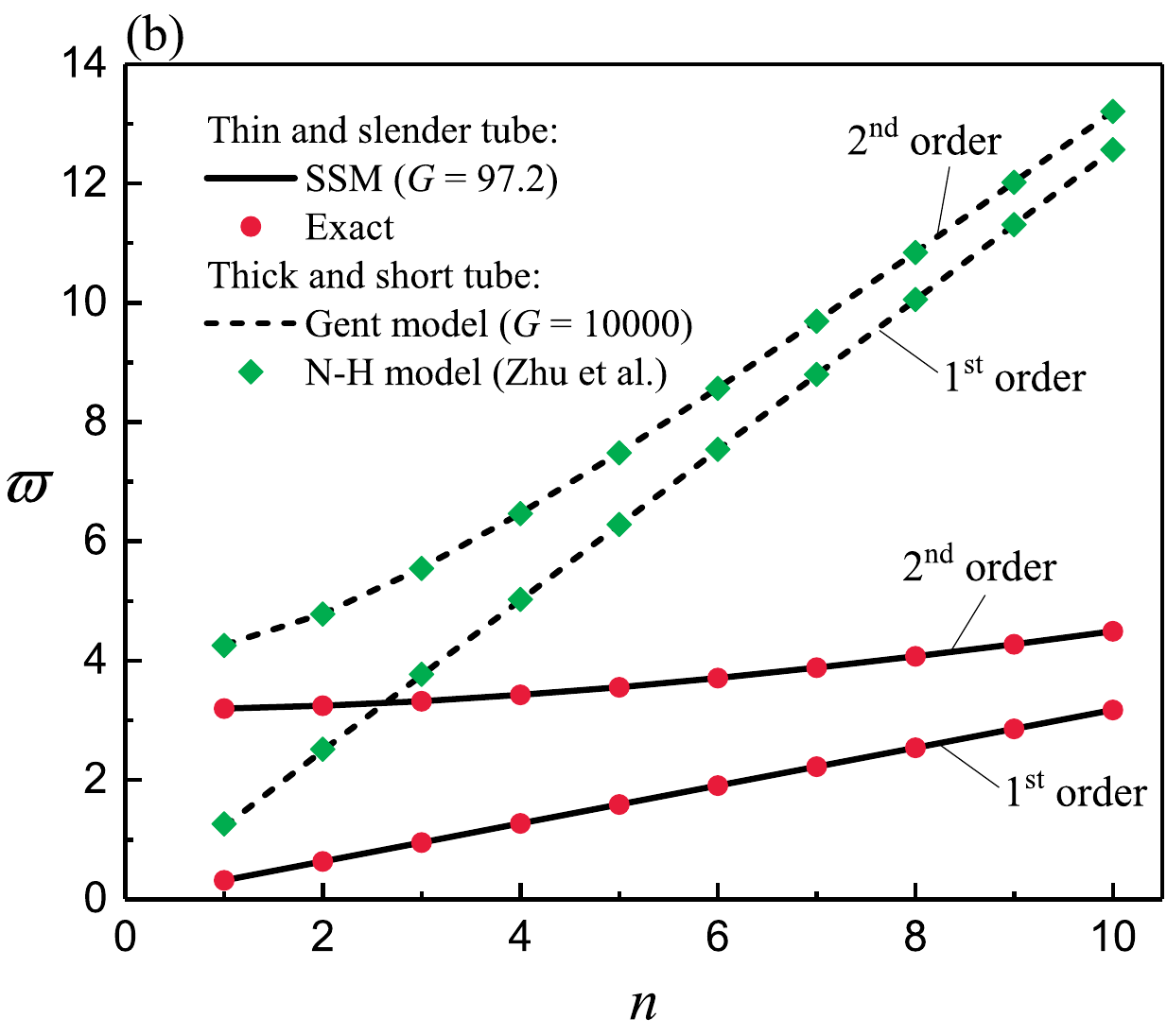}}
	\subfigure{
		\label{Validation.sub.3}
		\includegraphics[width=0.48\textwidth]{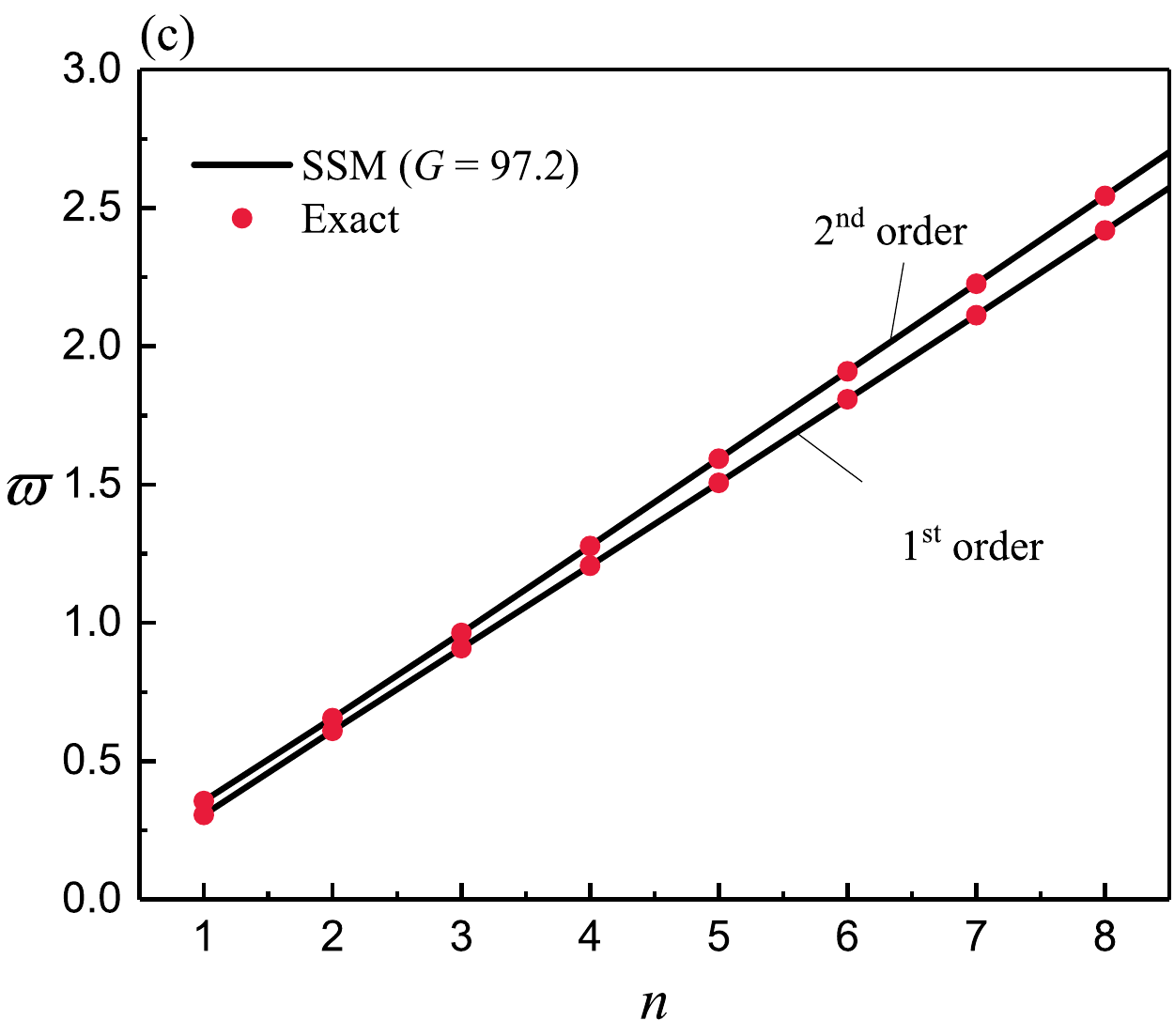}}
	\subfigure{
		\label{Validation.sub.4}
		\includegraphics[width=0.48\textwidth]{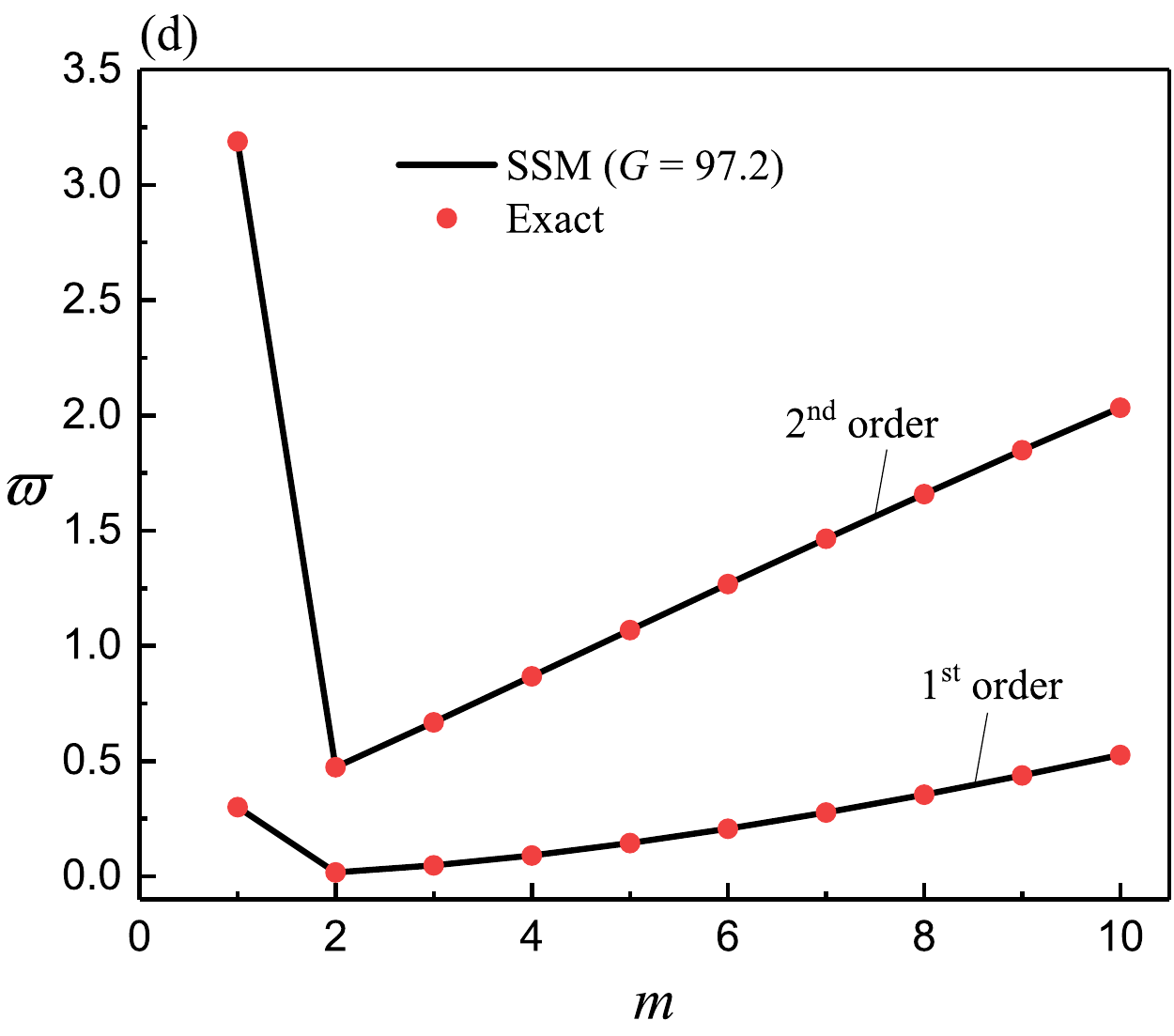}}
	\caption{Accuracy analysis of the first two dimensionless vibration frequencies $\varpi$ obtained by the exact solutions and the SSM for the pre-stretched ($\lambda_{z}$=2) \emph{thin and slender} $(\eta=0.9$ and $L/H=10)$ SEA tube and \emph{thick and short} $(\eta=0.2$ and $L/H=2.5)$ SEA tube of the Gent and neo-Hookean models, without applied radial electric voltage: (a) L vibrations including the breathing mode $(m=0, n=0)$; (b) T vibrations $(m=0)$; (c) Non-axisymmetric vibrations $(m=1)$; (d) Prismatic vibrations $(n=0)$.} 
	\label{Validation_figure}
\end{figure}

Based on the SSM and the exact solutions, we carry out an accuracy analysis of the first two dimensionless vibration frequencies $\varpi$ versus the axial mode number $n$ or the circumferential mode number $m$ for the pre-stretched ($\lambda_{z}=2$) \emph{thin and slender} $(\eta=0.9, L/H=10)$ SEA tube and for the \emph{thick and short} $(\eta=0.2, L/H=2.5)$ SEA tube with no applied voltage, for the following four kinds of vibrations: Fig.~\ref{Validation.sub.1} for L vibrations including the breathing mode $(m=n=0)$, Fig.~\ref{Validation.sub.2} for T vibrations $(m=0)$, Fig.~\ref{Validation.sub.3} for non-axisymmetric vibrations $(m=1)$, and Fig.~\ref{Validation.sub.4} for prismatic vibrations $(n=0)$. For the Gent material model, the dimensionless Gent constant $G$ is chosen to be 97.2.

In Fig.~\ref{Validation.sub.1}, regardless of geometric sizes, there is only one radial vibration frequency for the breathing mode $m=n=0$ as a result of the tube incompressibility. Moreover, the natural frequency of breathing mode in a thin and slender SEA tube is lower than those of other first-order L vibration modes $(n \geq 1)$, while in the case of thick and short tube, the natural frequency of the breathing mode is larger than those of the three L vibration modes $(n=1,2,3)$ of the first order. In addition, we make the link with the data in the case of a thick and short neo-Hookean SEA tube from \cite{zhu2020electrostatically} and that of the Gent model with a large  $G$ ($G=10^4$)  in both Figs.~\ref{Validation.sub.1} and \ref{Validation.sub.2}, denoted by the black-dashed line and green diamond, respectively. As expected, the natural frequencies of the Gent tube with a large $G$ coincide with those of the neo-Hookean tube. For the four kinds of vibrations shown in Fig.~\ref{Validation_figure}, it is clear that the vibration frequencies calculated by the SSM agree with those predicted from the exact solutions in the entire axial and circumferential mode number range.

In conclusion, the SSM-based numerical results are highly accurate for the 3D free vibration analysis because of the great convergence rate and the excellent agreement with the exact solutions.

\subsection{Strain-stiffening effect on axisymmetric and prismatic vibrations}

In this subsection, we investigate how the strain-stiffening effect affects the natural frequencies of axisymmetric vibrations (including the L vibrations and T vibrations) and prismatic vibrations. If not otherwise stated, we consider a \emph{thin and slender} tube with the geometric sizes set as $\eta=0.9$ and $L/H=10$. 

\begin{figure}[h!] 
	\centering  
	\subfigure{
		\label{L_vibration_fre.sub.1}
		\includegraphics[width=0.48\textwidth]{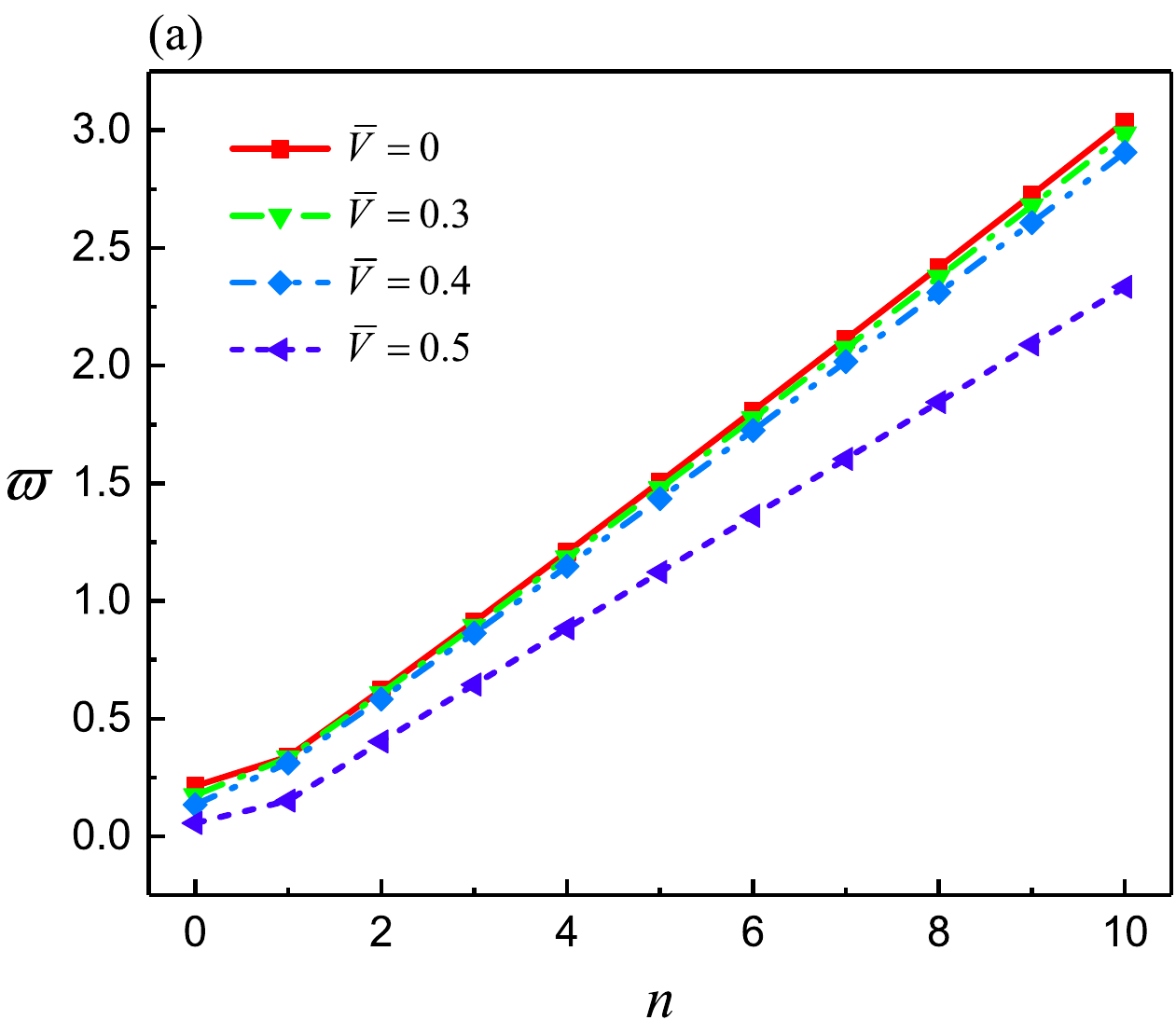}}
	\subfigure{
		\label{T_vibration.sub.2}
		\includegraphics[width=0.48\textwidth]{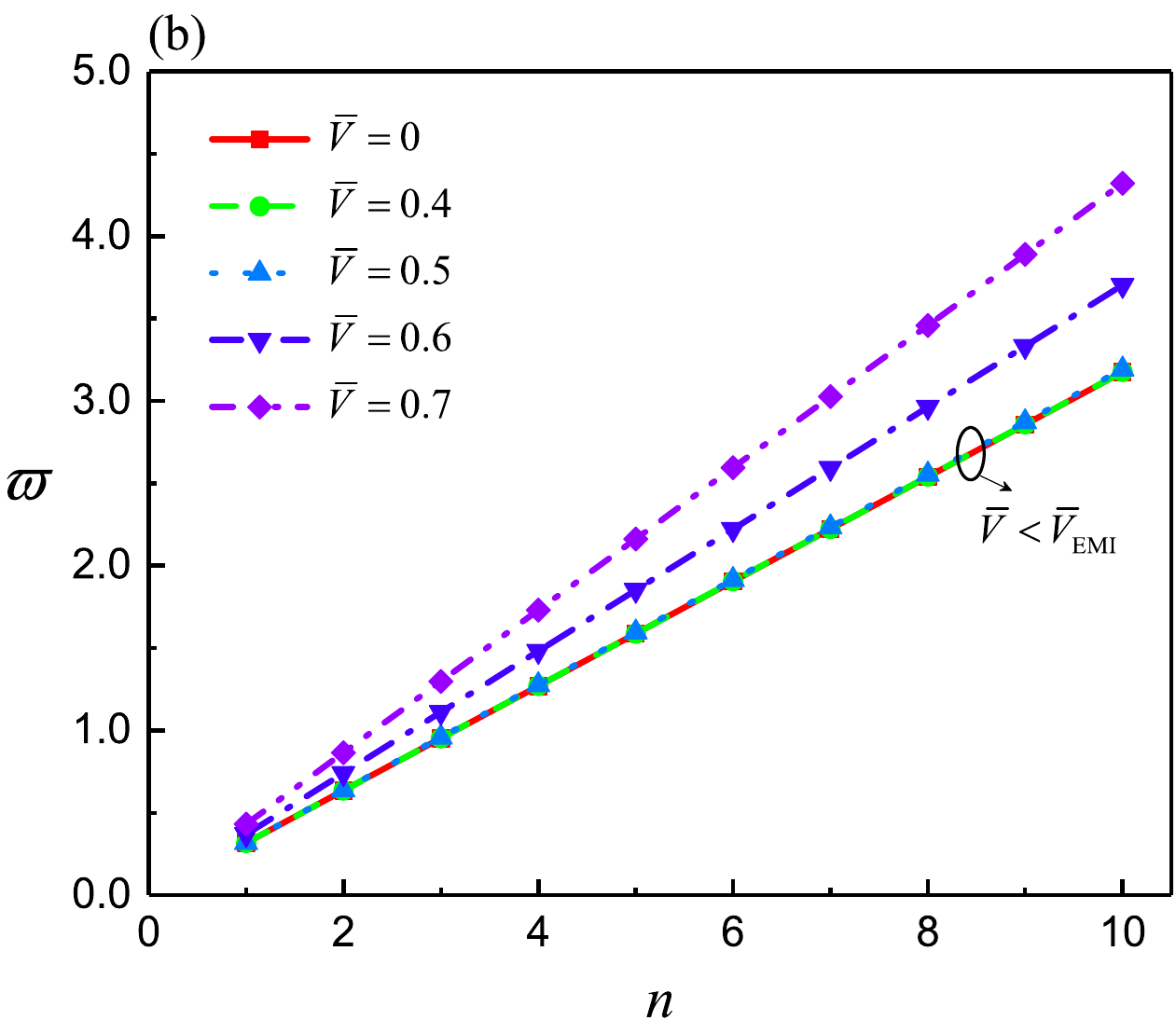}}
	\subfigure{
		\label{P_vibration.sub.3}
		\includegraphics[width=0.48\textwidth]{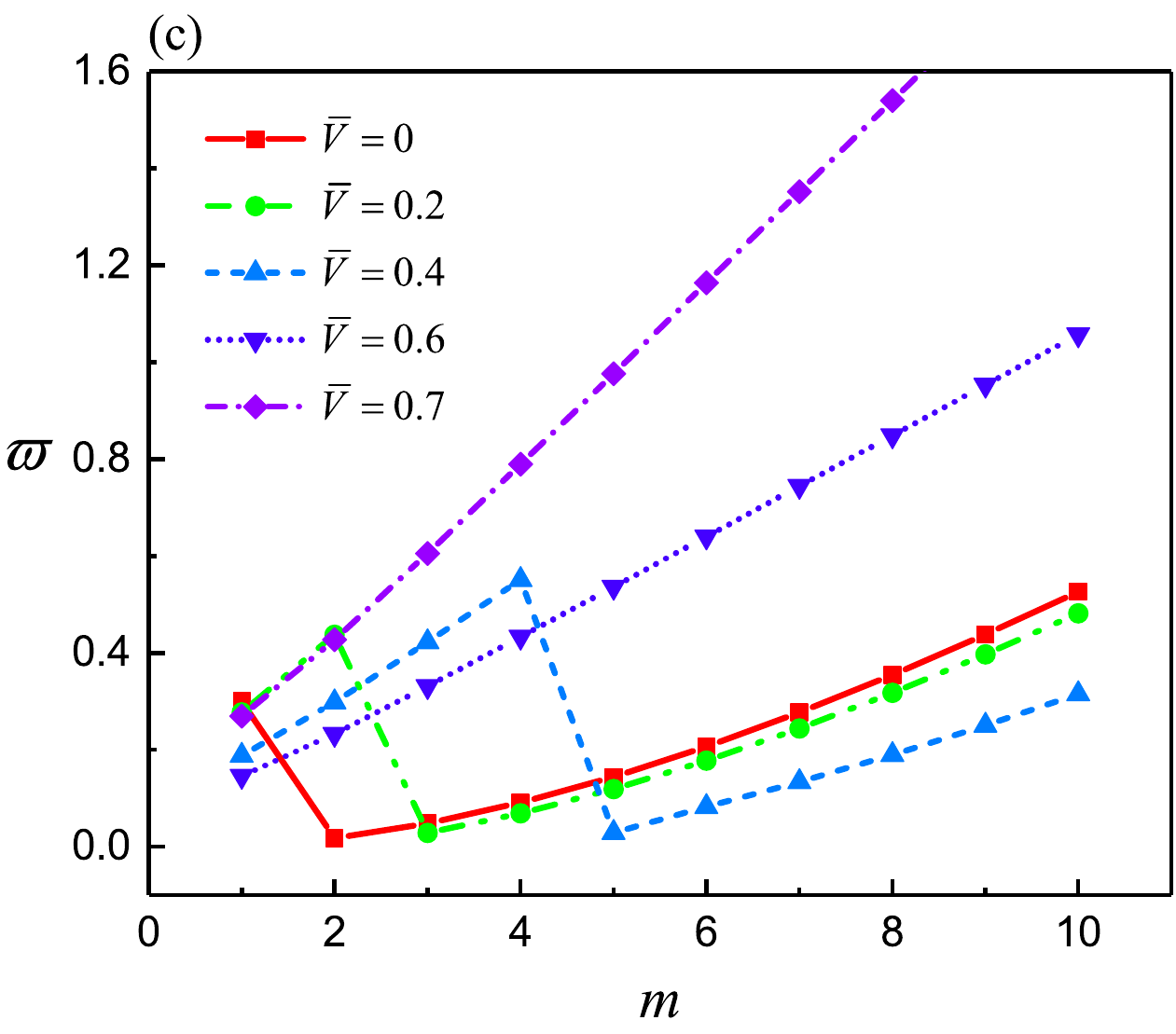}}
	\caption{Dimensionless frequency spectra in a pre-stretched ($\lambda_{z}=2$) thin and slender SEA tube $(\eta=0.9, L/H=10)$ employing the Gent model ($G=97.2$) for different values of radial electric voltage: (a) the first-order frequency of L vibrations ($\varpi$ versus $n$); (b) the first-order frequency of T vibrations ($\varpi$ versus $n$); (c) the first-order frequency of prismatic vibrations ($\varpi$ versus $m$). } 
	\label{Axisymmetric_and_P_frequency_spectra}
\end{figure}

First, for a pre-stretched ($\lambda_{z}=2$) thin and slender Gent ideal SEA tube with $G=97.2$, the variations of the first-order dimensionless natural frequency $\varpi$ with the axial mode number $n$ are displayed in Figs.~\ref{L_vibration_fre.sub.1} and \ref{T_vibration.sub.2} under different radial electric voltages for the L and T vibrations, respectively. In Fig.~\ref{P_vibration.sub.3}, we exhibit the variation trend of the first-order natural frequency $\varpi$ with the circumferential mode number $m$ for the prismatic vibrations.

It is seen from Fig.~\ref{L_vibration_fre.sub.1} that the first-order frequency of L vibrations goes up monotonously with increasing $n$. As the voltage $\overline{V}$ increases, the natural frequency $\varpi$ decreases in the entire axial mode number range. Specifically, we find that a relatively low voltage (e.g., $\overline{V}<0.3$) barely affects the L vibration frequency, but as $\overline{V}$ continues growing, there exists a big gap between the two frequency spectra stimulated by $\overline{V}=0.4$ and $\overline{V}=0.5$. Such a dramatic change occurs due to the tube's rapid expansion, resulting in the sharp decrease of the global stiffness when $\overline{V} < \overline{V}_{\mathrm{EMI}}$.

For the T vibrations shown in Fig.~\ref{T_vibration.sub.2}, the vibration frequency $\varpi$ increases monotonously and linearly with the axial mode number $n$ for an arbitrary applied radial voltage $\overline{V}$. Taking the electro-mechanical instability voltage $\overline{V}_{\mathrm{\mathrm{EMI}}}$ exclusive to the neo-Hookean SEA tube as the demarcation point, the change of voltage makes no difference to the vibration frequency for this kind of thin and slender pre-stretched SEA tube for $\overline{V} < \overline{V}_{\mathrm{EMI}}$. However, for $\overline{V}> \overline{V}_{\mathrm{EMI}}$, the natural frequency has a remarkable rise with the applied voltage because of the strain-stiffening effect.

For the prismatic vibrations depicted in Fig.~\ref{P_vibration.sub.3}, it is found that the circumferential mode number $m$ corresponding to the lowest natural frequency depends on the applied voltage when $\overline{V}<\overline{V}_{\mathrm{EMI}}$. For example, the lowest vibration frequency is obtained at $m=2$ for $\overline{V}=0$, while for $\overline{V}=0.2$, it occurs at $m=3$. For $\overline{V}> \overline{V}_{\mathrm{EMI}}$ (such as the cases $\overline{V}=0.6$ and $\overline{V}=0.7$), the natural frequency increases monotonically with the increasing circumferential mode number and the lowest frequency is achieved at $m=1$. Moreover, the natural frequency for $\overline{V}> \overline{V}_{\mathrm{EMI}}$ goes up monotonically with the applied voltage in the entire circumferential mode number range.

\begin{figure}[h!] 
	\centering  
	\subfigure{
		\label{axisymmetric_breathing.sub.1}
		\includegraphics[width=0.48\textwidth]{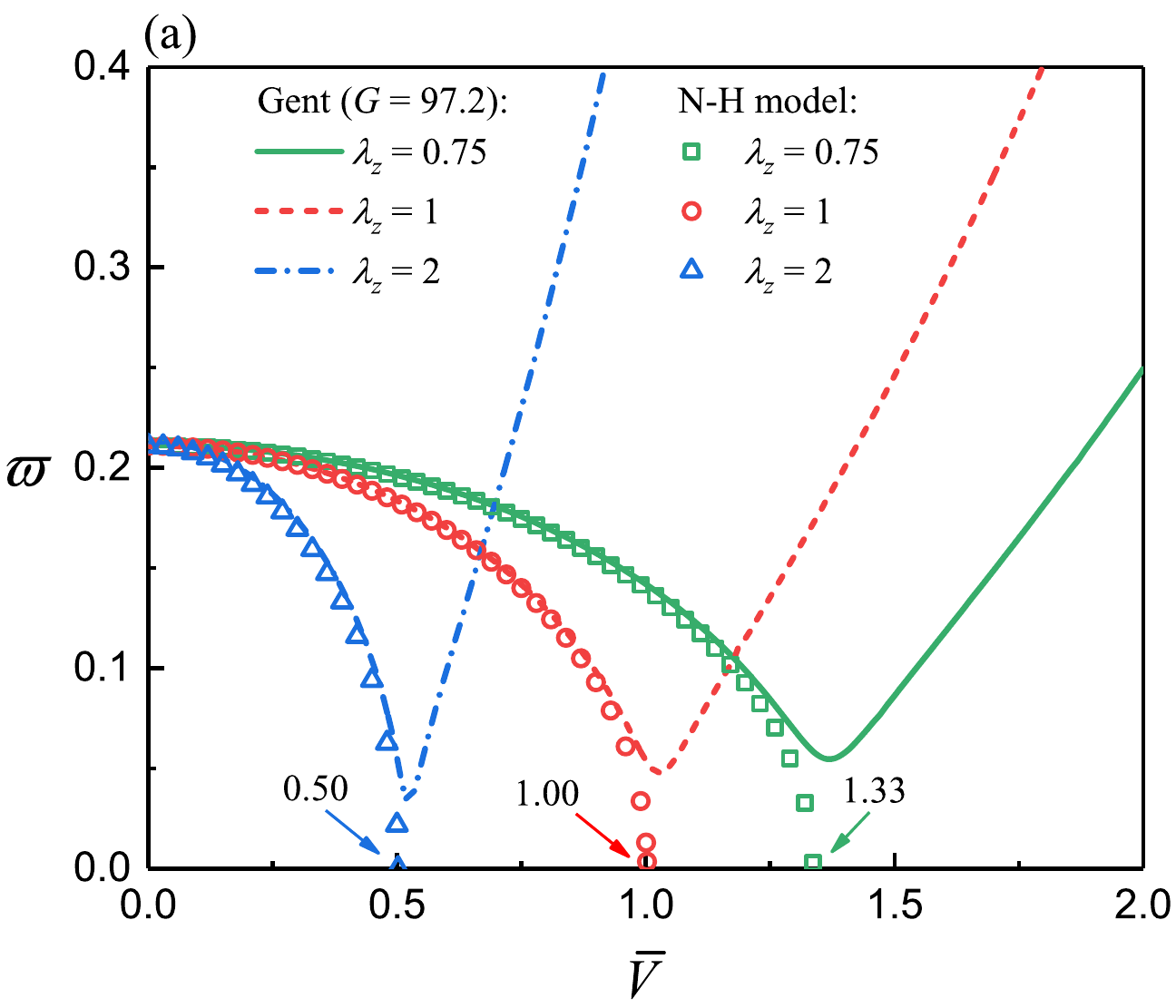}}
	\subfigure{
		\label{axisymmetric_L_vibration.sub.2}
		\includegraphics[width=0.48\textwidth]{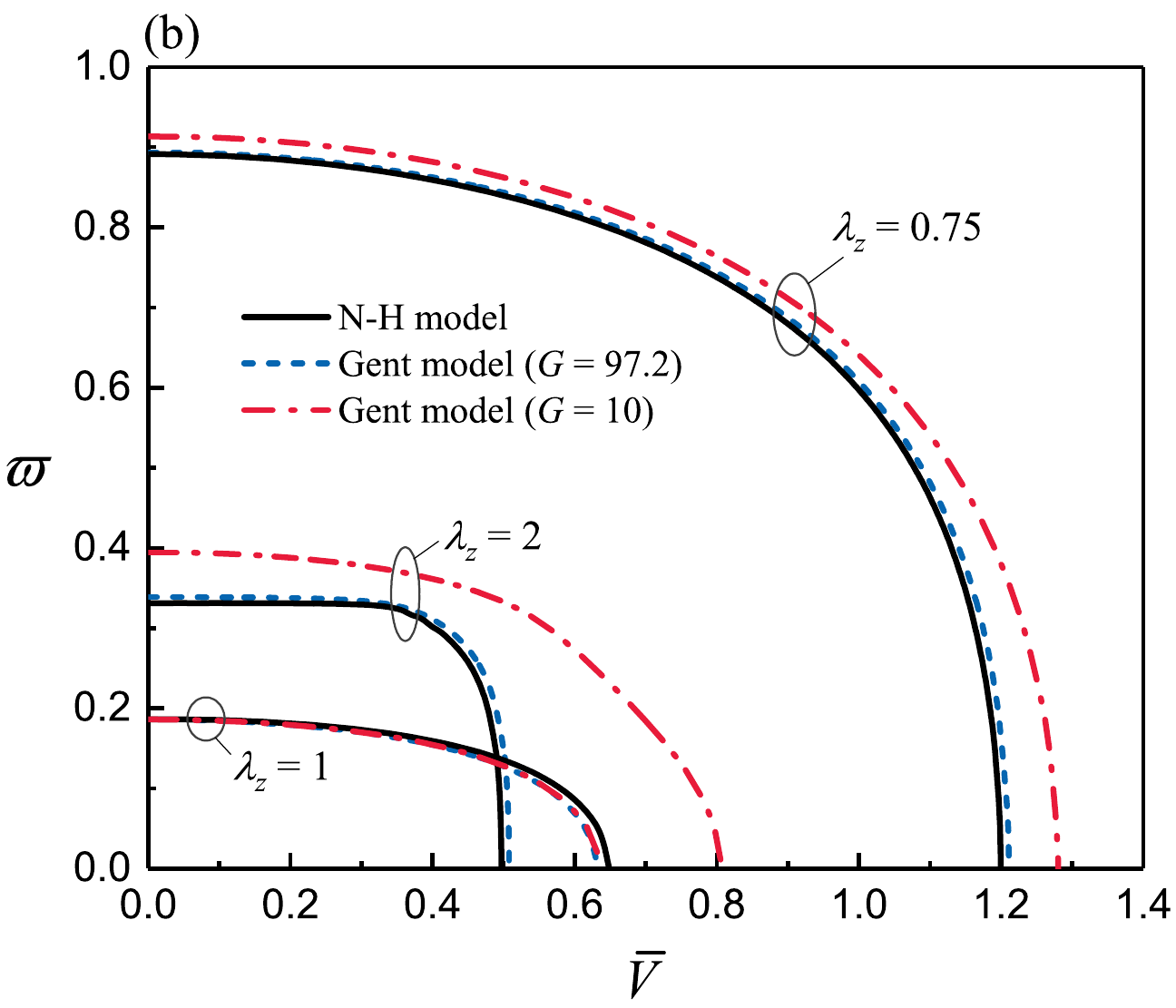}}
	\subfigure{
		\label{axisymmetric_T_vibration.sub.3}
		\includegraphics[width=0.48\textwidth]{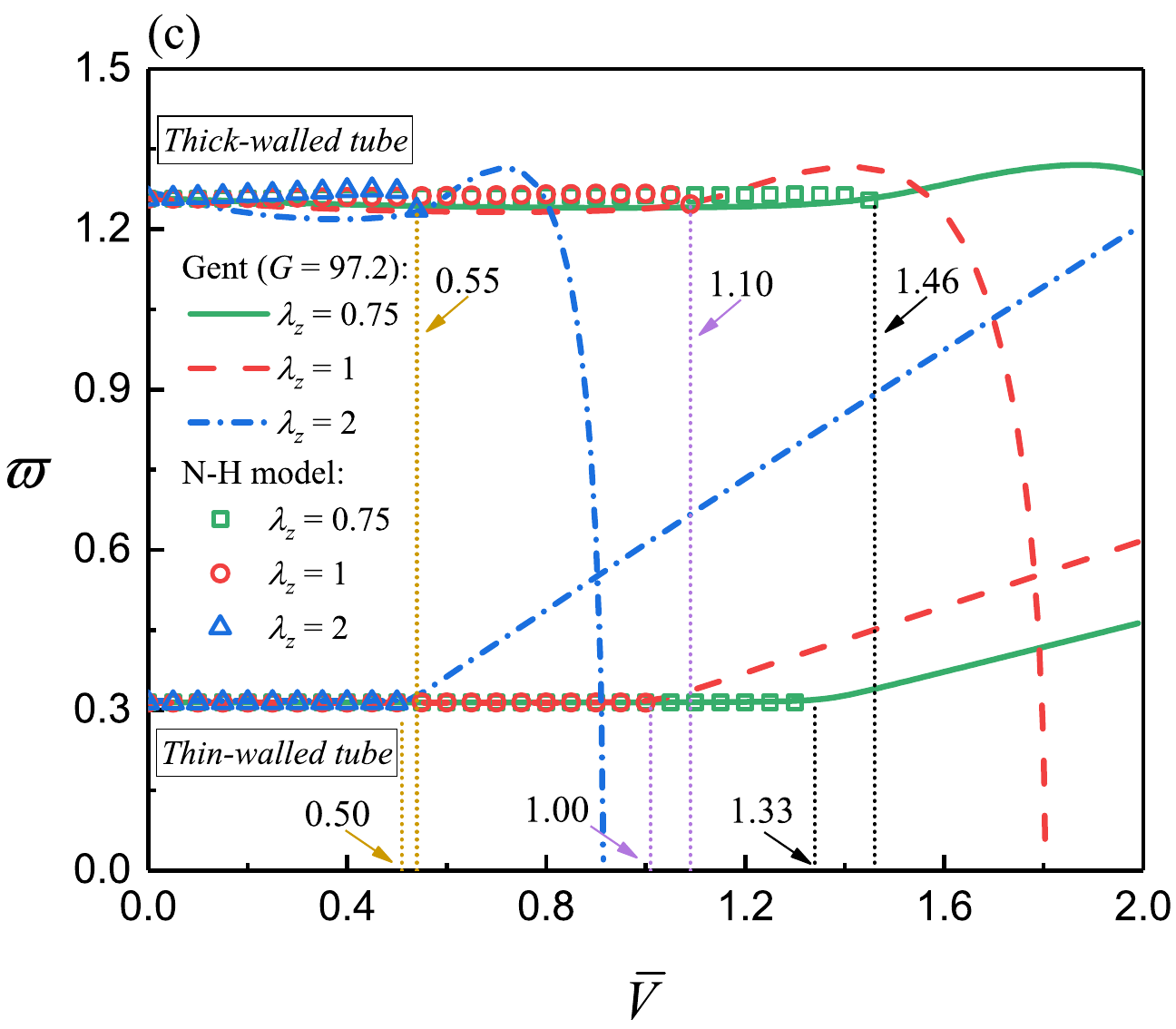}}
	\subfigure{
		\label{Prismatic_vibration.sub.4}
		\includegraphics[width=0.48\textwidth]{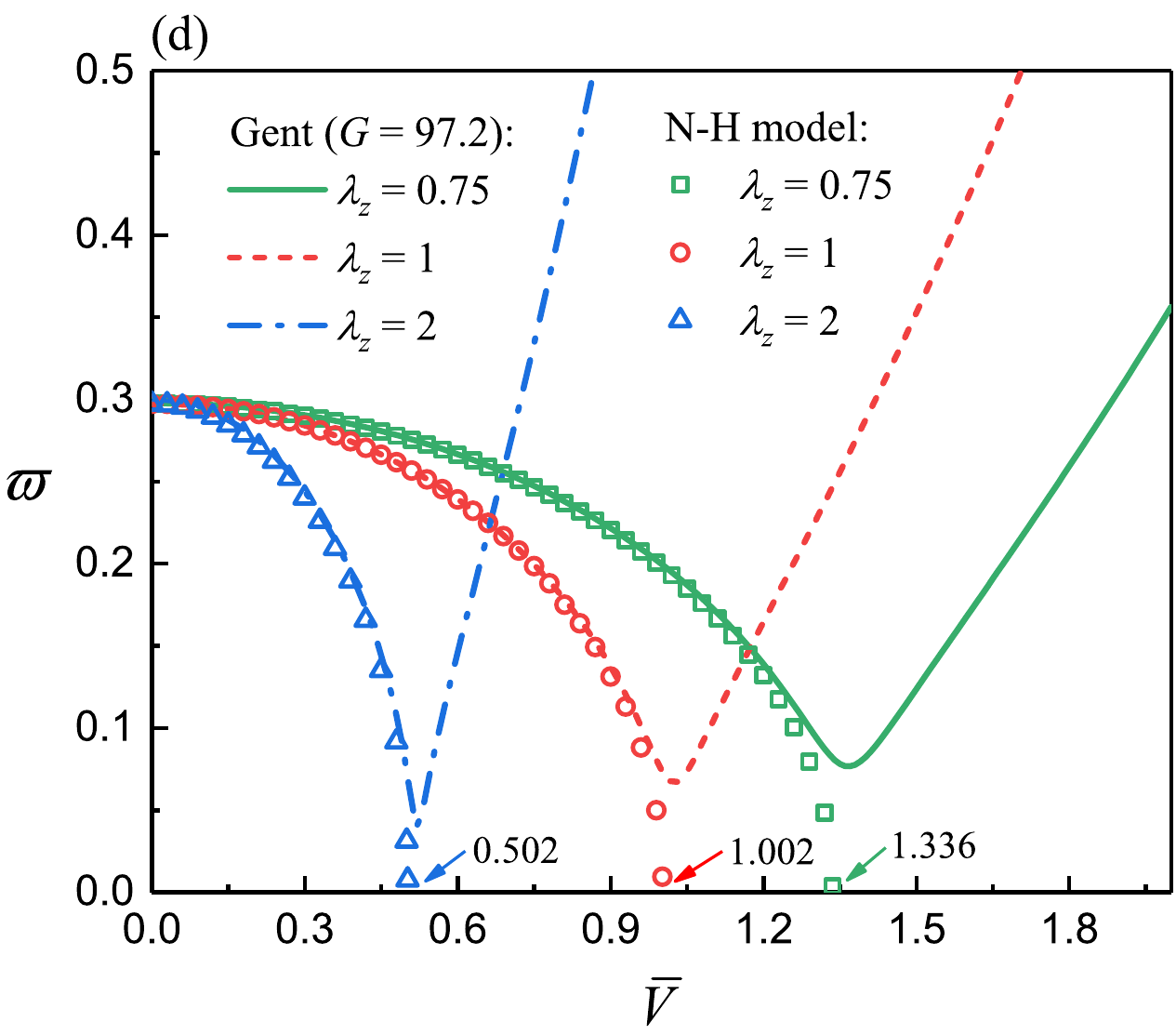}}
	\caption{Variation curves of the first-order dimensionless vibration frequency $\varpi$ as functions of the dimensionless radial voltage $\overline{V}$ in a \emph{thin and slender} SEA tube $(\eta=0.9, L/H=10)$ for the Gent and neo-Hookean models under different axial pre-stretches: (a) breathing mode $(m=n=0)$; (b) L vibration mode $(m=0, n=1)$; (c) T vibration mode $(m=0, n=1)$; (d) prismatic vibration mode $(m=1, n=0)$. Note: The results for a \emph{thick and short} SEA tube $(\eta=0.2, L/H=2.5)$ are also displayed in Fig.~\ref{axisymmetric_T_vibration.sub.3} for comparison purpose.} 
	\label{Axisymmetric_vibration_figure}
\end{figure}

Then, to clearly reveal the influences of voltage and strain-stiffening effect on the vibration behaviors, we exhibit in Fig.~\ref{Axisymmetric_vibration_figure} the variations of the natural frequency $\varpi$ with the dimensionless radial voltage $\overline{V}$ for an SEA tube characterized by the Gent $(G=97.2)$ and neo-Hookean models at three different axial pre-stretches. Four specific cases are included in Fig.~\ref{Axisymmetric_vibration_figure}: breathing mode $(m=n=0)$ in Fig.~\ref{axisymmetric_breathing.sub.1}; L vibration mode $(m=0, n=1)$ in Fig.~\ref{axisymmetric_L_vibration.sub.2}; T vibration mode $(m=0,n=1)$ in Fig.~\ref{axisymmetric_T_vibration.sub.3} and prismatic vibration mode $(m=1,n=0)$ in Fig.~\ref{Prismatic_vibration.sub.4}.  

For the breathing mode (which can be seen as a special case of L vibrations) shown in Fig.~\ref{axisymmetric_breathing.sub.1}, the lines correspond to the Gent model while the symbols denote the neo-Hookean model, and different line and symbol styles represent different axial pre-stretches. 
Clearly, the natural frequencies for the neo-Hookean SEA tube decrease nonlinearly to zero with the increase in voltage. The voltage corresponding to $\varpi=0$ is called the critical voltage $\overline{V}_{\mathrm{cr}}$ and this threshold is referred to as the breathing mode instability. 
Axisymmetric barreling instabilities \citep{simpson1984barrelling, chen2017bifurcation,wu2020nonlinear} in the SEA tube arise from the global stiffness rapidly decreasing when the applied voltage gets closer to the critical value $\overline{V}_{\mathrm{cr}}$. 
The critical voltages of the breathing mode for the neo-Hookean SEA tube are $\overline{V}_{\mathrm{cr}}=0.50$, $1.00$, $1.33$ at the axial pre-stretches $\lambda_z=2$, $1$, $0.75$, respectively, which are identical to the electro-mechanical instability voltages $\overline{V}_{\mathrm{EMI}}$ mentioned in Section \ref{section4.1}. Meanwhile, we find that a larger axial pre-stretch results in a lower critical voltage $\overline{V}_{\mathrm{cr}}$, which means the axial pre-stretch destabilizes the neo-Hookean SEA tube. 
Moreover, the curves for the neo-Hookean model all start from an identical point when $\overline{V}=0$, because the natural frequency of the breathing mode is independent of axial pre-stretch in the absence of voltage \citep{zhu2020electrostatically}. 
For the Gent SEA tube, the natural frequency of the breathing mode goes down at first due to the decrease of the global stiffness and then increases conversely and rapidly because of the strain-stiffening effect when the voltage gradually grows, and the breathing mode instability is eliminated. In addition, the analytical solution Eq.~(\ref{special_case_for_breathing_mode}) to the breathing mode of the Gent SEA tube when $\overline{V}=0$ demonstrates that although its natural frequency varies with the axial pre-stretch $\lambda_{z}$, the chosen value of Gent constant $G=97.2$ is large enough to eliminate the effect of the pre-stretch considered; that is why the curves for the Gent model almost start from an identical point when $\overline{V}=0$, which is similar to the result of the neo-Hookean model.

For the L vibration mode $(m=0, n=1)$ depicted in Fig.~\ref{axisymmetric_L_vibration.sub.2}, the variation curves of the first-order vibration frequency $\varpi$ with the applied voltage $\overline{V}$ are presented for the Gent and neo-Hookean models ($G=10, 97.2, \infty$) at three cases of axial pre-stretch. Generally, the natural frequencies $\varpi$ of this L vibration mode for both neo-Hookean and Gent models all decline nonlinearly to zero with an increase in $\overline{V}$. The phenomenon that a larger axial pre-stretch results in a lower critical voltage $\overline{V}_{\mathrm{cr}}$ exists for both the neo-Hookean model and Gent model with $G=97.2$, which is qualitatively similar to that of the breathing mode for the neo-Hookean model. However, for the Gent model with $G=10$, the critical voltage of an SEA tube subject to either axial pre-extension $(\lambda_z = 2)$ or pre-compression $(\lambda_z = 0.75)$ is larger than that without axial pre-stretch $(\lambda_z = 1)$ because of the earlier strain-stiffening effect. 

For the T vibration mode $(m=0,n=1)$, we plot the first-order natural frequency of a \emph{thin and slender} or \emph{thick and short} SEA tube for three different axial pre-stretches in Fig.~\ref{axisymmetric_T_vibration.sub.3}. The Gent model $(G=97.2)$ and neo-Hookean model are all presented for comparison. The neo-Hookean numerical data for a \emph{thick and short} SEA tube is taken from \cite{zhu2020electrostatically}. Lines represent the Gent model while symbols denote the neo-Hookean model. For the neo-Hookean model, it is obvious that the horizontal first-order natural frequencies for the SEA tubes in two geometries are independent of the applied voltage and axial pre-stretch within the range of $\overline{V}_{\mathrm{EMI}}$, as explained in \emph{Appendix D} of \cite{zhu2020electrostatically}. 
As for the Gent model, the natural frequency curves of the \emph{thin and slender} tube are also horizontal when $\overline{V}< \overline{V}_{\mathrm{EMI}}$, but as $\overline{V}$ exceeds $\overline{V}_{\mathrm{EMI}}$, its natural frequency curves rise with the increase of voltage. However, when it comes to the \emph{thick and short} Gent tube, it is apparent that the frequency curves change nonlinearly and finally reduce to zero when the applied voltage gradually increases from zero. 
Thus, we can find that the increase of the Gent tube thickness strengthens its sensitivity of natural frequency to the voltage. Moreover, although the analytical solution Eq.~(\ref{special_case_for_T_vibration}) to the first-order T vibration mode of the Gent SEA tube shows that its natural frequency at $\overline{V}=0$ relies on the axial pre-stretch $\lambda_{z}$, the Gent constant value $G=97.2$ is large enough to weaken the effect from $\lambda_{z}$, making the natural frequencies at different axial pre-stretches almost start from an identical point when $\overline{V}=0$.

In Fig.~\ref{Prismatic_vibration.sub.4}, we display the variation curves of prismatic vibration mode $(m=1,n=0)$ in a thin and slender SEA tube for the first-order vibration frequency in response to the applied voltage. A similar variation trend to the breathing mode in Fig.~\ref{axisymmetric_breathing.sub.1} is observed for both the neo-Hookean and Gent models. For example, the natural frequency for the neo-Hookean SEA tube reduces nonlinearly to zero when the voltage increases. 
The point where $\varpi=0$ for the neo-Hookean model in Fig.~\ref{Prismatic_vibration.sub.4} corresponds to the prismatic diffuse instability \citep{haughton1979bifurcation2, bortot2018prismatic}. 
We observe from Fig.~\ref{mode_shape_of_P}(a) that the shape of tube cross-section of the prismatic vibration mode for $m=1$ is analogous to that of the breathing mode; at the same time, we are considering the case of a thin-walled tube, which explains why the critical voltages of prismatic vibration mode $m=1$ for the neo-Hookean model in Fig.~\ref{Prismatic_vibration.sub.4} are very close to those of breathing mode for the neo-Hookean model in Fig.~\ref{axisymmetric_breathing.sub.1}. 
For the Gent SEA tube, when the applied voltage increases, the natural frequency first goes down to a small value because of the decrease of the global stiffness and then increases due to the strain-stiffening effect. Therefore, for this specific prismatic vibration mode $(m=1,n=0)$, the Gent SEA tube with strain-stiffening effect eliminates the prismatic diffuse instability that can occur in the neo-Hookean tube. However, for other prismatic vibration modes with $m \geq 2$, the prismatic diffuse instabilities might occur in the Gent SEA tube \citep{bortot2018prismatic}, as described below.

\begin{figure}[H] 
	\centering  
	\subfigure{
		\label{P_vibrations_N-H.sub.1}
		\includegraphics[width=0.48\textwidth]{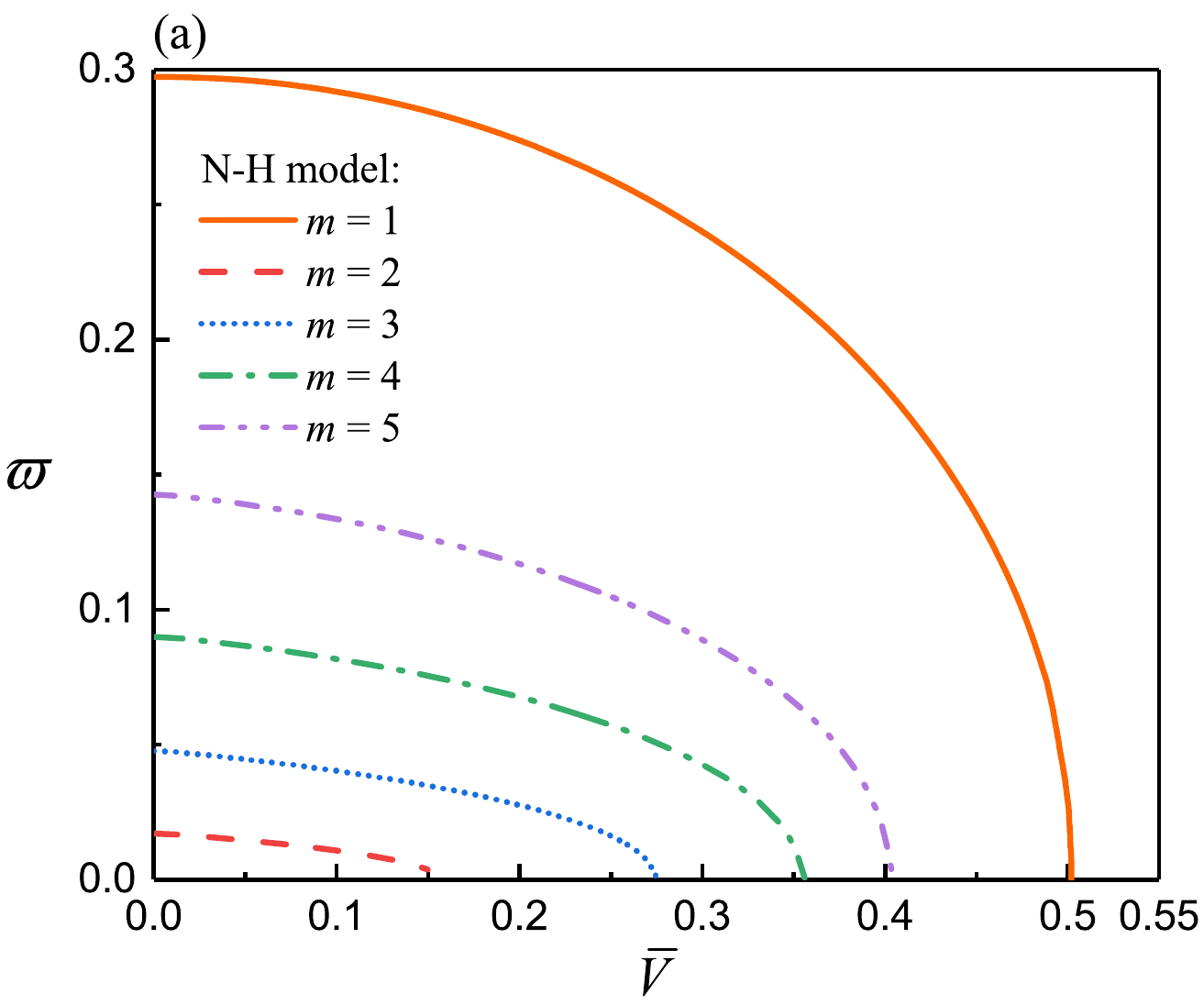}}
	\subfigure{
		\label{P_vibrations_Gent.sub.2}
		\includegraphics[width=0.48\textwidth]{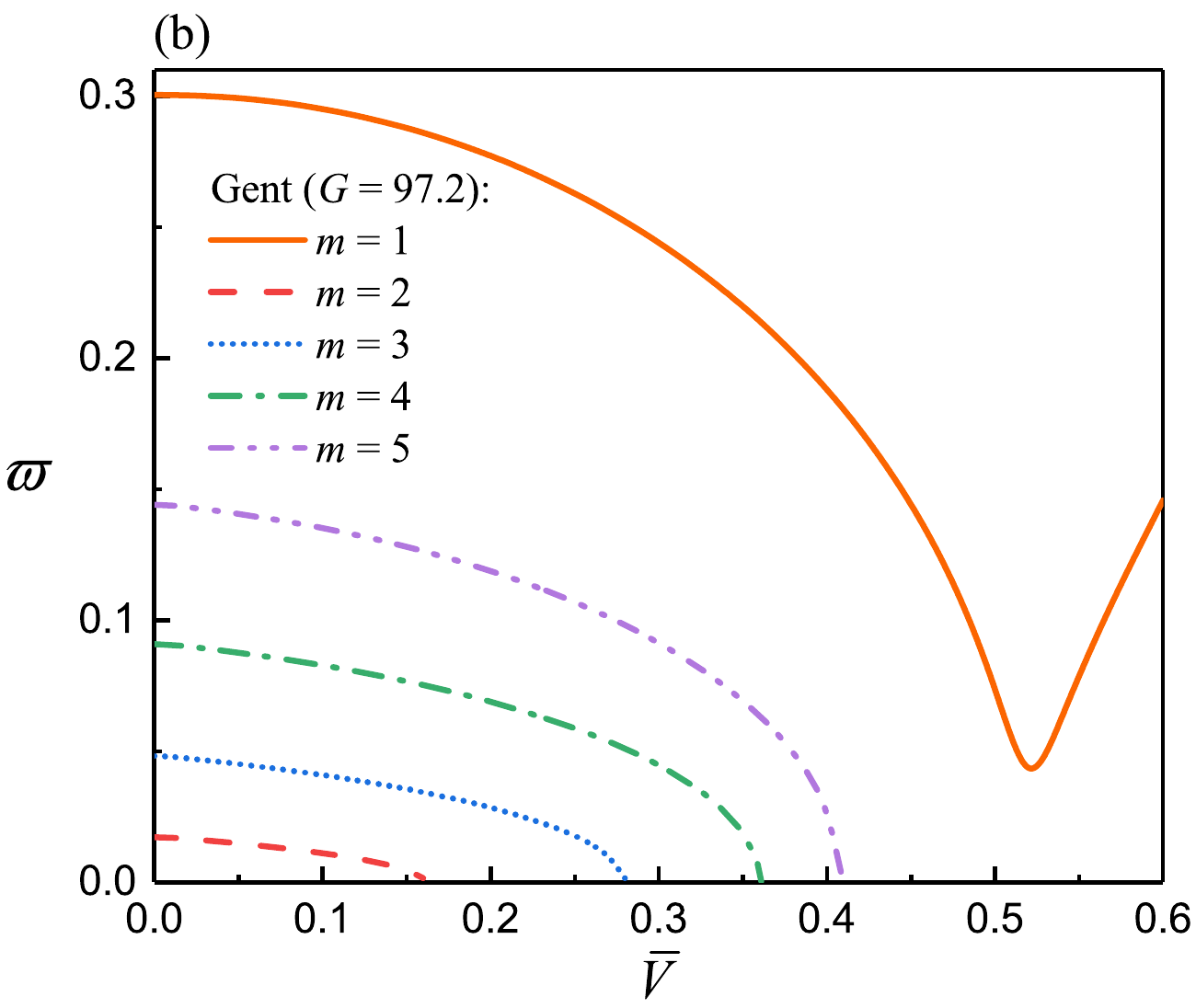}}
	\caption{The first-order dimensionless frequencies $\varpi$ of prismatic vibration modes $m=1\sim5$ as functions of the dimensionless radial voltage $\overline{V}$ in a pre-stretched ($\lambda_{z}=2$) thin and slender SEA tube $(\eta=0.9, L/H=10)$: (a) the neo-Hookean model; (b) the Gent model $(G=97.2)$.} 
	\label{P_vibrations_Gent_N-H}
\end{figure}

We further plot the curves of prismatic vibration frequency as functions of the radial voltage for $m=1\sim5$ in Fig.~\ref{P_vibrations_Gent_N-H} for both the neo-Hookean and Gent SEA tubes. 
We observe from Fig.~\ref{P_vibrations_N-H.sub.1} for the neo-Hookean model that all of the prismatic vibration frequencies decrease nonlinearly to zero in response to the continuously increasing radial voltage. 
The critical voltage for $m=1$ is higher than those for $m=2\sim5$ and the critical voltage goes up with $m$ for $m\geq2$. 
Specifically, the critical voltages $\overline{V}_{\mathrm{cr}}$ of prismatic vibrations for different circumferential mode numbers are: $\overline{V}_{\mathrm{cr}}=0.502$ for $m=1$, $\overline{V}_{\mathrm{cr}}=0.157$ for $m=2$, $\overline{V}_{\mathrm{cr}}=0.275$ for $m=3$, $\overline{V}_{\mathrm{cr}}=0.356$ for $m=4$, and $\overline{V}_{\mathrm{cr}}=0.404$ for $m=5$. However, we notice from Fig.~\ref{P_vibrations_Gent.sub.2} that prismatic instability of $m=1$ for the neo-Hookean model does not exist for the Gent model. Similar to Fig.~\ref{P_vibrations_N-H.sub.1}, as the radial voltage gradually increases, the first-order vibration frequency decreases to zero with $\overline{V}_{\mathrm{cr}}=0.160$ for $m=2$, $\overline{V}_{\mathrm{cr}}=0.280$ for $m=3$, $\overline{V}_{\mathrm{cr}}=0.361$ for $m=4$, and $\overline{V}_{\mathrm{cr}}=0.409$ for $m=5$, respectively. The critical voltage for the Gent model increases monotonically when increasing the circumferential mode number $m$ for $m\geq2$. Moreover, the critical voltage of the same circumferential mode number for the Gent model is slightly larger than that for the neo-Hookean model. These phenomena agree well with the results obtained by \cite{bortot2018prismatic}
\textcolor{black}{and their Figs.~4(a-c) illustrated the critical voltage for the prismatic diffuse instability as a function of thickness-to-mean radius ratio of the SEA tubes clamped without axial pre-stretch or clamped after pre-stretch, where cases for both the Gent and neo-Hookean models were presented for various circumferential mode numbers.}

For a better illustration, we present in Fig.~\ref{mode_shape_of_P} four prismatic vibration mode shapes for $m=1\sim4$ in a pre-stretched ($\lambda_{z}=2$) thin and slender Gent SEA tube subject to a radial voltage $\overline{V}=0.2$. 
The intensity of circumferential gridlines represents the distribution of radial displacement, while the radially scattered gridlines denote the distribution of circumferential displacement. 
According to Eq.~(\ref{Assumed_solution_Pris}), the prismatic vibrations only have two coupled components: the radial displacement component $u_{r}$ and the circumferential displacement component $u_{\theta}$. Thus, the SEA tube remains prismatic in axial direction while its cross-section loses its circular shape for $m\neq0$. The circumferential mode number $m$ plays an important role in shaping the prismatic-vibrating mode shape. It can be seen from Fig.~\ref{mode_shape_of_P} that $m$ vibration crests and troughs are distributed in the prismatic mode shapes for the $m$th order circumferential mode number.

\begin{figure}[H] 
	\centering
	\includegraphics[width=1.0\textwidth]{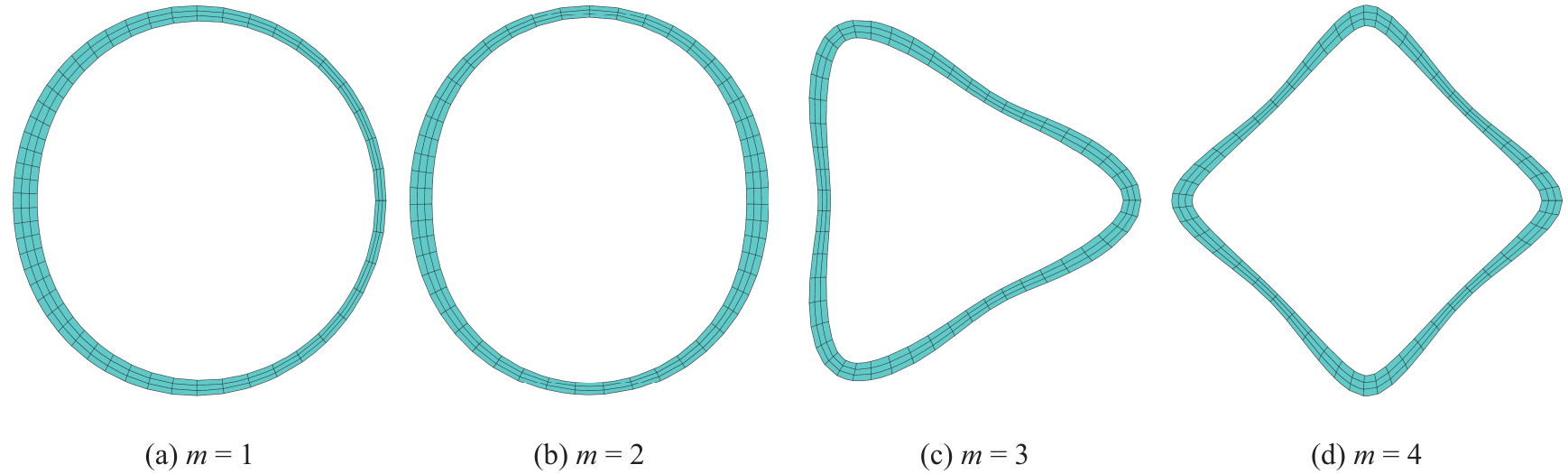} 
	\caption{Prismatic vibration mode shapes in a pre-stretched $(\lambda_{z}=2)$ thin and slender Gent SEA tube $(\eta=0.9, L/H=10)$ subject to $\overline{V}=0.2$ for different circumferential mode numbers: (a) $m=1$; (b) $m=2$; (c) $m=3$; (d) $m=4$.} 
	\label{mode_shape_of_P} 
\end{figure}
\begin{figure}[H] 
	\centering  
	\subfigure{
		\label{w-g_breathing.sub.1}
		\includegraphics[width=0.48\textwidth]{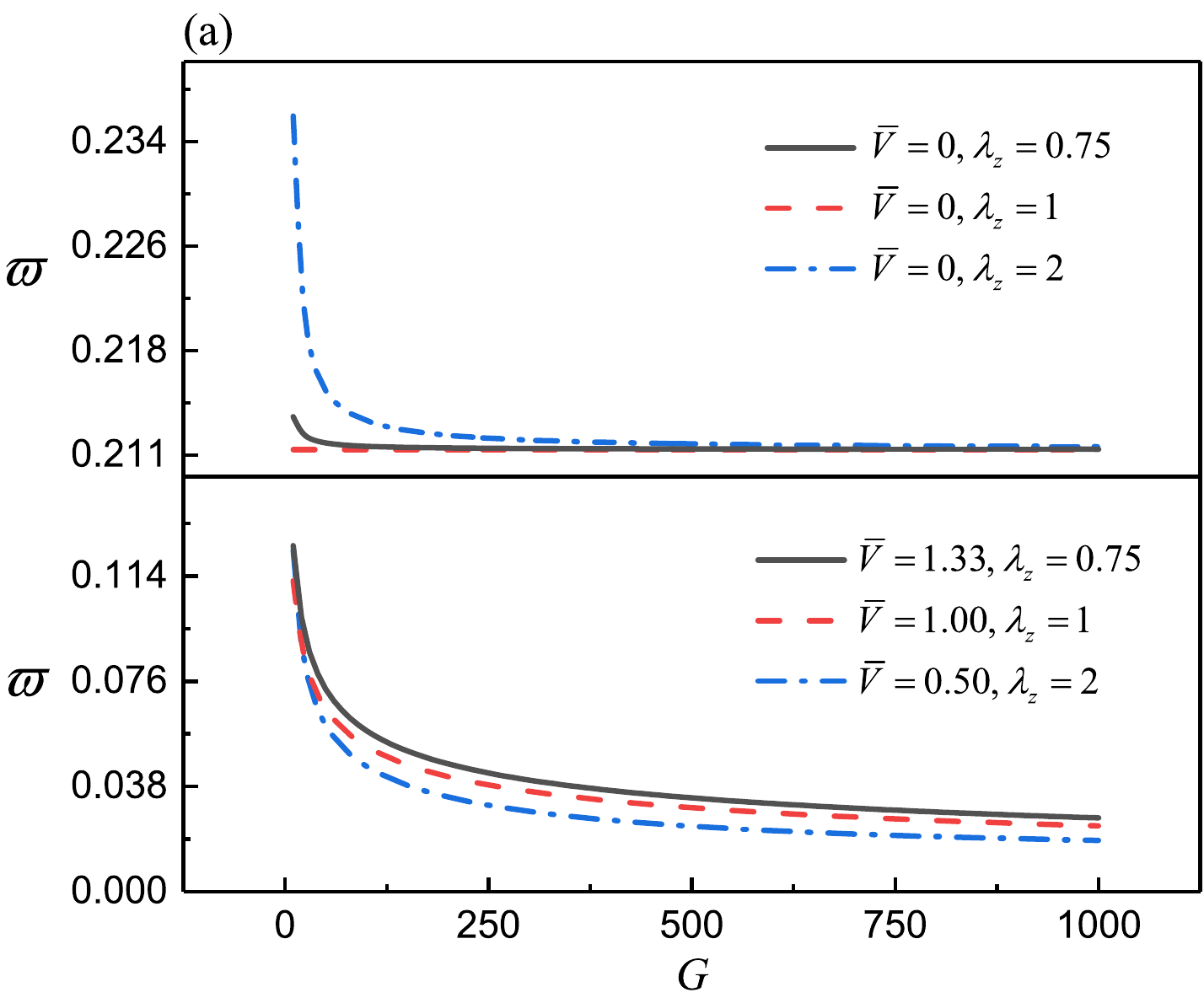}}
	\subfigure{
		\label{w-g_L_vibration.sub.2}
		\includegraphics[width=0.48\textwidth]{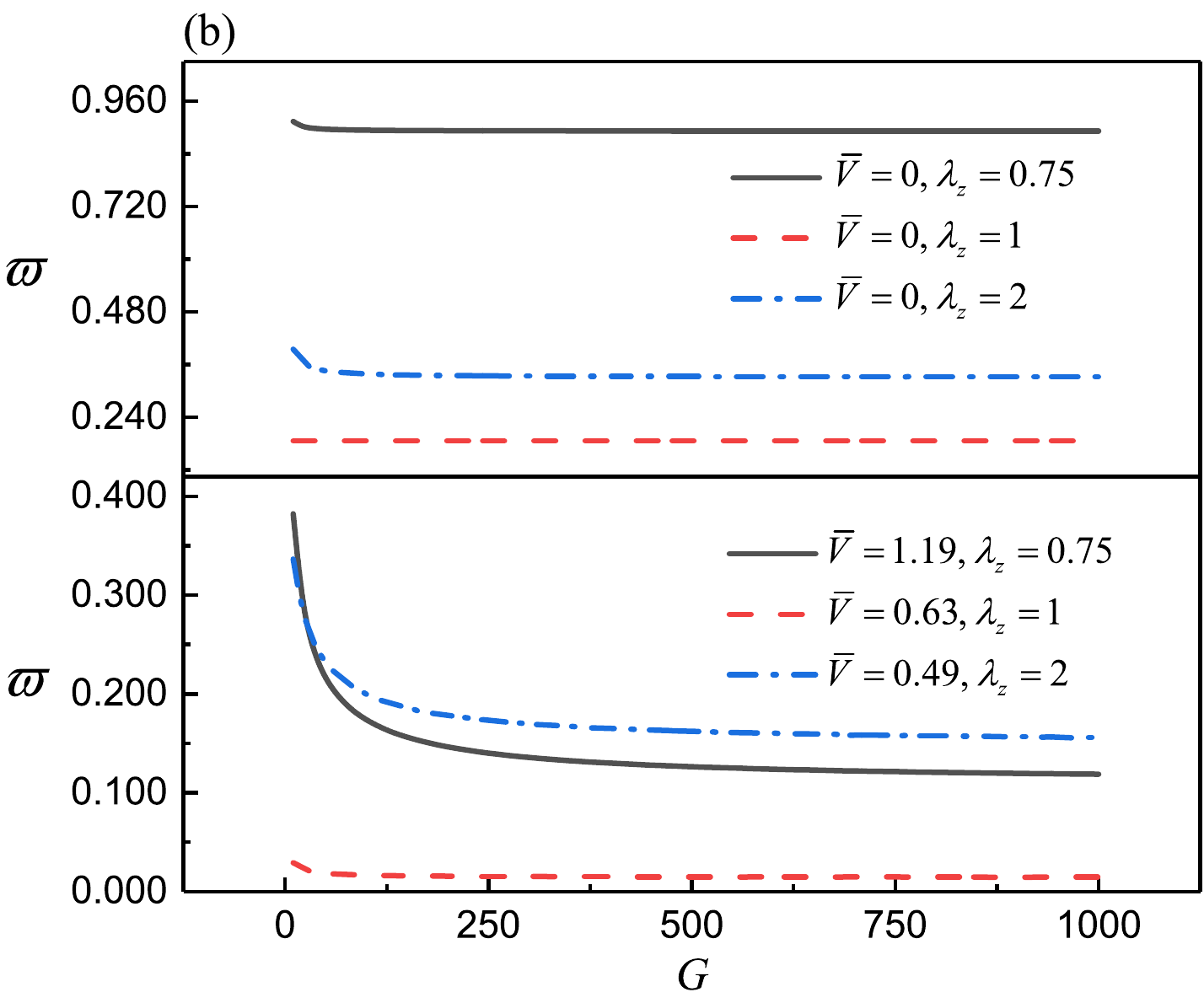}}
	\subfigure{
		\label{w-g_T_vibration.sub.3}
		\includegraphics[width=0.48\textwidth]{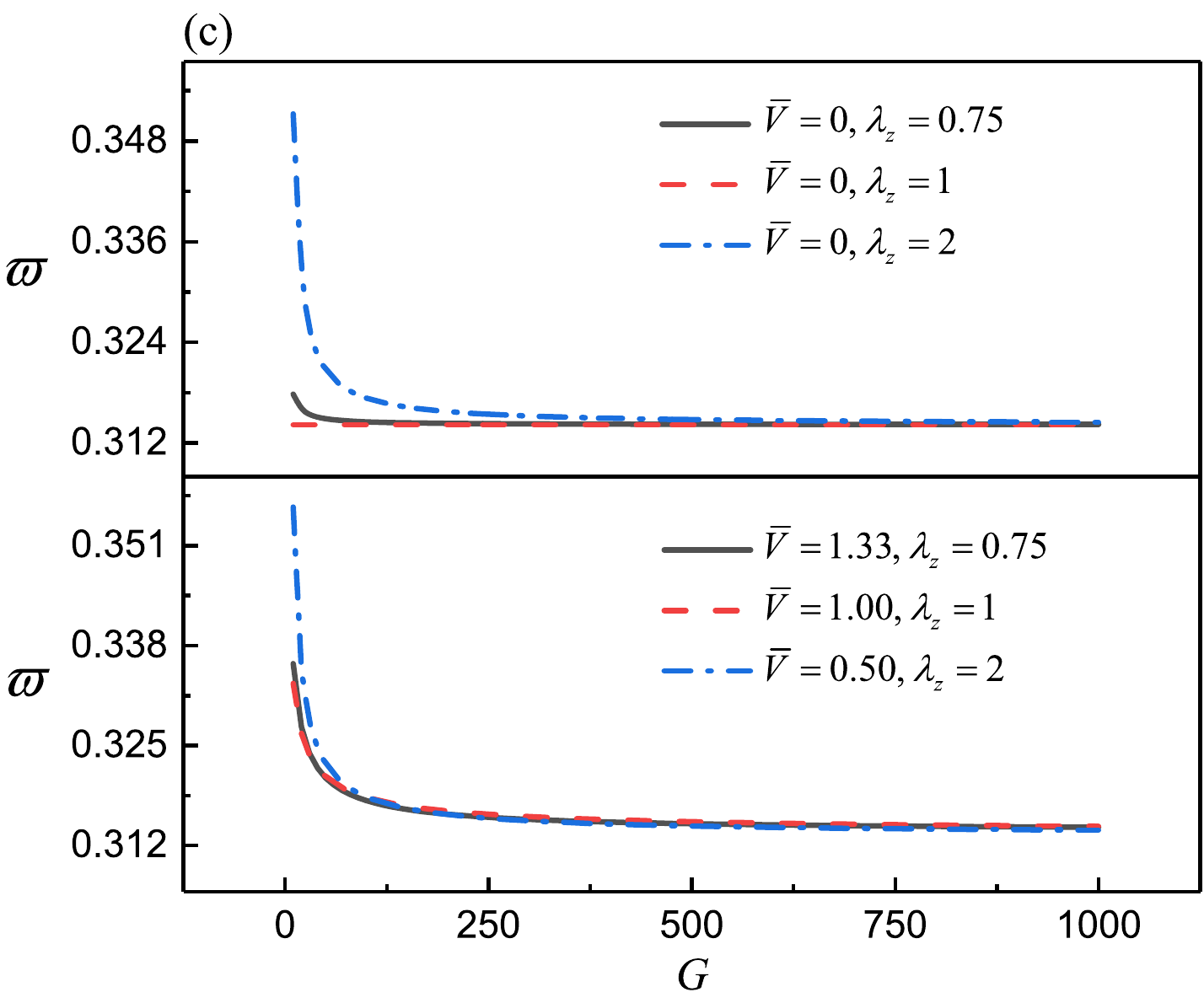}}
	\subfigure{
		\label{w-g-P_vibration.sub.4}
		\includegraphics[width=0.48\textwidth]{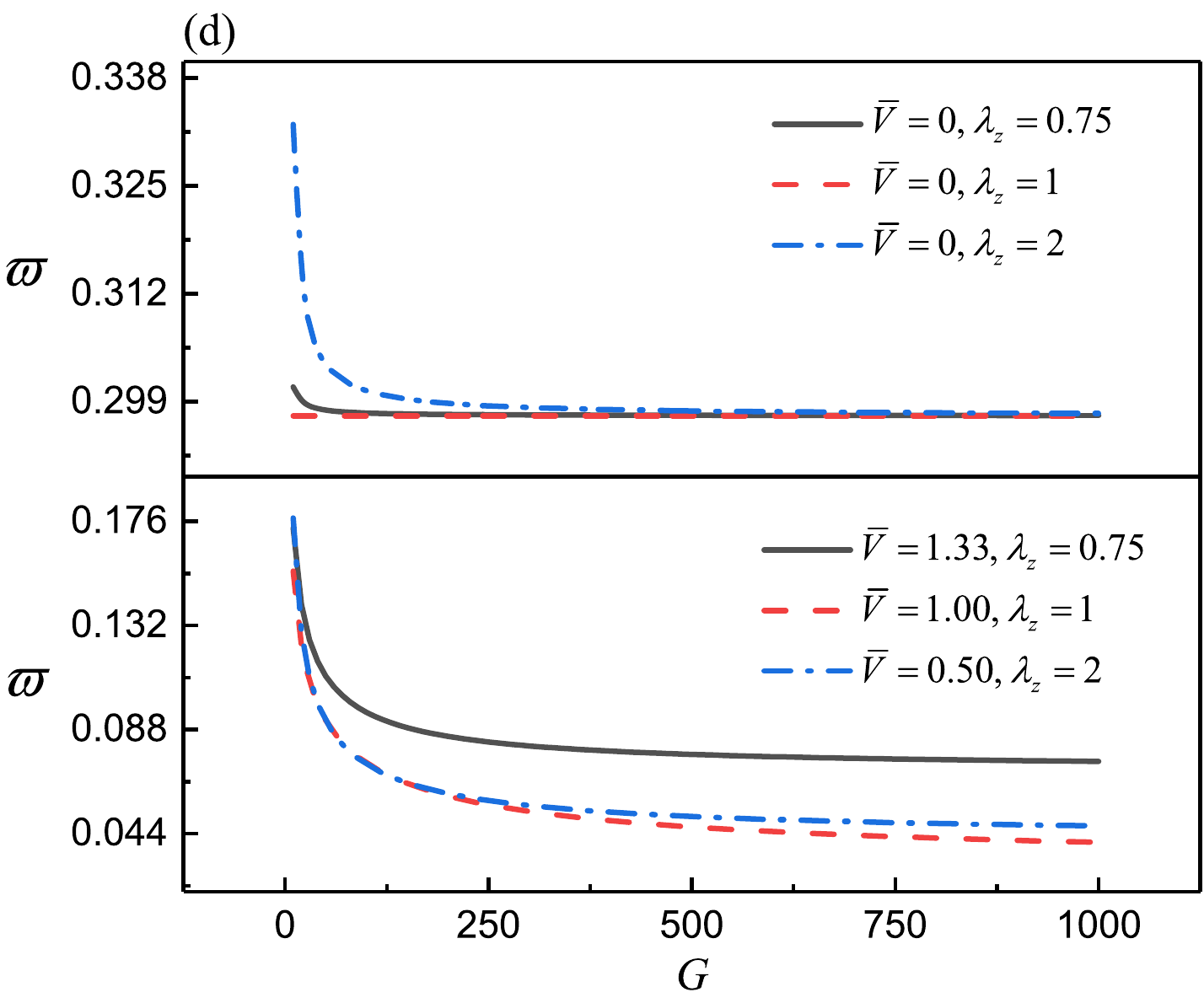}}
	\caption{Variation curves of the first-order dimensionless vibration frequency $\varpi$ with the Gent parameter $G$ in a pre-stretched thin and slender $(\eta=0.9, L/H=10)$ SEA tube under different axial pre-stretches and voltages: (a) breathing mode $(m=0, n=0)$; (b) L vibration mode $(m=0, n=1)$; (c) T vibration mode $(m=0, n=1)$; (d) prismatic vibration mode $(m=1, n=0)$. } 
	\label{omega_vs_G_figure}
\end{figure}

To better understand how strain stiffening affects axisymmetric and prismatic vibrations, we present variation curves of the first-order natural frequency $\varpi$ with the Gent parameter $G$ in Fig.~\ref{omega_vs_G_figure} for a thin and slender SEA tube under different axial pre-stretches and voltages. We examine the strain-stiffening effect under two circumstances: no radial voltage $\overline{V}=0$ and high radial voltage close to the $\overline{V}_{\mathrm{EMI}}$ of the neo-Hookean model. We see in Fig.~\ref{omega_vs_G_figure} that as the Gent parameter $G$ increases to infinity, the frequency tends to converge to that of the neo-Hookean model, as expected. 
Recalling Figs.~\ref{Axisymmetric_vibration_figure}(a), \ref{Axisymmetric_vibration_figure}(b) and \ref{Axisymmetric_vibration_figure}(d), the natural frequencies approach zero as the applied radial voltage approaches $\overline{V}_{\mathrm{EMI}}$ in a neo-Hookean SEA tube. Therefore, for the high radial voltage close
to the $\overline{V}_{\mathrm{EMI}}$, the vibration frequency $\varpi$ decreases monotonically to zero with the increase of $G$, as shown in the bottom subplots of Figs.~\ref{omega_vs_G_figure}(a), \ref{omega_vs_G_figure}(b) and \ref{omega_vs_G_figure}(d).
However, the natural frequency for the T vibration mode in Fig.~\ref{omega_vs_G_figure}(c) decreases to a finite value when increasing $G$ to infinity, because the first-order T vibration frequency in a thin and slender SEA tube is independent of radial voltage and axial pre-stretch within the range of $\overline{V}_{\mathrm{EMI}}$, regardless of whether the neo-Hookean or Gent model is considered, as displayed in Fig.~\ref{Axisymmetric_vibration_figure}(c). When there is no radial voltage in each top subplot of Figs.~\ref{omega_vs_G_figure}(a)-(d), the SEA tube subject to pre-extension or pre-compression gets stiffened rapidly with the strain as $G$ becomes relatively small, which explains that the natural frequency increases drastically when $G$ decreases further from the small value. 
In other words, the stronger the strain-stiffening effect is, the easier it is to achieve a strain-stiffened state under a small deformation. Additionally, it can be seen from the top subplots of Figs.~\ref{omega_vs_G_figure}(a)-(d) that for a fixed pre-stretch $\lambda_{z}=1$ and zero radial voltage, the natural frequencies for all vibration modes are independent of the Gent parameter $G$, which physically means that the strain-stiffening effect has no influence on the vibration frequency of a SEA tube without electro-mechanical biasing fields.

\subsection{Effect of electro-mechanical biasing fields on non-axisymmetric vibrations.}

In this subsection, we expound on how the electro-mechanical biasing fields influence the non-axisymmetric vibration characteristics in a thin and slender $(\eta=0.9, L/H=10)$ SEA tube.

\begin{figure}[H] 
	\centering  
    \subfigure{
	\label{w-n_fre.sub.1}
	\includegraphics[width=0.48\textwidth]{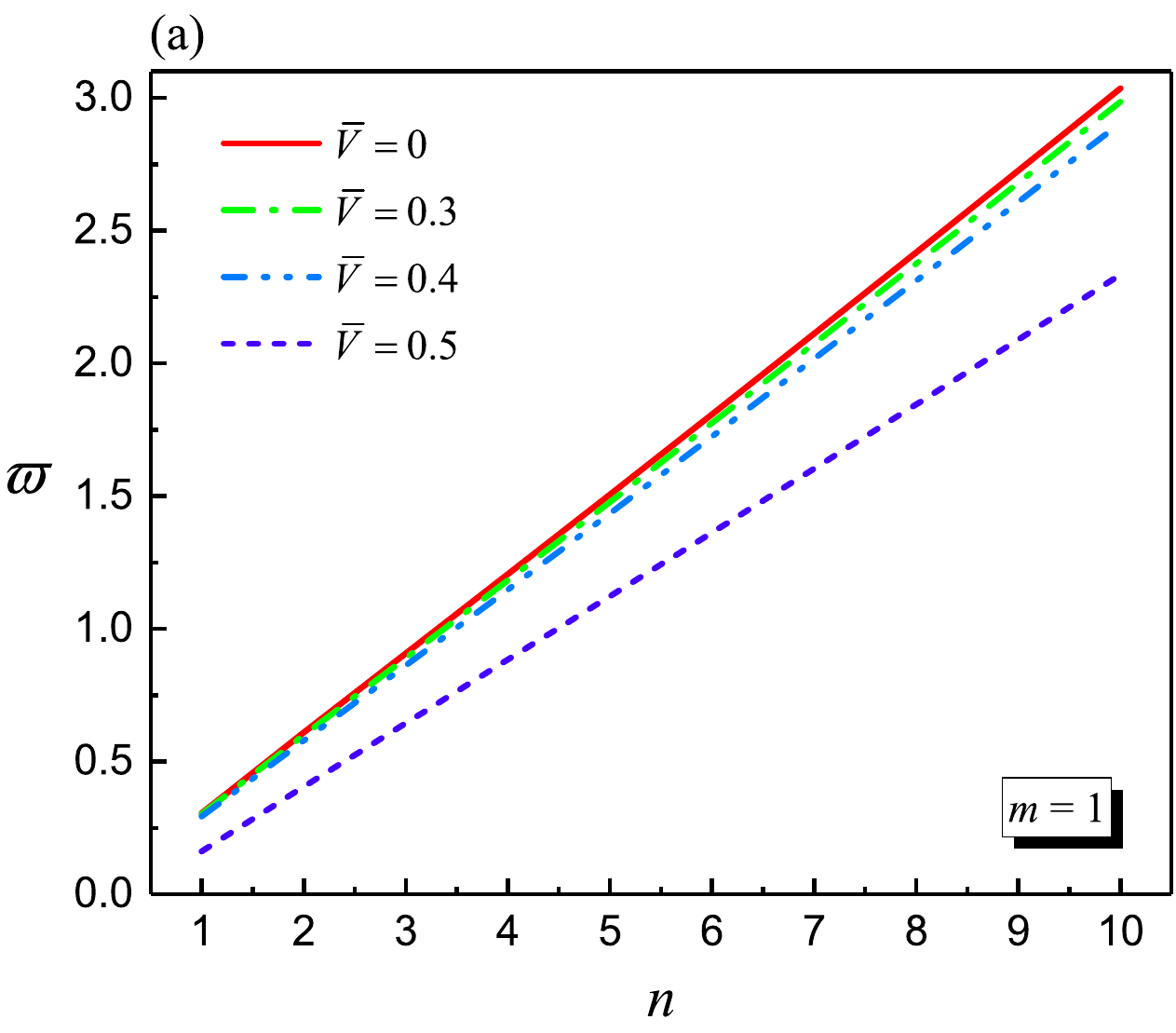}}
    \subfigure{
	\label{w-n_fre.sub.2}
	\includegraphics[width=0.48\textwidth]{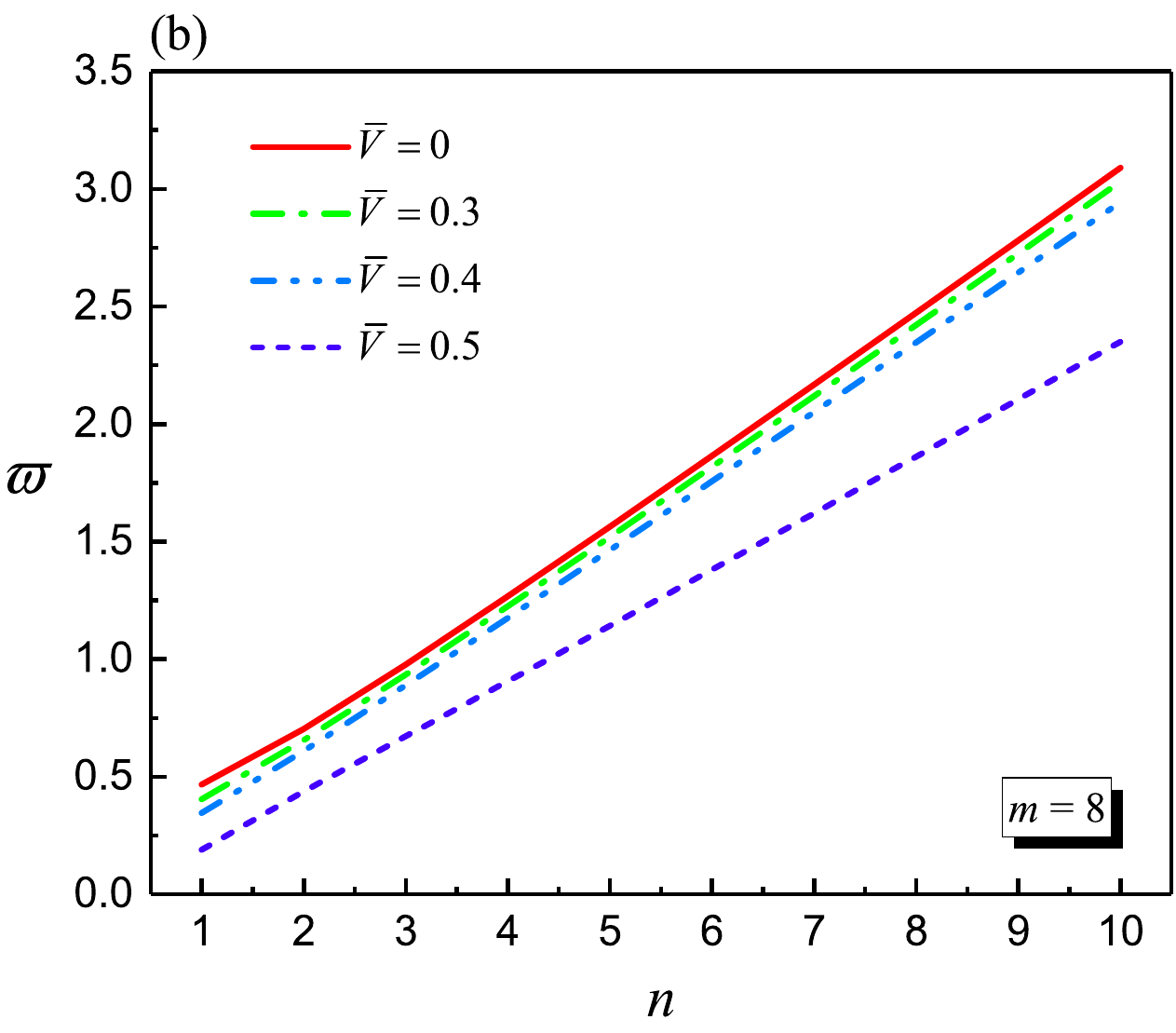}}
	\caption{The first-order dimensionless frequency $\varpi$ of non-axisymmetric vibrations versus the axial mode number $n$ in a pre-stretched ($\lambda_{z}=$2) thin and slender Gent SEA tube $(\eta=0.9, L/H=10, G=97.2)$ for different values of radial electric voltage and two circumferential mode numbers: (a) $m=1$; (b) $m=8$.} 
	\label{Non_axi_frequency_spectra}
\end{figure}

First, we display in Fig.~\ref{Non_axi_frequency_spectra} the variation of the first-order dimensionless frequency $\varpi$ with the axial mode number $n$ in a pre-stretched ($\lambda_{z}=2$) Gent SEA tube for different values of radial voltage. Two representative circumferential mode numbers $m=1$ and $m=8$ are selected in Figs.~\ref{w-n_fre.sub.1} and \ref{w-n_fre.sub.2}, respectively. It can be seen that the vibration frequency increases almost linearly with the increase of axial mode number $n$. For $m=1$, a low radial voltage (e.g., $\overline{V}<0.4$) barely affects the natural frequency,  whereas the natural frequency decreases rapidly when the SEA tube is subject to a relatively higher radial voltage (e.g., $\overline{V}=0.5$). Moreover, when comparing Fig.~\ref{w-n_fre.sub.2} with $m=8$ to Fig.~\ref{w-n_fre.sub.1}, we observe a more noticeable frequency change within the range of small $n$ (e.g., $n=1 \sim 4$) when increasing the radial voltage with $\overline{V} \leq 0.4$.

\begin{figure}[H] 
	\centering  
	\subfigure{
		\label{w-m_fre.sub.1}
		\includegraphics[width=0.48\textwidth]{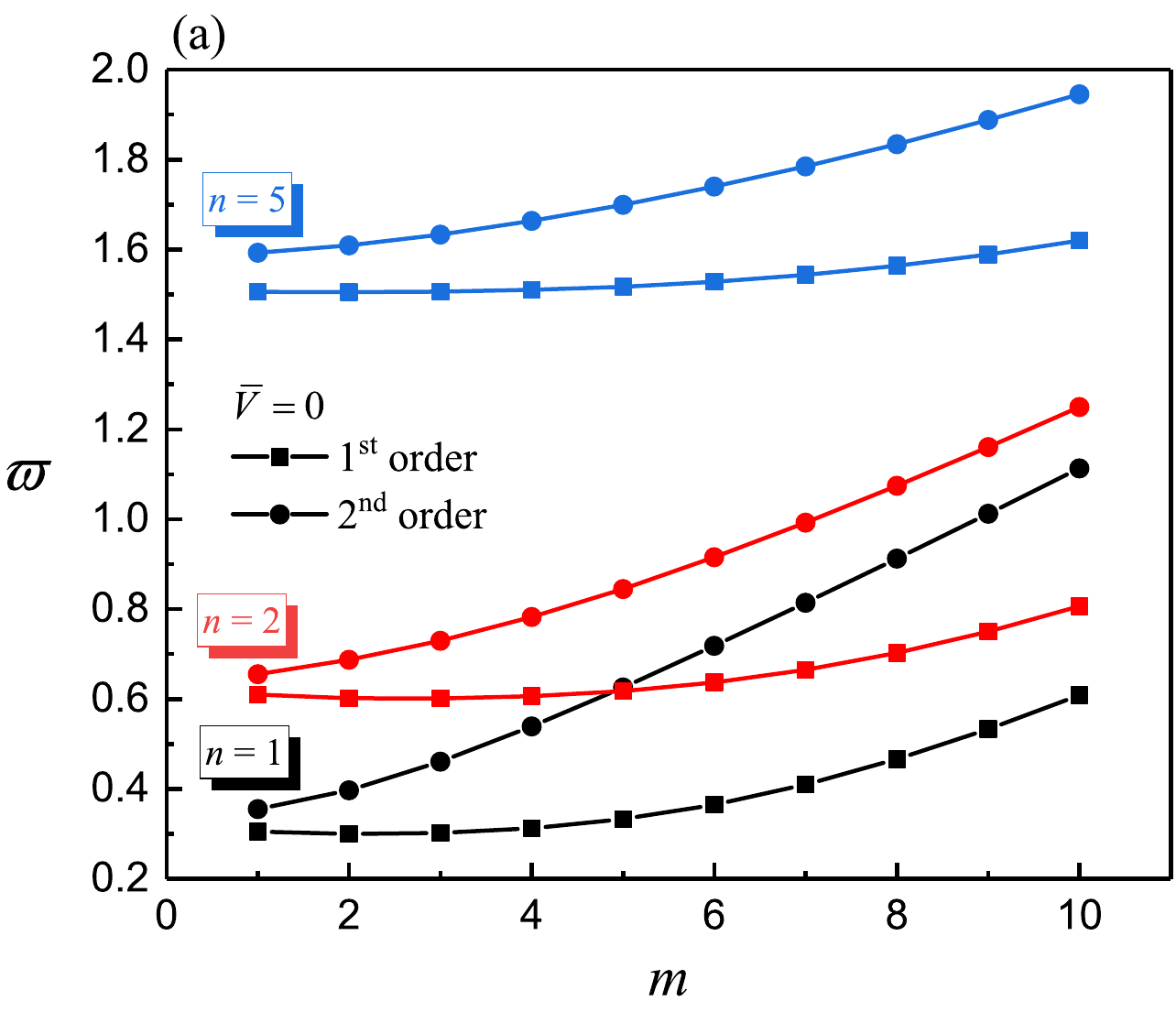}}
	\subfigure{
		\label{w-m_vibration_fre.sub.2}
		\includegraphics[width=0.48\textwidth]{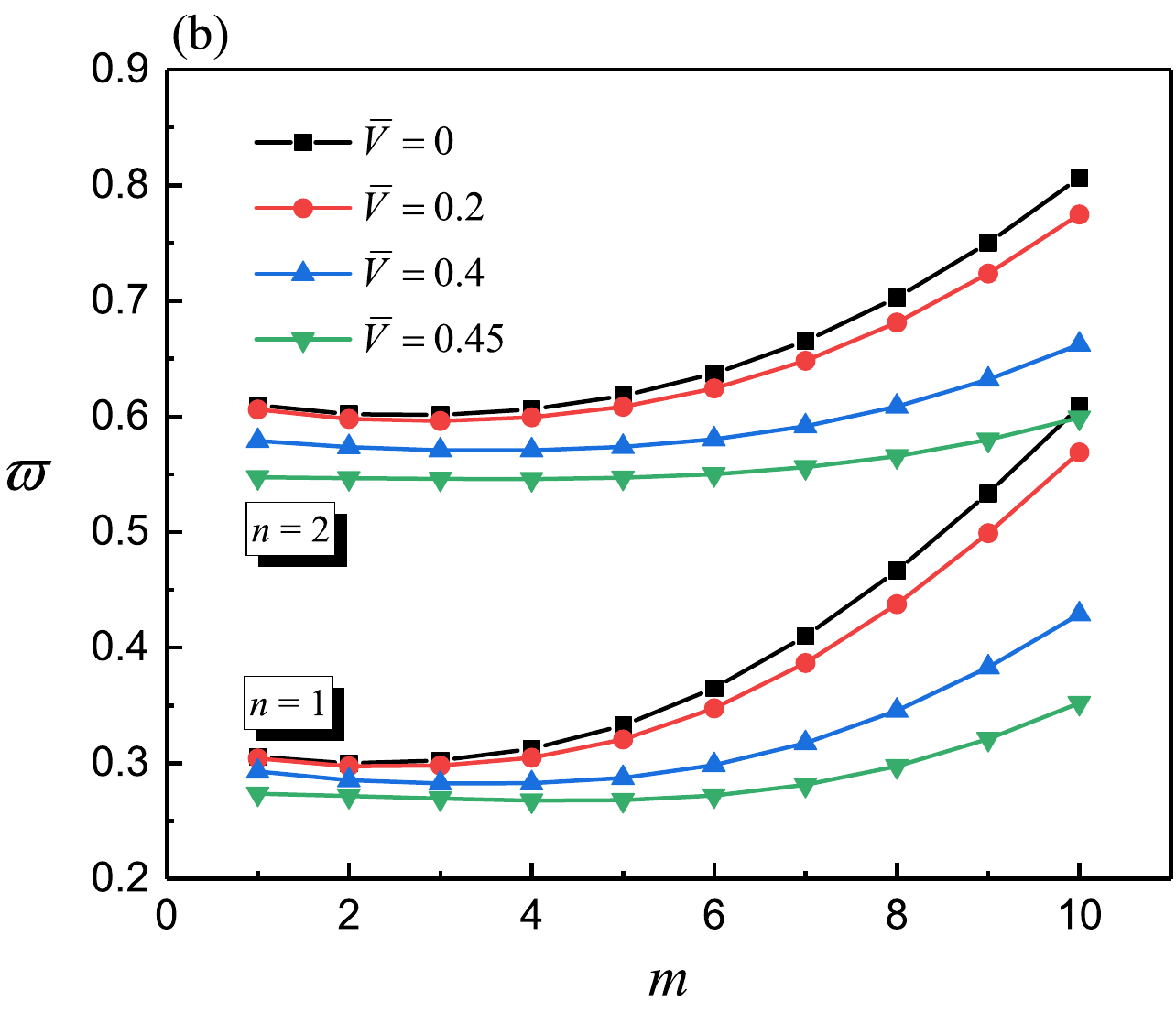}}
	\caption{The dimensionless frequency $\varpi$ of non-axisymmetric vibrations versus the circumferential mode number $m$ in a pre-stretched ($\lambda_{z}=2$) thin and slender Gent SEA tube $(\eta=0.9, L/H=10, G=97.2)$: (a) the first two frequencies for three axial mode numbers $(n=1, 2$ and $5)$ with $\overline{V}=0$; (b) the first-order frequency for different radial voltages $(\overline{V}=0$, $0.2$, $0.4$ and $0.45)$ and two axial mode numbers $(n=1$ and $2)$.} 
	\label{w_m_frequency_spectra}
\end{figure}

Then, to investigate the influence of circumferential mode number $m$, we plot in Fig.~\ref{w_m_frequency_spectra} the dimensionless frequency spectra ($\varpi$ versus $m$) in a pre-strecthed ($\lambda_{z}=2$) Gent SEA tube for different values of axial mode number and radial voltage. 
Specifically, in Fig.~\ref{w-m_fre.sub.1}, we illustrate the first two vibration frequencies without radial voltage for three axial mode numbers $n=1,2,5$. 
We observe that the first-order frequency change is very small when $m\leq4$, and that the frequency then goes up gradually with the increase of $m$. In addition, the second-order frequency increases significantly and monotonically with the axial mode number $m$. In Fig.~\ref{w-m_vibration_fre.sub.2}, we consider two axial mode numbers $n=1,2$ and four radial voltages $\overline{V}=0, 0.2, 0.4, 0.45$. 
We see that the overall frequency curves of non-axisymmetric vibration decline when increasing the radial voltage across the entire range of circumferential mode number. In particular, the voltage-induced frequency reduction of a large circumferential mode number is more obvious than that of a small one.

\begin{figure}[h!] 
	\centering  
	\subfigure{
		\label{nonaxisymmetric.sub.3}
		\includegraphics[width=0.44\textwidth]{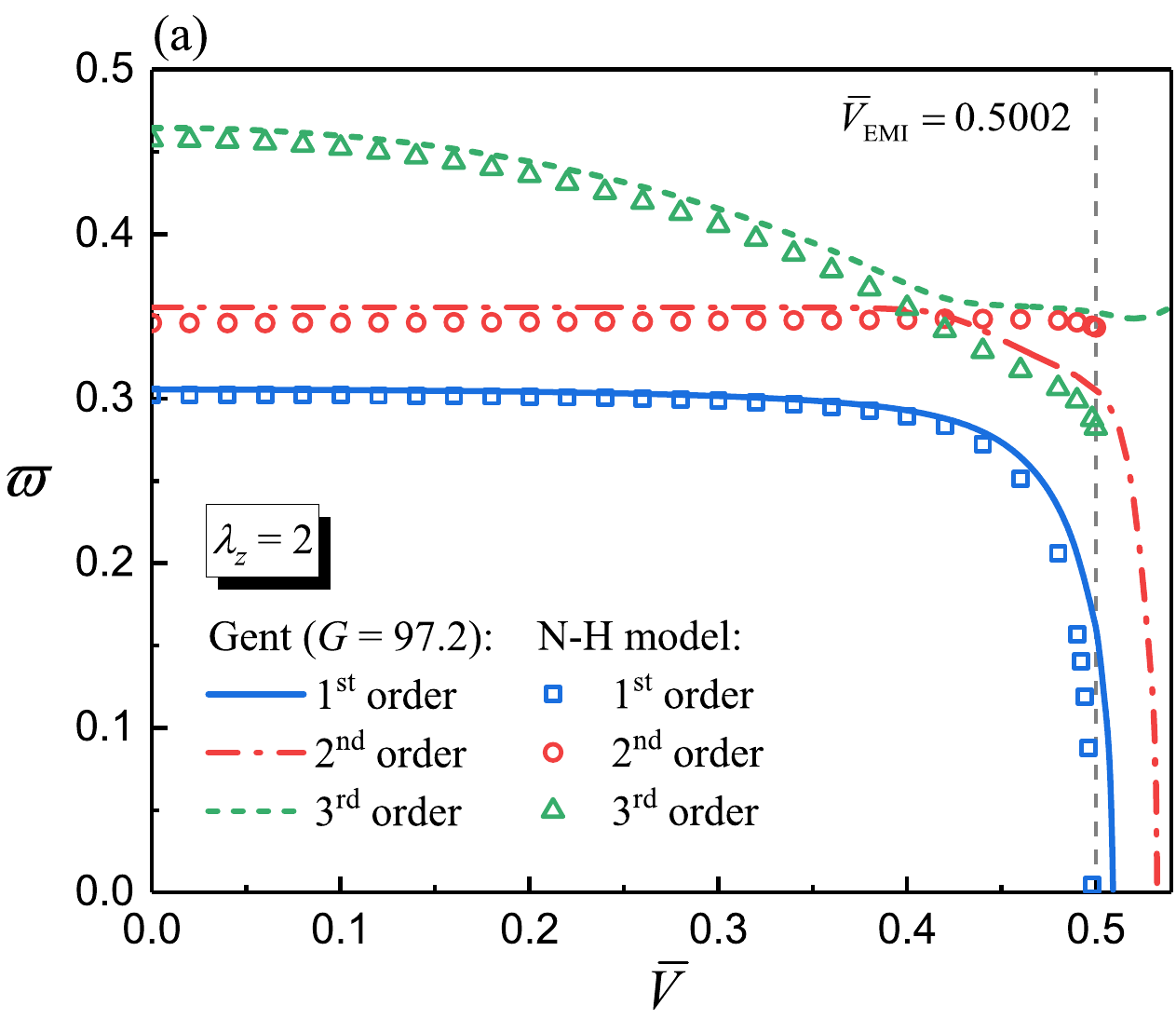}}
	\subfigure{
		\label{nonaxisymmetric.sub.2}
		\includegraphics[width=0.44\textwidth]{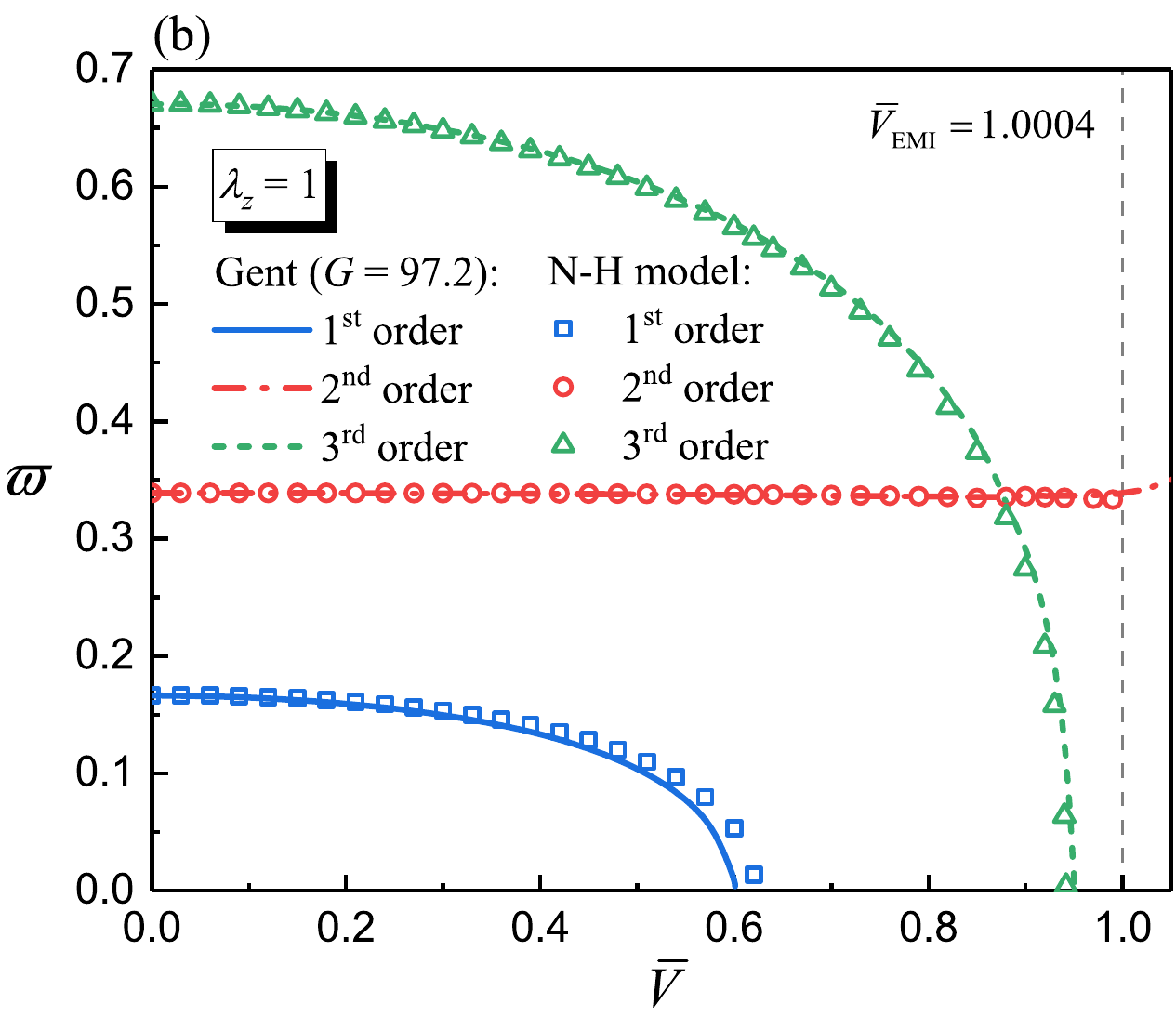}}
	\subfigure{
		\label{nonaxisymmetric.sub.4}
		\includegraphics[width=0.44\textwidth]{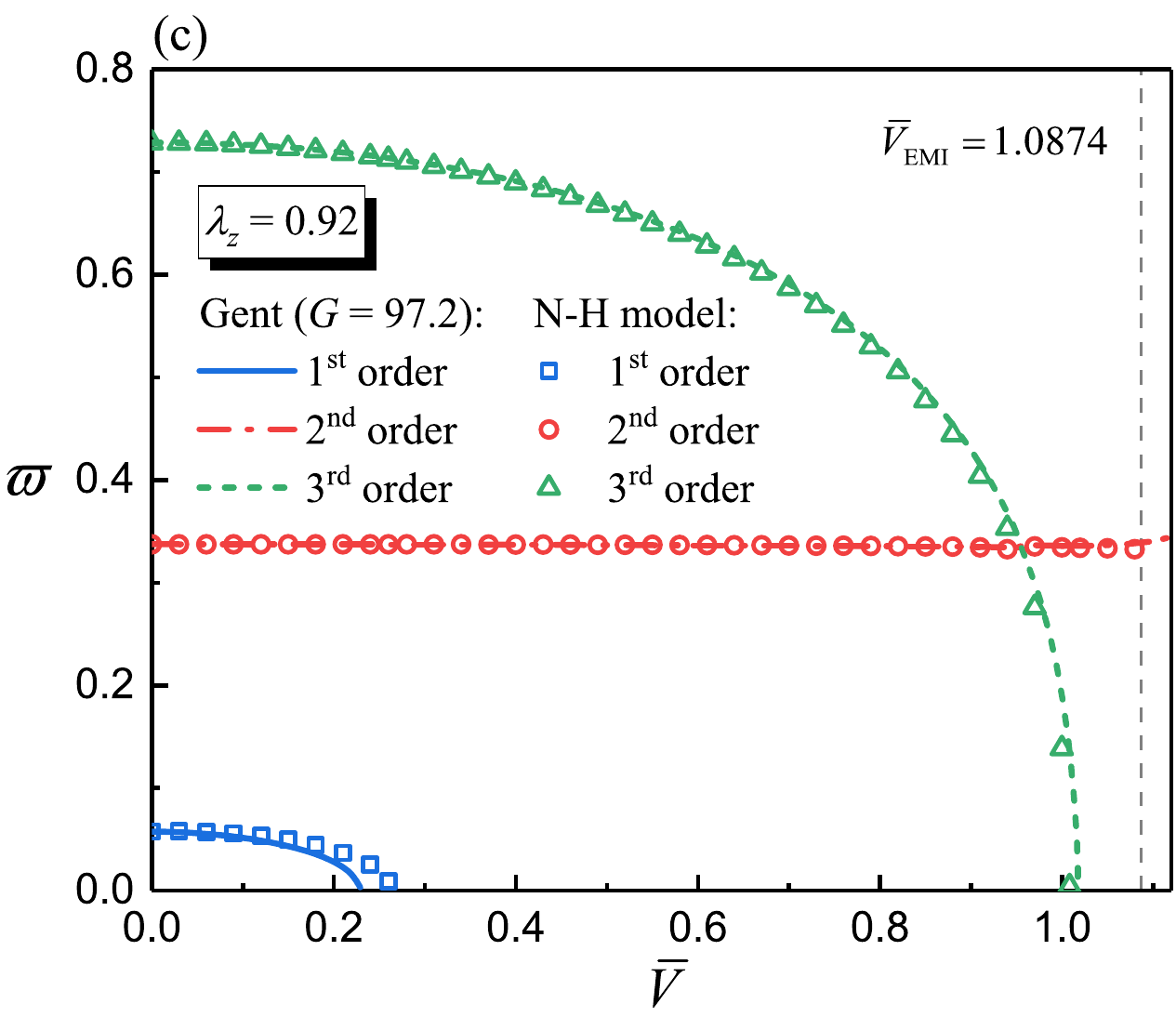}}
	\subfigure{
		\label{nonaxisymmetric.sub.1}
		\includegraphics[width=0.44\textwidth]{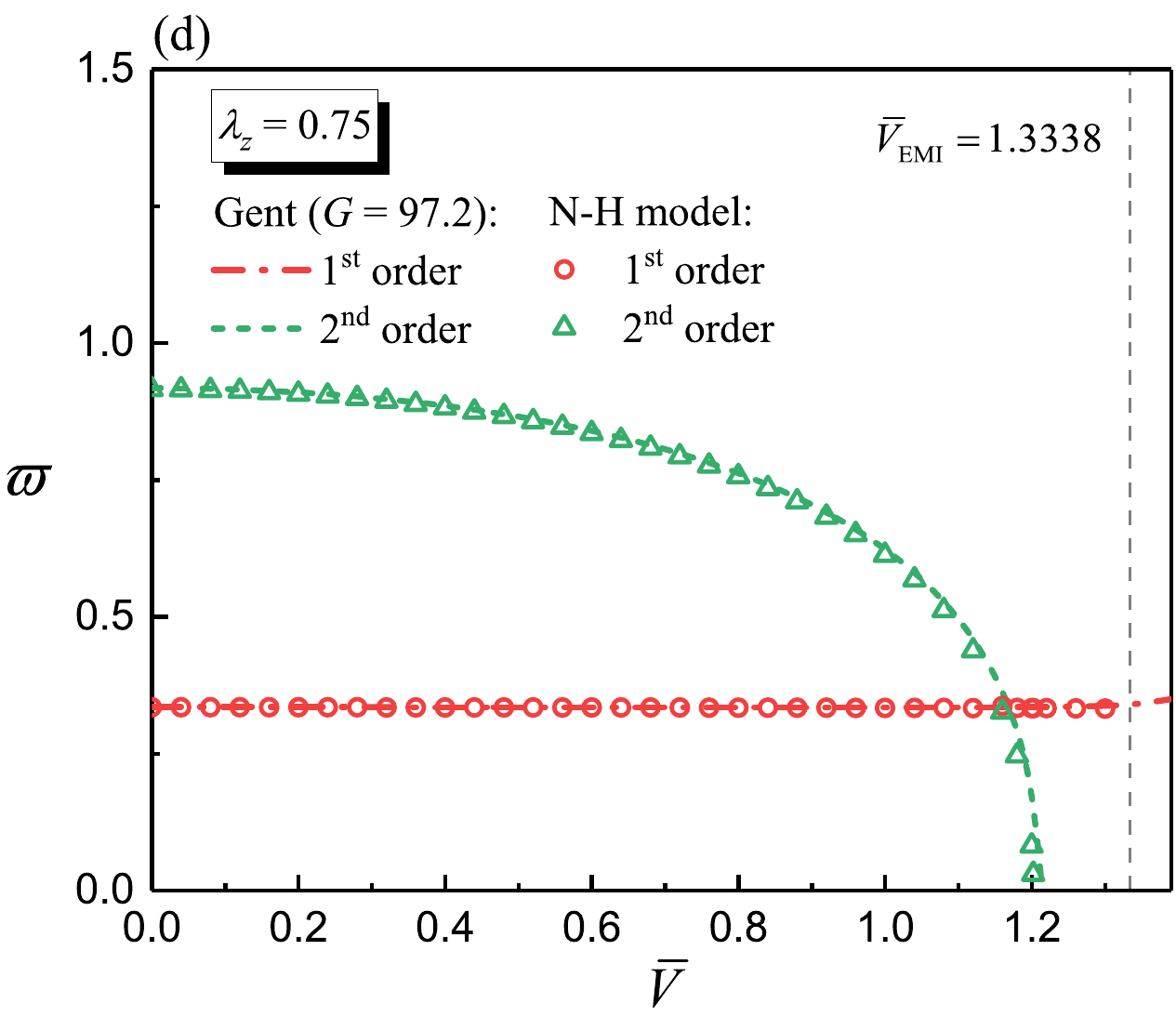}}
	\caption{The first three or two vibration frequencies $\varpi$ of non-axisymmetric mode with $m=n=1$ as functions of radial voltage $\overline{V}$ in a thin and slender $(\eta=0.9, L/H=10)$ SEA tube for both the neo-Hookean model and Gent model $(G=97.2)$ under four different axial pre-stretches: (a) $\lambda_z=2$; (b) $\lambda_z=1$; (c) $\lambda_z=0.92$; (d) $\lambda_z=0.75$.} 
	\label{nonaxisymmetric_vibration_figure}
\end{figure}

In Fig.~\ref{nonaxisymmetric_vibration_figure}, we depict the variation curves of the natural frequency $\varpi$ of non-axisymmetric vibration mode $m=n=1$ with the radial voltage $\overline{V}$ for four representative axial pre-stretches $(\lambda_{z}=2, 1, 0.92, 0.75)$. Both the Gent model $(G=97.2)$ and neo-Hookean model are shown for comparison. Lines represent the results based on the Gent model, while symbols indicate those predicted by the neo-Hookean model. 

Fig.~\ref{nonaxisymmetric.sub.3} shows the variation curves of $\varpi$ versus $\overline{V}$ with the SEA tube subject to a fixed pre-extention $\lambda_{z}=2$. 
The first-order frequency goes down to zero monotonically when increasing the voltage for both the neo-Hookean and Gent models. 
The point where $\varpi=0$ for the non-axisymmetric vibrations corresponds to the 3D non-axisymmetric buckling instability \citep{haughton1979bifurcation2, su2020voltage}. 
\textcolor{black}{In the context of voltage-controlled non-axisymmetric instabilities, \cite{su2020voltage} illustrated the critical stretch and pattern shapes associated with the buckling of SEA tubes in his Fig.~10 and considered specific radius ratios and length aspect ratios, under various applied voltages and torsions.}
We note from Fig.~\ref{nonaxisymmetric.sub.3} that the veering phenomenon occurs between the second-order mode (Mode 2) and third-order mode (Mode 3) for the Gent model, while mode crossing happens for the neo-Hookean model. The term `veering' here refers to the situation where two branches approach each other and then veer away and diverge instead of crossing. Mathematically, veering is accompanied by rapid changes in the eigenvectors \citep{mace2012wave, wu2017guided}. The sequence of mode order needs to be clarified here: for the mode veering phenomenon, since there is no mode crossing among branches, the sequence is still defined based on the frequency, from low to high. However, for the mode-crossing phenomenon, the modes that remain the same before and after the crossing point are defined as modes with the same order, rather than being based on their frequency.

\begin{figure}[h!] 
	\centering 
	\includegraphics[width=0.85\textwidth]{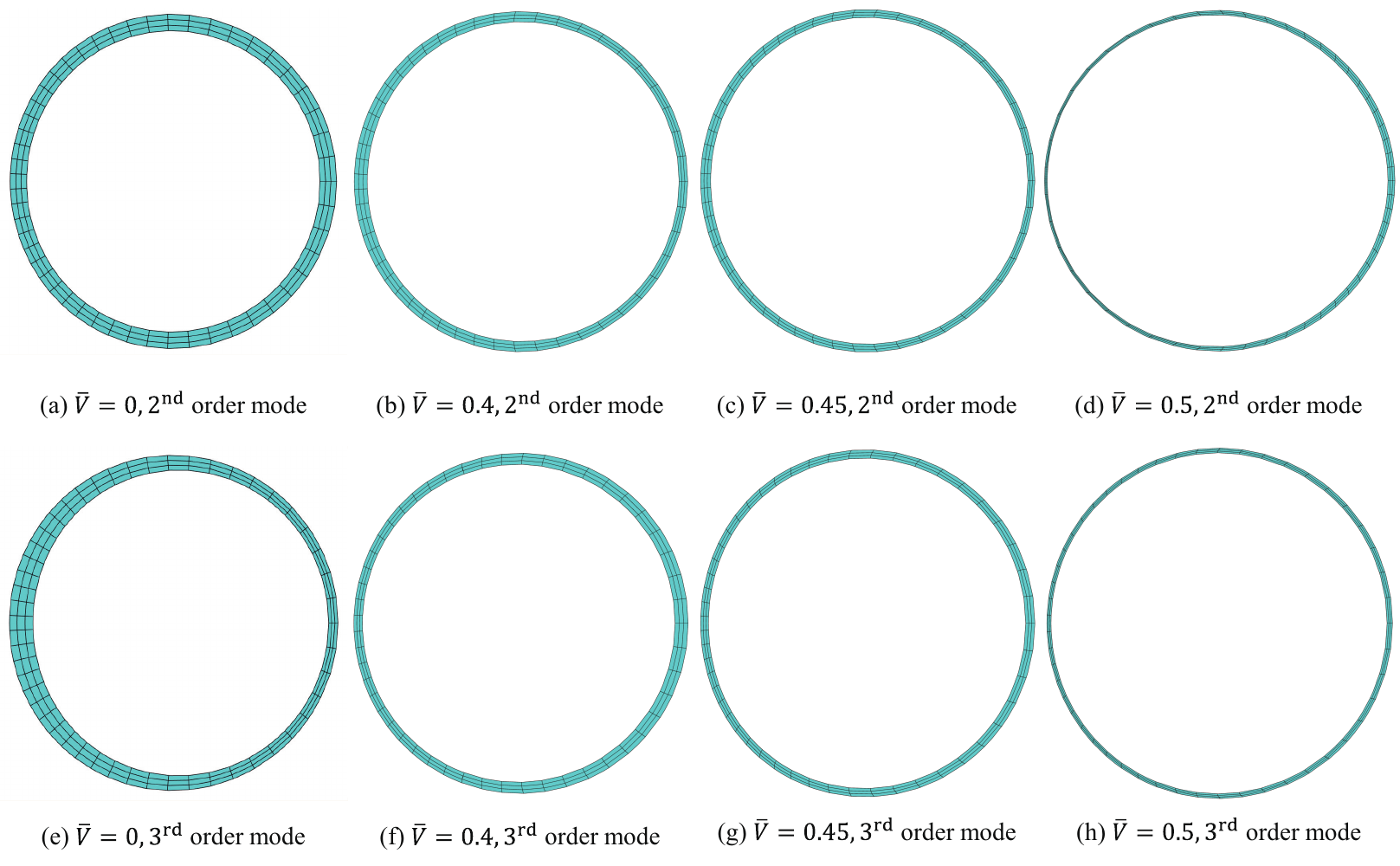} 
	\caption{Cross-section view of mode veering validation of non-axisymmetric vibration with $m=n=1$ for the Gent model in a pre-stretched $(\lambda_{z}=2)$ SEA tube for Fig.~\ref{nonaxisymmetric.sub.3} between Mode 2 (a-d) and Mode 3 (e-h): (a, e) $\overline{V}=0$; (b, f) $\overline{V}=0.4$; (c, g) $\overline{V}=0.45$; (d, h) $\overline{V}=0.5$.} 
	\label{Veering_2.0} 
\end{figure}

\begin{figure}[H] 
	\centering 
	\includegraphics[width=0.85\textwidth]{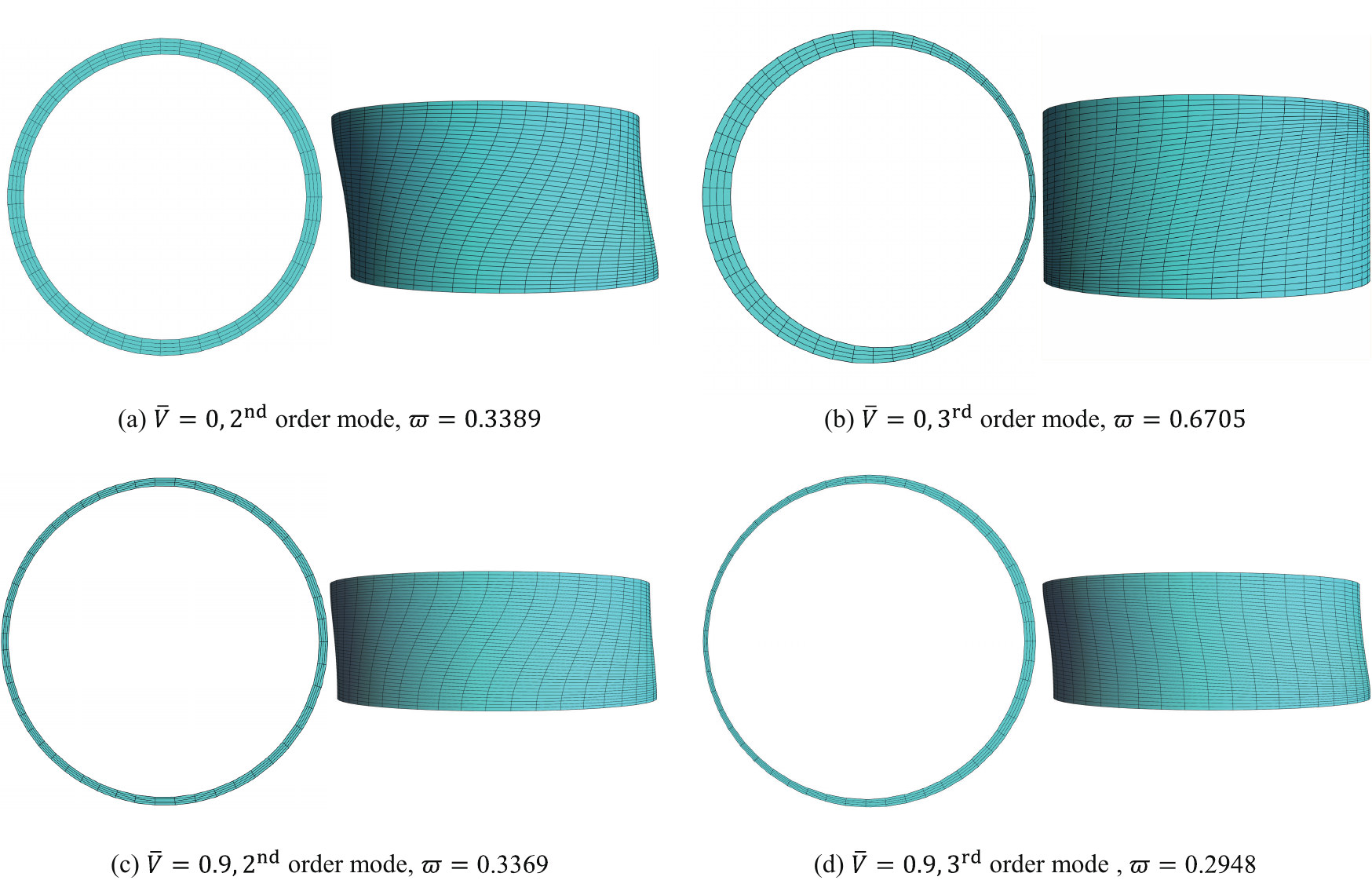} 
	\caption{Mode-crossing validation (top and front views) of non-axisymmetric vibration with $m=n=1$ for the Gent model in an SEA tube without pre-stretch $(\lambda_{z}=1)$ for Fig.~\ref{nonaxisymmetric.sub.2} between Mode 2 (a, c) and Mode 3 (b, d): (a, b) $\overline{V}=0$; (c, d) $\overline{V}=0.9$.} 
	\label{mode_shape_corssing_1.0} 
\end{figure}

\begin{figure}[H] 
	\centering 
	\includegraphics[width=0.85\textwidth]{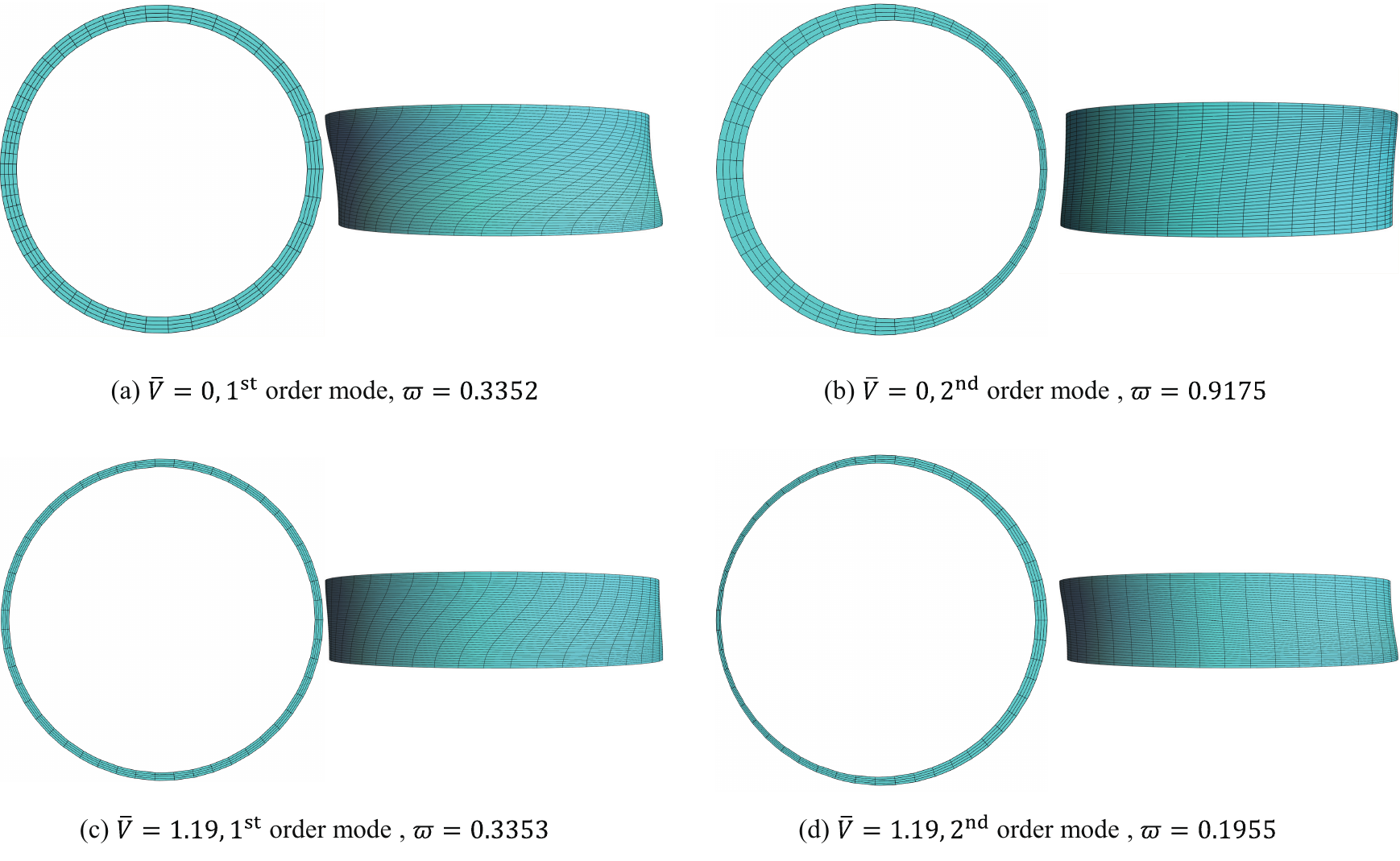} 
	\caption{Mode-crossing validation (top and front views) of non-axisymmetric vibration with $m=n=1$ for the Gent model in a pre-stretched $(\lambda_{z}=0.75)$ SEA tube for Fig.~\ref{nonaxisymmetric.sub.1} between Mode 1 (a, c) and Mode 2 (b, d): (a, b) $\overline{V}=0$; (c, d) $\overline{V}=1.19$.} 
	\label{mode_shape_corssing_0.75} 
\end{figure}

To proceed, we provide the mode veering validation for the Gent model with top view of the vibration modes in the veering progress in Fig.~\ref{Veering_2.0}. Four representative radial voltages ($\overline{V}=0, 0.4, 0.45, 0.5$) are chosen. The second- and third-order vibration modes at $\overline{V}=0$ are illustrated in Figs.~\ref{Veering_2.0}(a) and \ref{Veering_2.0}(e) for reference as the initial vibration modes. A circular end surface and evenly distributed thickness in Fig.~\ref{Veering_2.0}(a) demonstrate that the circumferential variations of radial displacement $u_{r}$ are hardly noticeable, while large circumferential changes in $u_{r}$ are illustrated in Fig.~\ref{Veering_2.0}(e), where the tube thickness is not distributed uniformly. 
As the voltage grows to 0.4 (before the mode veering point), the second- and third-order vibration mode shapes keep consistent with the initial vibration modes with $\overline{V}=0$. However, when the voltage is further increased to 0.45 or 0.5, the mode shapes between Mode 2 and Mode 3 switch. 
In Fig.~\ref{Veering_2.0}(d), the circumferential change in $u_{r}$ is exhibited, which is consistent with the mode shape in Fig.~\ref{Veering_2.0}(e). 
In Fig.~\ref{Veering_2.0}(h), the tube end surface remains circular and its thickness is evenly distributed, which corresponds to the mode shape shown in Fig.~\ref{Veering_2.0}(a). 
Therefore, when the applied voltage is close to triggering the veering, a slight change in voltage can significantly alter the vibration mode shape. 
Usually, it leads to a flow of mechanical energy between adjacent vibration modes, which physically expresses a modal transformation. 
The validation of mode crossing predicted by the neo-Hookean model follows the results in Figs.~\ref{mode_shape_corssing_1.0} and \ref{mode_shape_corssing_0.75} for $\lambda_{z}=1$ and $\lambda_{z}=0.75$, respectively.

In Fig.~\ref{nonaxisymmetric.sub.2}, there is no axial pre-stretch, $\lambda_{z}=1$. As the radial voltage increases, the first- and third-order natural frequencies decrease monotonically toward zero (i.e., the 3D non-axisymmetric buckling instability occurs \citep{haughton1979bifurcation2, su2020voltage}), while the second-order natural frequency remains independent of the voltage within the range of $\overline{V}_{\mathrm{EMI}}$ for both the neo-Hookean and Gent models. Moreover, a mode-crossing phenomenon appears between Mode 2 and Mode 3 for both energy models, and its validation for the Gent model is illustrated in Fig.~\ref{mode_shape_corssing_1.0}, where we provide the vibration mode shapes of Mode 2 and Mode 3 stimulated by two selected radial voltages $\overline{V}=0$ and $\overline{V}=0.9$ before and after the mode-crossing point. 
When examining the second-order mode shape at $\overline{V}=0$ in Fig.~\ref{mode_shape_corssing_1.0}(a), the top view reveals that the circumferential changes in $u_r$ are barely noticeable and the cross-section remains circular, while the sharp spiral gridlines observed from the front view indicate that $u_{\theta}$ undergoes significant changes along the axial direction. 
Then, considering the third-order mode shape at $\overline{V}=0$ in Fig.~\ref{mode_shape_corssing_1.0}(b), the top view reveals significant circumferential change in $u_{r}$ leading to uneven thickness distribution in the cross-section, while from the front view, only slight change in $u_{\theta}$ along the axial direction can be observed. For $\overline{V}=0.9$ (after the mode-crossing point), the vibration mode shapes of Mode 2 in Fig.~\ref{mode_shape_corssing_1.0}(c) and Mode 3 in Fig.~\ref{mode_shape_corssing_1.0}(d) remain the same as those at $\overline{V}=0$. Therefore, we refer to the branch crossing between Mode 2 and Mode 3 in Fig.~\ref{nonaxisymmetric.sub.2} as the mode crossing, where it is important to emphasize again that the definition of the same order modes is based on the similarity of modes before and after the crossing point, rather than the frequency values of modes. Numerical calculations (not reproduced here) show that a similar phenomenon of mode crossing also exists in the neo-Hookean model.

In Figs.~\ref{nonaxisymmetric.sub.4} and \ref{nonaxisymmetric.sub.1}, the axial pre-stretch is compressive, as $\lambda_{z}=0.92$ and $\lambda_{z}=0.75$, respectively. 
Clearly, the vibration frequencies of the second-order mode in Fig.~\ref{nonaxisymmetric.sub.4} and of the first-order mode in Fig.~\ref{nonaxisymmetric.sub.1} are all independent of the radial voltage within the range of $\overline{V}_{\mathrm{EMI}}$. However, as the radial voltage increases, the frequencies of other vibration modes decrease to zero monotonically, where the 3D non-axisymmetric buckling instabilities occur. In particular, the critical voltage of the first vibration mode initially increases and then decreases until it reaches zero as the axial pre-stretch changes from axial pre-extension to axial pre-compression. Similar to Fig.~\ref{nonaxisymmetric.sub.2}, the mode-crossing phenomenon happens between Mode 2 and Mode 3 in Fig.~\ref{nonaxisymmetric.sub.4} and between Mode 1 and Mode 2 in Fig.~\ref{nonaxisymmetric.sub.1} for both the neo-Hookean and Gent models. This mode-crossing phenomenon can be validated by comparing the vibration modes before and after the crossing point. 
As shown in Fig.~\ref{mode_shape_corssing_0.75}, the mode shape variations of Mode 1 and Mode 2 for the Gent model and $\lambda_{z}=0.75$,  before ($\overline{V}=0$) and after ($\overline{V}=0.9$) the crossing point, are essentially the same as those depicted in Fig.~\ref{mode_shape_corssing_1.0}.

\begin{figure}[h!] 
	\centering 
	\includegraphics[width=1.0\textwidth]{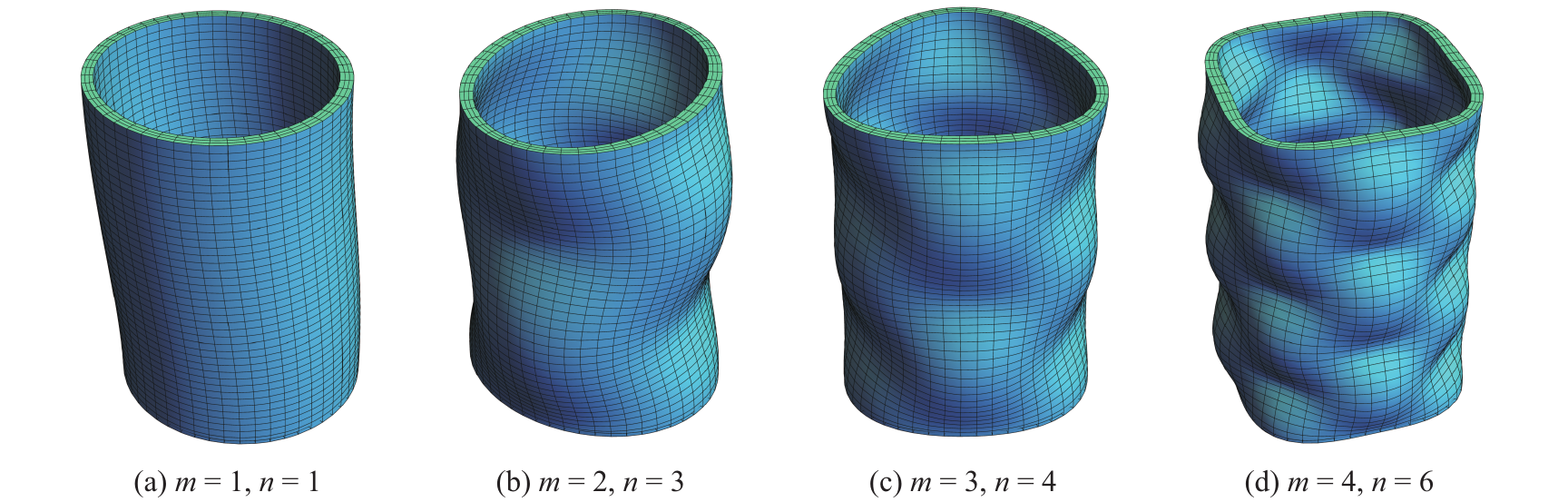} 
	\caption{Non-axisymmetric vibration mode shapes in a pre-stretched $(\lambda_{z}=2)$ thin and slender Gent SEA tube $(\eta=0.9, L/H=10, G=97.2)$ subject to $\overline{V}=0.2$ for different circumferential and axial mode numbers: (a) $m=1, n=1$; (b) $m=2, n=3$; (c) $m=3, n=4$; (d) $m=4, n=6$.} 
	\label{mode_shape_of_nonaxi} 
\end{figure}

To demonstrate the diversity of modes, we present in Fig.~\ref{mode_shape_of_nonaxi} the 3D mode shapes of non-axisymmetric vibrations with different combinations of circumferential and axial mode numbers for $\lambda_{z}=2$ and $\overline{V}=0.2$. The illustrated 3D mode shapes are composed of three coupled displacement components: the radial displacement $u_{r}$, circumferential displacement $u_{\theta}$, and axial displacement $u_{z}$. The SEA tube vibrates sinusoidally in the axial and circumferential directions following Eq.~(\ref{Assumed_solution}). Moreover, the axial and circumferential mode numbers are integer multiples of the half-wavelength in the axial direction and the wavelength in the circumferential direction, respectively.

\begin{figure}[H] 
	\centering 
	\includegraphics[width=0.5\textwidth]{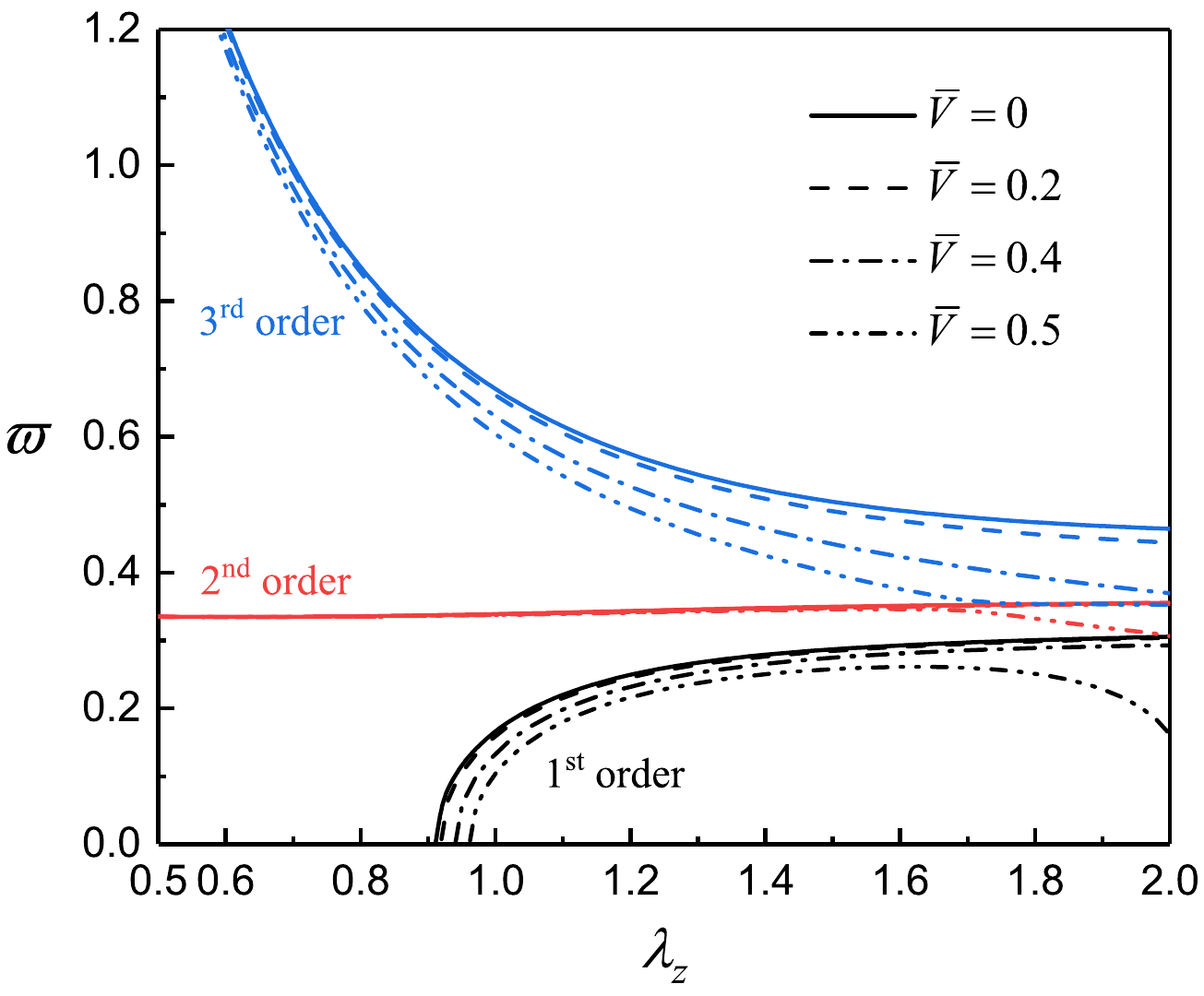} 
	\caption{The first three frequencies $\varpi$ of non-axisymmetric vibration mode with $m=n=1$ as functions of axial pre-stretch $\lambda_{z}$ in a thin and slender Gent SEA tube $(\eta=0.9, L/H=10, G=97.2)$ subject to four different radial voltages $\overline{V}=0$, $0.2$, $0.4$, and $0.5$.} 
	\label{effect_of_lambdaZ} 
\end{figure}

To clearly reveal the effect of axial pre-stretch on the non-axisymmetric vibration characteristics, the variation curves of the first three natural frequencies $\varpi$ with the axial pre-stretch $\lambda_{z}$ in a Gent SEA tube are plotted in Fig.~\ref{effect_of_lambdaZ} for $m=n=1$ and under four different radial voltages $\overline{V}=0$, $0.2$, $0.4$, and $0.5$. Taking Fig.~\ref{nonaxisymmetric_vibration_figure} into consideration, it is clear from Fig.~\ref{effect_of_lambdaZ} that the second-order natural frequency is barely affected by the axial pre-stretch when $\overline{V} \leq 0.4$. But when $\overline{V} > 0.4$ (e.g., $\overline{V}=0.5$), a noticeable frequency veering phenomenon occurs near $\lambda_{z} \approx 1.7$ between Mode 2 and Mode 3, which explains that the second-order frequency for $\overline{V}=0.5$ decreases with the axial pre-stretch for $\lambda_{z} > 1.7$, as shown in Fig.~\ref{effect_of_lambdaZ}. Additionally, we observe that the first-order frequency for $\overline{V} = 0.5$ also decreases when the axial pre-stretch increases from 1.7 to 2. This is because the geometric sizes of the tube increase and the overall stiffness declines as a result of the high voltage close to $\overline{V}_{\mathrm{EMI}}$. Nevertheless, the first- and third-order natural frequency curves diverge from the second-order one in opposite directions when decreasing the axial pre-stretch for $\lambda_{z}<1.7$. To be specific, the third-order natural frequency continues to increase as $\lambda_{z}$ decreases, while the first-order frequency goes down gradually as $\lambda_{z}$ decreases from axial pre-extension to axial pre-compression. In particular, the first-order frequency curve disappears (i.e., the first-order frequency drops to $\varpi = 0$ and the 3D non-axisymmetric buckling instability happens) when the axial pre-stretch reaches a critical value $\lambda_{z}^{\mathrm{cr}}$. For example, the critical axial pre-stretches are $\lambda_{z}^{\mathrm{cr}}=0.911$, $0.918$, $0.941$, and $0.962$ for $\overline{V}=0$, $0.2$, $0.4$, and $0.5$, respectively. Thus, the critical axial pre-stretch has a rise when increasing the radial voltage, which destabilizes the SEA tube. This disappearance can also be observed in Fig.~\ref{nonaxisymmetric_vibration_figure}, where the first-order frequency curve gradually disappears when $\lambda_{z}$ decreases from axial pre-extension to axial pre-compression.

\begin{figure}[H] 
	\centering  
	\subfigure{
		\label{G_vs_omega_2.sub.1}
		\includegraphics[width=0.48\textwidth]{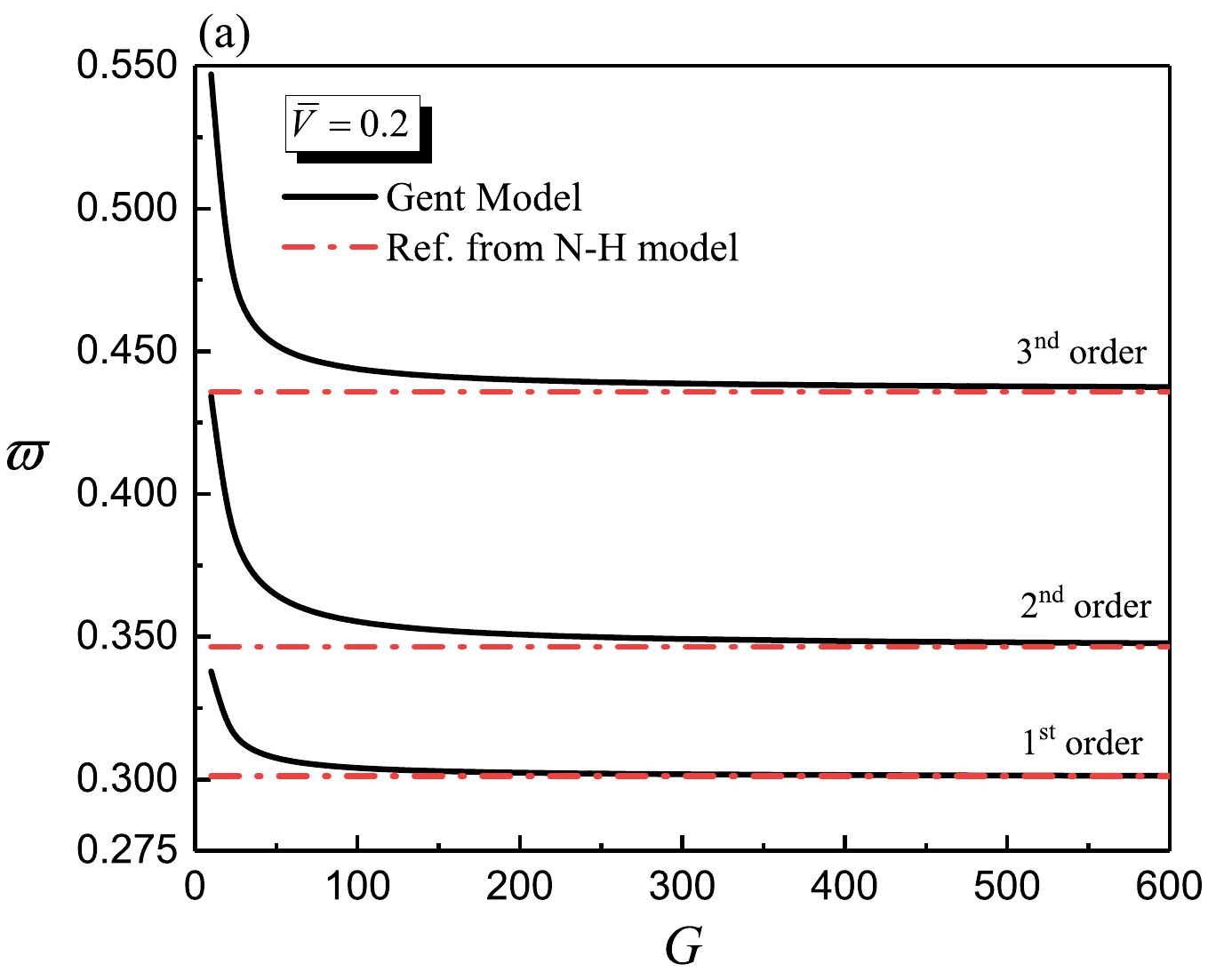}}
	\subfigure{
		\label{G_vs_omega_2.sub.2}
		\includegraphics[width=0.48\textwidth]{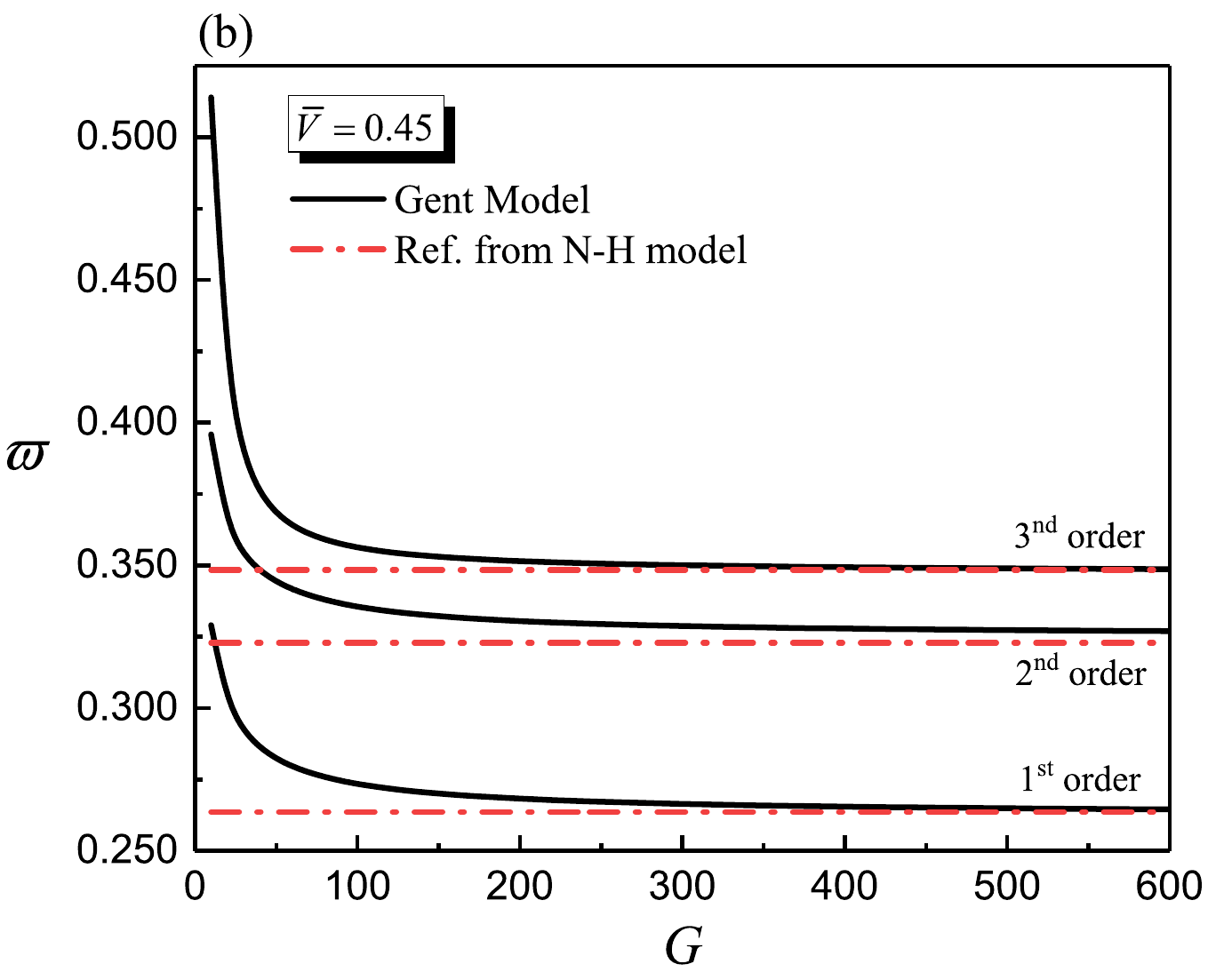}}
	\caption{Variation curves of the first three frequencies $\varpi$ of non-axisymmetric vibration mode $m=n=1$ with the Gent parameter $G$ in a pre-stretched $(\lambda_{z}=2)$ SEA tube subject to (a) $\overline{V}=0.2$ (a low radial voltage) and (b) $\overline{V}=0.45$ (a high radial voltage). } 
	\label{G_vs_omega_2_figure}
\end{figure}

Moreover, in order to study the strain-stiffening effect on the non-axisymmetric vibration behaviors, particularly when the SEA tube is subject to low and high radial voltages, the first three natural frequencies $\varpi$ of the non-axisymmetric vibration mode $m=n=1$ are depicted in Fig.~\ref{G_vs_omega_2_figure} as functions of the Gent parameter $G$ for a pre-stretched $(\lambda_{z}=2)$ SEA tube. It can be seen that for both low $(\overline{V}=0.2)$ and high $(\overline{V}=0.45)$ radial voltages, the first three natural frequencies decrease rapidly at first with the increase of $G$, and then gradually tend to the straight lines representing the results predicted by the neo-Hookean model. Furthermore, comparing Fig.~\ref{G_vs_omega_2_figure}(a) with Fig.~\ref{G_vs_omega_2_figure}(b), the corresponding Gent parameter $G$, for which the frequency predictions based on the Gent model start to deviate from those by the neo-Hookean model, increases from approximately $G\approx100$ for $\overline{V}=0.2$ to $G\approx200$ for $\overline{V}=0.45$. In other words, the strain-stiffening effect manifests itself earlier at relatively high voltages. 
We note that, as the Gent parameter $G$ further decreases from the deviation point between the Gent and neo-Hookean models, the strain-stiffening effect becomes increasingly pronounced, resulting in a rapid increase in frequency. Similar phenomena can be observed for axial pre-compression and are not shown here for brevity.

When there is no axial pre-stretch (i.e., $\lambda_{z}=1$), it can be found from Fig.~\ref{nonaxisymmetric.sub.2} that the 3D non-axisymmetric buckling instability voltage ($\overline{V}_{\mathrm{cr}}=0.6008$) of the first mode for the Gent model is far less than the electro-mechanical instability voltage  $\overline{V}_{\mathrm{EMI}} =1.0004$. According to the results when $\lambda_{z}=1$ in Fig.~\ref{NL_res.sub.2}, the Gent SEA tube does not get strain-stiffened for $\overline{V}<\overline{V}_{\mathrm{EMI}}$, even when the Gent parameter $G$ decreases to 10. Thus, the first-order natural frequency of the SEA tube for the mode $m=n=1$ does not change with the Gent parameter $G$ unless it is chosen to be extremely small (e.g., $G<0.5$). 

\textcolor{black}{In order to more clearly demonstrate the voltage-controlled vibration frequency of a specific material and discuss the failure possibility of electric breakdown (EB) under the combination of extreme deformation and high voltage conditions for the Gent model, we have selected Silicone CF19-2186 \citep{shmuel2016manipulating} by the manufacturer Nusil as a specific numerical example, with its material properties being $\rho=1100 \mathrm{kg/m^3}$, $\mu=333 \mathrm{kPa}$, $\varepsilon_r=2.8$, $E_{\mathrm{EB}}=235 \mathrm{MV/m}$ and $G=46.3$, where $\varepsilon_{r}$ and $E_{\mathrm{EB}}$ are the relative permittivity and dielectric strength, beyond which the EB phenomenon occurs. 
Using the material properties of Silicone CF19-2186, the dimensionless breakdown electric field is calculated as $\overline{E}_{\mathrm{EB}}=E_{\mathrm{EB}} \sqrt{\varepsilon/\mu}=2.027$, where $\varepsilon=\varepsilon_{0}\varepsilon_{r}$ is the material permittivity with $\varepsilon_{0}=8.85\mathrm{pF/m}$. For the static axisymmetric deformation of SEA tubes subject to axial pre-stretch and radial voltage, the radial electric field can be derived as $E_{r}=Q(a)/(2 \pi r \varepsilon \lambda_{z} L)$ based on the ideal dielectric energy model (\ref{Gent_model})$_2$ and constitutive equation (\ref{relations_from_Omega_star})$_3$. Thus, based on Eq.~(\ref{voltage_charge}) the radial electric field of the SEA tube is expressed as $E_{r}=-V/(r \ln \overline{\eta})$, which is inversely proportional to the radial coordinate $r$. Clearly, the maximum value of the radial electric field is obtained at inner surface of the SEA tube and its dimensionless form is $\overline{E}_{r}^{\mathrm{max}}=E_{r}^{\mathrm{max}} \sqrt{\varepsilon/\mu}=- \overline{V}(1-\eta)/(\lambda_{a} \eta \ln\overline{\eta})$.
}

\begin{figure}[H] 
	\centering  
	\subfigure{
		\label{omega_vs_electric.sub.1}
		\includegraphics[width=0.47\textwidth]{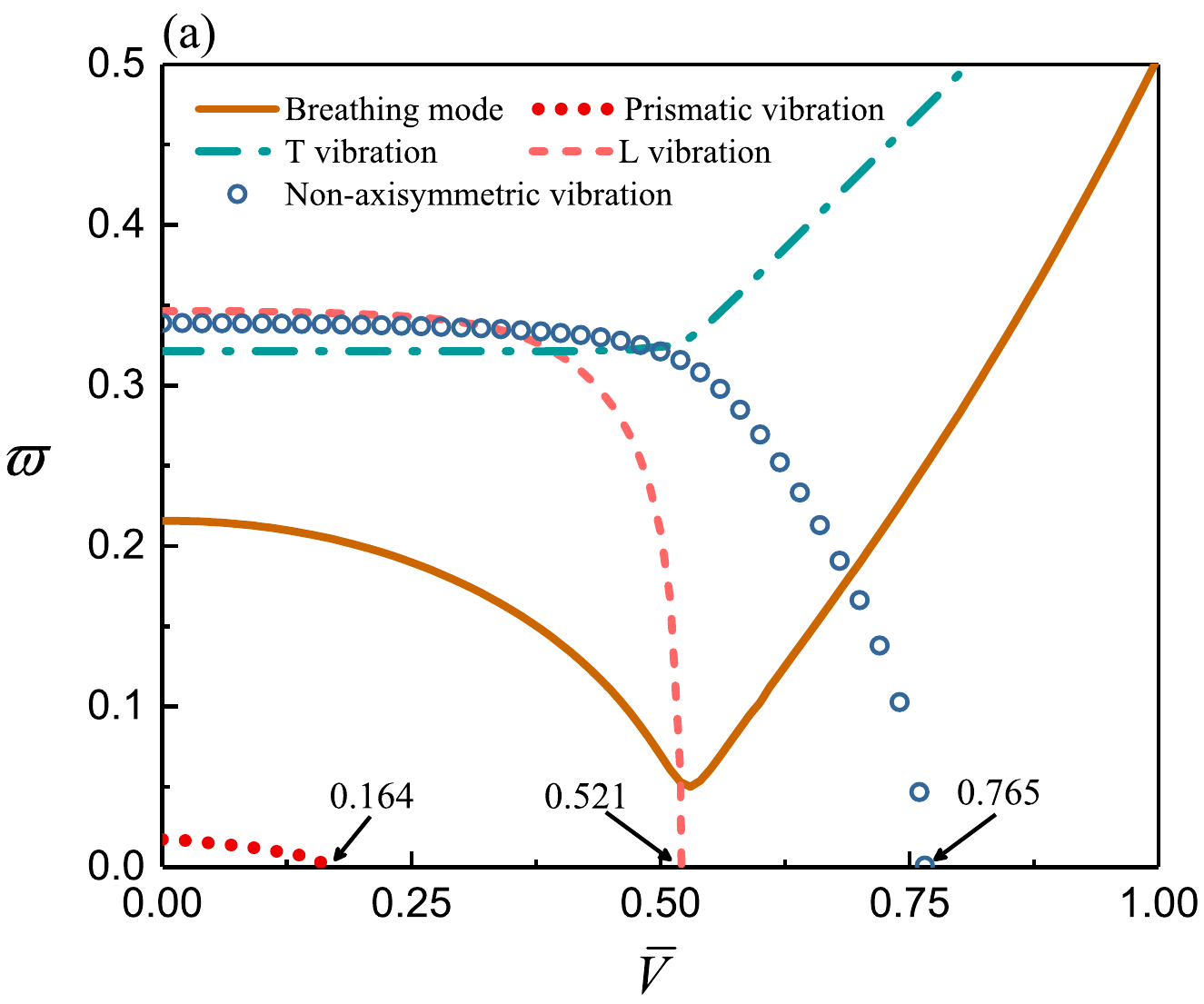}}
	\subfigure{
		\label{electric_breakdown.sub.2}
		\includegraphics[width=0.49\textwidth]{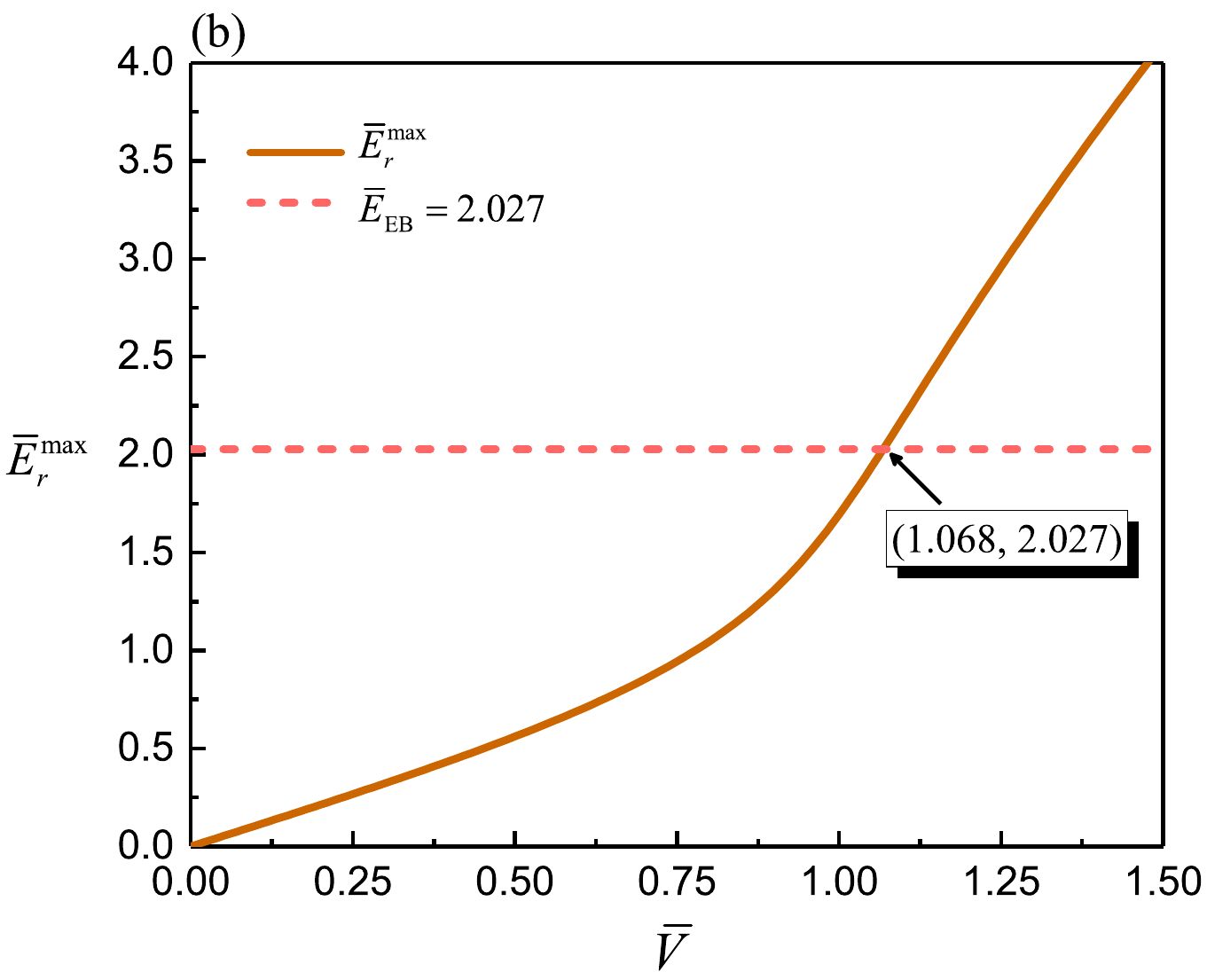}}
	\caption{\textcolor{black}{(a) The lowest dimensionless vibration frequencies $\varpi$ of the breathing mode $(m=n=0)$, L vibration mode $(m=0, n=1)$, T vibration mode $(m=0, n=1)$, prismatic vibration mode $(m=2, n=0)$ and non-axisymmetric vibration mode $(m=n=1)$ as functions of the dimensionless radial voltage $\overline{V}$ and (b) the comparison between the dimensionless breakdown electric field and maximum radial electric field in a pre-stretched ($\lambda_{z}=2$) thin and slender Gent SEA tube $(\eta=0.9, L/H=10, G=46.3)$ made of Silicone CF19-2186.}} 
	\label{possibility of failure}
\end{figure}

\textcolor{black}{For a pre-stretched $(\lambda_{z}=2)$ thin and slender Gent SEA tube $(\eta=0.9, L/H=10, G=46.3)$ made of Silicone CF19-2186, we plotted in Fig.~\ref{omega_vs_electric.sub.1} the lowest dimensionless vibration frequencies of five different vibration modes as functions of dimensionless radial voltage. It is seen that similar to the previous results, the vibration frequency of the breathing mode first decreases and then increases because of the strain-stiffening effect, while that of the T vibration mode remains unchanged at first and then rises quickly. The frequencies of the L vibration, prismatic vibration and non-axisymmetric vibration modes decrease monotonically to zero, corresponding to the relevant instabilities. The dimensionless instability voltages are marked in Fig.~\ref{omega_vs_electric.sub.1}, which are $\overline{V}_{\mathrm{cr}}=0.164$ for the prismatic vibration mode $(m=2, n=0)$, $\overline{V}_{\mathrm{cr}}=0.521$ for the L vibration mode $(m=0, n=1)$, and $\overline{V}_{\mathrm{cr}}=0.765$ for the non-axisymmetric vibration mode $(m=1, n=1)$. The EB failure possibility can be determined by comparing $\overline{E}_{\mathrm{r}}^{\mathrm{max}}$ with $\overline{E}_{\mathrm{EB}}$ for different applied voltages, as shown in Fig.~\ref{electric_breakdown.sub.2}. It is apparent from Fig.~\ref{possibility of failure} that the dimensionless EB voltage is larger than the instability voltages of the relevant vibration modes, which means that the buckling instability or prismatic diffuse instability of a SEA tube made of Silicone CF19-2186 occur before its electric breakdown. In addition, Fig.~\ref{possibility of failure} demonstrates that electrostatically tunable vibration characteristics of SEA tubes made of Silicone CF19-2186 is feasible over a wide voltage range.}

\section{Conclusions} \label{section5}

We investigated the electrostatically tunable non-axisymmetric vibration characteristics of an SEA cylindrical tube with strain-stiffening effect under inhomogeneous biasing fields induced by an axial pre-stretch and a radial electric voltage. First, we used the finite electro-elasticity theory to derive the governing equations of nonlinear axisymmetric static response and radially inhomogeneous biasing fields of an incompressible SEA tube characterized by the Gent ideal dielectric model. Next, based on the relevant linearized theory for incremental fields proposed by Dorfmann and Ogden, we employed the State-Space Method (SSM) to tackle the inhomogeneous biasing fields and obtain the frequency equations for small-amplitude prismatic vibrations and non-axisymmetric vibrations of the activated SEA tube. Finally, we conducted numerical calculations to verify the accuracy and convergence of the SSM in dealing with the non-axisymmetric vibrations, and to thoroughly study the influence of the strain-stiffening effect on the axisymmetric and prismatic vibrations, as well as the influences of electro-mechanical biasing fields and strain-stiffening effect on the non-axisymmetric vibrations. We can summarize our main results as follows:

\begin{enumerate}[(1)]
	\item The SSM is a highly efficient and accurate method to study the superimposed non-axisymmetric vibrations of SEA tubes subject to inhomogeneous biasing fields.
	
	\item As the radial voltage increases, the vibration frequency of the breathing mode predicted by the neo-Hookean model monotonically decreases to zero due to the reduction in global stiffness, while that anticipated by the Gent model first reduces to a small value owing to the attenuation of global stiffness, and then increases conversely because of the strain-stiffening effect.
	
	\item For the L vibrations, the vibration frequencies predicted by both the neo-Hookean and Gent models exhibit a nonlinear decline towards zero with the increase of radial voltage and the axisymmetric barreling instabilities occur.	
	
	\item For the T vibrations of thin SEA tubes characterized by the Gent model, the lowest vibration frequency does not change with the radial voltage when it is less than the electro-mechanical instability voltage $\overline{V}_{\mathrm{EMI}}$. However, as the voltage exceeds $\overline{V}_{\mathrm{EMI}}$, the lowest vibration frequency starts to increase significantly as a result of the strain-stiffening effect.
	
	\item For the prismatic vibration with a circumferential mode number $m\geq2$, its lowest vibration frequencies for both the neo-Hookean and Gent models decrease nonlinearly to zero when continuously increasing the voltage and prismatic diffuse instabilities appear in the SEA tube. The critical instability voltage increases monotonically with the circumferential mode number.
	
	\item For the non-axisymmetric vibrations, the lowest vibration frequencies monotonically go down to zero when increasing the voltage for both the neo-Hookean and Gent models and 3D non-axisymmetric buckling instabilities happen. The lowest frequency curve gradually disappears when the axial pre-stretch goes from pre-extension to pre-compression. Mode veering or crossing phenomena exist in two higher-order vibration modes resulting in modal conversion, which can be judged by analyzing the modal shape evolution.
	
	\item As the axial pre-stretch reduces from axial pre-extension to axial pre-compression, the lowest vibration frequency of non-axisymmetric vibrations continues to go down and eventually reaches zero, indicating the 3D non-axisymmetric buckling instability. The critical axial pre-stretch increases with the radial voltage, which destabilizes the SEA tube.
	
	\item Regardless of the vibration type, the vibration frequencies at first decrease rapidly with an increase of the value of the Gent parameter, and then gradually tend to the results predicted by the neo-Hookean model. The strain-stiffening effect manifests itself earlier at relatively high voltages.
	
\end{enumerate}

Our results indicate the possibility of exploiting the electro-mechanical biasing fields to realize on-demand tunability of small-amplitude vibration behaviors of SEA tubes exhibiting the strain-stiffening effect. The present study provides guidelines for further experimental research and design of SEA tube-based electrostatically tunable resonant systems, which may find a wide range of potential applications, including active vibration isolators, tunable energy harvesters, tunable sound generators, as well as biomedical actuators and sensors.

\textcolor{black}{It is emphasized that the main focus of this work is to investigate the influence of strain-stiffening effect on the vibration behaviors of SEA tubes. Note that \cite{arora2022deformation} explored the influence of strain-stiffening effect on the stability of soft periodic laminates and achieved deformation-activated negative group velocity by customizing the stiffening behavior of the non-Gaussian soft phases. Thus, how the strain-stiffening effect modulates the buckling instability and wave characteristics of negative group velocity for SEA structures is a topic worthy of further investigation.}


\section*{Acknowledgments}


The work is supported by the National Natural Science Foundation of China (Nos. 11872329, 12192210, 12192211, 12272339, and 12072315), the 111 Project, PR China (No. B21034), the Natural Science Foundation of Zhejiang Province, PR China (No. LD21A020001), the research project from Huanjiang Laboratory (Zhuji, Zhejiang Province), and the China Scholarship Council (No. 202006320363). WB gratefully acknowledges the support of the European Union Horizon 2020 Research and Innovation Programme under the Marie Skłodowska-Curie Actions (Grant No. 896229).



\appendix

\section{Frequency equations of free vibrations in a pre-stretched hyperelastic tube characterized by the Gent model} \label{AppendixA}

In this appendix, we use the \emph{conventional displacement method} to derive the frequency equations of four kinds of vibrations (i.e. non-axisymmetric vibrations, L vibrations (including the breathing mode), T vibrations and prismatic vibrations) in a pre-stretched hyperelastic tube characterized by the Gent model. For arbitrary energy function models, the detailed derivations of the frequency equations of the first three kinds of vibrations in a pre-stretched hyperelastic tube have been provided in our previous work \citep{zhu2020electrostatically}. 

Three displacement functions $\psi$, $F$, and $K$ to express the displacement components are introduced as:
\begin{equation} \label{general displacement functions}
	u_r = \dfrac{1}{r} \dfrac{\partial \psi}{\partial \theta} - \dfrac{\partial F}{\partial r},
	\qquad
	u_{\theta} = -\dfrac{\partial \psi}{\partial r} - \dfrac{1}{r}\dfrac{\partial F}{\partial \theta},
	\qquad
	u_z = K,
\end{equation}
where $\psi$, $F$, $K$ and $\dot{p}$ have the following assumed formal solutions:
\begin{equation} \label{assumed_solutions}
	\begin{array}{l}
	\psi =  \overline{\psi}\left( r \right) \sin\left( m \theta \right) \cos \left( n \pi \zeta \right) e^{\mathrm{ i} \omega t},
	\qquad
	F = \overline{F}\left( r \right) \cos\left( m \theta \right) \cos\left( n \pi \zeta \right) e^{\mathrm{ i} \omega t} ,
	\\\\
	K = \overline{K}\left( r \right) \cos\left( m \theta \right) \sin\left( n \pi \zeta \right) e^{\mathrm{ i} \omega t},
	\qquad
	\dot{p} = \overline{p} \left( r \right)  \cos\left( m \theta \right) \cos \left( n \pi \zeta \right) e^{\mathrm{ i} \omega t}. 
	\end{array}    
\end{equation}

The specific derivation process for the frequency equation of the non-axisymmetric and axisymmetric vibrations can be found in the \emph{Appendix C} in the paper of \cite{zhu2020electrostatically} and is omitted here for brevity. Specifically, we can refer to the frequency equations (C.11) and (C.12) for the non-axisymmetric vibrations and (C.13) for the axisymmetric vibrations (including the L vibrations and T vibrations) in that paper.

Now consider an axially pre-stretched hyperelastic tube characterized by the incompressible Gent model with strain-energy function Eq.~(\ref{Gent_model}) with $I_4=I_5=0$. We  obtain the required effective material parameters as:
\begin{equation} \label{effective_material_parameters}
	{\small
\begin{array}{l}
    	c_{11} = {\mathcal{A}}_{01111} + p = 2 \lambda^{2}_{\theta} \Omega_{1} + 4 \lambda^{4}_{\theta} \Omega_{11} + p ,
    	\quad
    	c_{12} = {\mathcal{A}}_{01122} = 4 \lambda^{4}_{\theta} \Omega_{11},
    	\quad
    	c_{13} = {\mathcal{A}}_{01133} = 4 \lambda^{-2}_{\theta} \Omega_{11},
    	\\\\
    	c_{33} = {\mathcal{A}}_{03333} + p =4 \lambda^{-8}_{\theta} \Omega_{11} + 2 \lambda^{-4}_{\theta} \Omega_{1} + p, 
        \quad
    	c_{58} = {\mathcal{A}}_{01331} + p = p, 
    	\quad
    	c_{55} = {\mathcal{A}}_{01313} = 2 \lambda^{2}_{\theta} \Omega_{1},
    	\\\\
    	c_{66} = c_{55}, \quad c_{69} = {\mathcal{A}}_{01221} + p =p, 
    	\quad
    	c_{77} = {\mathcal{A}}_{03131} = 2 \lambda^{-4}_{\theta}\Omega_{1}, 
	\end{array}}
\end{equation}
\noindent
where $\Omega_{1} = \mu G/[2 \left( G - I_{1} +3 \right)]$, $\Omega_{11} = \mu G/[2 \left( G - I_{1} +3 \right)^2]$, and $\lambda_{\theta}=\lambda^{-1/2}_{z}$ for the homogeneous deformation in the hyperelastic tube without the electro-mechanical coupling. According to Eq.~(\ref{Eulerian-electric_field})$_1$, the Lagrange multiplier $p$ can be obtained from $\tau_{rr} = 0$ as
\begin{equation} \label{pp}
	\begin{array}{l}
		p = 2 \lambda^{-1}_{z} \Omega_{1}.  
	\end{array}
\end{equation}

\subsection{Breathing mode}

For the breathing mode with $m = n =0$ and $u_{\theta} = u_{z} = 0$, we recall the frequency equation from Eq.~(C.20) in the paper of \cite{zhu2020electrostatically},
\begin{equation} 
	\rho \omega^{2} = \left( c_{12} - c_{11} \right)\left( \dfrac{1}{a^{2}} - \dfrac{1}{b^{2}} \right) / \ln\dfrac{a}{b},
\end{equation}
which, when combined with Eqs.~(\ref{effective_material_parameters}) and (\ref{pp}),
results in the following frequency equation of the breathing mode for the pre-stretched Gent hyperelastic tube:
\begin{equation} \label{special_case_for_breathing_mode}
	\varpi^2 = \rho \omega^{2} H^{2} / \mu = -  \dfrac{2G}{G-I_{1}+3} \left( 1-\eta \right)^{2} \left( 1 - \eta^{2} \right) / \left( \eta^{2} \ln \eta \right),
\end{equation}
where $\eta = A/B$ and $I_{1} = 2\lambda^{-1}_{z} + \lambda^{2}_{z}$. Thus, the vibration frequency of the breathing mode of the Gent hyperelastic tube depends on the inner-to-outer radius ratio $\eta$ and axial pre-stretch $\lambda_{z}$ appearing in $I_1$, but is independent of the length-to-thickness ratio $L/H$. When $G$ is large enough, the effect of axial pre-stretch can be counteracted, as with the result of the neo-Hookean model \citep{zhu2020electrostatically}.

\subsection{Purely torsional vibrations (T vibrations)}

For the T vibrations with $m =0$ and $u_{r} = u_{z} = \dot{p} =0$, the only non-zero displacement component is $u_{\theta} = \overline{v} \left( r \right) \cos \left( n \pi \zeta \right) e^{\rm i \omega t}$. Inserting Eq.~(\ref{effective_material_parameters}) into Eq.~(C.13) of \cite{zhu2020electrostatically}, the elements ($\overline{d}_{23}$, $\overline{d}_{26}$, $\overline{d}_{53}$, $\overline{d}_{56}$) related to the T vibrations can be rewritten as 
\begin{equation} \label{elements}
		\begin{array}{l}
			\overline{d}_{23} = G_{1} \overline{\alpha}_{3 \lambda}\left[   \overline{\alpha}_{3 \lambda} J_{0} \left( \dfrac{\overline{\alpha}_{3 \lambda}}{1-\eta} \right) - 2 \left( 1 - \eta \right) \overline{\alpha}_{3 \lambda} J_{1} \left( \dfrac{\overline{\alpha}_{3 \lambda}}{1-\eta}\right)   \right],
			\\\\
			\overline{d}_{26} = G_{1} \overline{\alpha}_{3 \lambda}\left[   \overline{\alpha}_{3 \lambda} Y_{0} \left( \dfrac{\overline{\alpha}_{3 \lambda}}{1-\eta} \right) - 2 \left( 1 - \eta \right) \overline{\alpha}_{3 \lambda} Y_{1} \left( \dfrac{\overline{\alpha}_{3 \lambda}}{1-\eta}\right)   \right], 
			\\\\
			\overline{d}_{53} = G_{1} \overline{\alpha}_{3 \lambda}\left[   \overline{\alpha}_{3 \lambda} J_{0} \left( \dfrac{ \eta \overline{\alpha}_{3 \lambda}}{1-\eta} \right) - 2 \left( \eta^{-1} -1 \right) \overline{\alpha}_{3 \lambda} J_{1} \left( \dfrac{\eta \overline{\alpha}_{3 \lambda}}{1-\eta}\right)   \right], 
			\\\\
			\overline{d}_{56} = G_{1} \overline{\alpha}_{3 \lambda}\left[   \overline{\alpha}_{3 \lambda} Y_{0} \left( \dfrac{ \eta \overline{\alpha}_{3 \lambda}}{1-\eta} \right) - 2 \left( \eta^{-1} -1 \right) \overline{\alpha}_{3 \lambda} Y_{1} \left( \dfrac{\eta \overline{\alpha}_{3 \lambda}}{1-\eta}\right)   \right], 
	\end{array} 
\end{equation}
where $G_{1}=G/\left( G-2 \lambda_{z}^{-1} - \lambda_{z}^2 +3 \right) $, $\overline{\alpha}_{3 \lambda}= \overline{\alpha}_{3} \lambda_{z}^{-1/2} = \sqrt{ \varpi^2 /G_{1} - \overline{\kappa}^2}$ with $\overline{\alpha}^2_{3}=[\rho \omega^{2}-(n\pi/l)^2 c_{77}]/c_{66}$ and $\overline{\kappa}=n \pi H/L$, and $J_m(\cdot)$ and $Y_m(\cdot)$ are the Bessel functions of the first and second kinds of order $m$, respectively. Thus, we obtain the frequency equation of the T vibrations for the Gent hyperelastic tube as
\begin{equation} \label{Freq_eq}
	\overline{\alpha}_{3 \lambda}^{2} \left[ J_{2} \left( \dfrac{\overline{\alpha}_{3 \lambda}}{1-\eta} \right) Y_{2} \left( \dfrac{\eta \overline{\alpha}_{3 \lambda}}{1-\eta} \right) - J_{2} \left( \dfrac{\eta \overline{\alpha}_{3 \lambda}}{1-\eta} \right) Y_{2} \left( \dfrac{ \overline{\alpha}_{3 \lambda}}{1-\eta} \right) \right] =0.
\end{equation}
It is clear from Eq.~(\ref{Freq_eq}) that the natural frequency of the T vibrations depends on the axial pre-stretch $\lambda_{z}$ in $G_1$, the length-to-thickness ratio $L/H$ in $\overline{\kappa}$ and the inner-to-outer radius ratio $\eta=A/B$. Furthermore, we note from Eqs.~(\ref{Freq_eq}) and (\ref{effective_material_parameters}) that $\overline{\alpha}^{2}_{3\lambda} 
=\varpi^2 /G_{1} - \overline{\kappa}^2=0 $ is one of the solutions to the frequency equation of the T vibrations, which yields
\begin{equation} \label{special_case_for_T_vibration}
	\varpi = \sqrt{G_{1}} \overline{\kappa} 
	=
	\sqrt{\dfrac{G}{G-2 \lambda_{z}^{-1} - \lambda_{z}^2 +3}} \dfrac{n \pi H}{L}.
\end{equation}
The solution (\ref{special_case_for_T_vibration}) represents the torsional displacement proportional to the radius (i.e., each cross-section of the tube rotates around its center during the vibration), and we see that its natural frequency depends on the axial pre-stretch and length-to-thickness ratio, but is independent of the inner-to-outer radius ratio.

\subsection{Prismatic vibrations}

For the case of prismatic vibrations with $m \neq 0, n=0$ and $u_{z} =0$, we have $K = 0$ and its corresponding assumed solutions can be rewritten from Eq.~(\ref{assumed_solutions}) as
\begin{equation}
	\begin{array}{l}
	\psi =  \overline{\psi}\left( r \right) \sin\left( m \theta \right)  e^{\rm i \omega t},
	\quad
	F = \overline{F}\left( r \right) \cos\left( m \theta \right)  e^{\rm i \omega t} ,
	\quad
	\dot{p} = \overline{p} \left( r \right)  \cos\left( m \theta \right) e^{\rm i \omega t}.
\end{array}    
\end{equation}
Moreover, its corresponding incremental governing equations (see Eq.~(C.4) of \cite{zhu2020electrostatically}) reduce to
\begin{equation} \label{incremental governing equations}
	\begin{array}{l}
		\left( \Lambda + \alpha_{3}^2  \right)  \overline{\psi} = 0, 
		\quad
		\Lambda \overline{F} =0,
		\\\\
		\left[ \left( c_{11} - c_{13} - c_{58} \right) \Lambda + \rho \omega^2  \right] \overline{F} + \overline{p} =0, 
	\end{array}
\end{equation}
where $\Lambda = {\mathrm{d}^2}/{\mathrm{d} r^2} + (1/r)\mathrm{d}/{\mathrm{d} r} - {m^2}/{r^2}$ and $\alpha_{3}^2 = \rho \omega^2 / c_{66}$. Obviously, Eq.~(\ref{incremental governing equations})$_{2}$ is a homogeneous Euler equation, with general solution
\begin{equation} \label{solution to G}
	\overline{F}\left( r \right)  = A_{1}r^{m} + B_{1} r^{-m},
\end{equation}
where $A_{1}$ and $B_{1}$ are the undetermined constants. Substituting Eq.~(\ref{incremental governing equations})$_{2}$ into Eq.~(\ref{incremental governing equations})$_{3}$ and using Eq.~(\ref{solution to G}), we obtain the solution of $\overline{p}\left(  r \right)$ as
\begin{equation} \label{solution to p}
	\overline{p} \left(  r \right)  = - \rho \omega ^2 \left( A_{1}r^{m} + B_{1} r^{-m} \right).
\end{equation}

In addition, Eq.~(\ref{incremental governing equations})$_{1}$ is clearly a Bessel equation of order $m$, and its solution is
\begin{equation} \label{solution to psi}
	\overline{\psi} \left(  r \right) = A_{2} J_{m} \left( \alpha_{3} r \right) + B_{2}Y_{m} \left( \alpha_{3} r \right),
\end{equation}
where $A_{2}$ and $B_{2}$ are arbitrary constants to be determined. Substituting Eqs.~(\ref{solution to G})-(\ref{solution to psi}) into Eqs.~(\ref{general displacement functions}) and (\ref{expansion_incre_cons})$_{1,4,8}$, we obtain the non-zero incremental transverse stress components as
\begin{equation}
	\begin{array}{l}
	\dot{T}_{0rr} = \Sigma_{rr} \cos \left( m \theta \right)  e^{\mathrm{i} \omega t},
	\quad
	\dot{T}_{0r \theta} =\Sigma_{r \theta} \sin \left( m \theta \right)  e^{\mathrm{i} \omega t},
\end{array}
\end{equation}
where
\begin{equation}
	\begin{array}{l}
		\Sigma_{rr} = c_{11} \left( \dfrac{m}{r} \overline{\psi}'  - \dfrac{m}{r^2} \overline{\psi} - \overline{F}'' \right) + c_{12} \dfrac{1}{r} \left[ m \left( - \overline{\psi}' + \dfrac{m}{r} \overline{F} \right) + \dfrac{m}{r} \overline{\psi} - \overline{F}'  \right] - \overline{p},
		\\\\
		\Sigma_{r \theta} = c_{66} \left( - \overline{\psi}'' + \dfrac{m}{r} \overline{F}' - \dfrac{m}{r^2} \overline{F} \right) - c_{69} \dfrac{1}{r} \left[ m \left( \dfrac{m}{r} \overline{\psi} - \overline{F}' \right) - \overline{\psi}' + \dfrac{m}{r} \overline{F}  \right],
	\end{array}
\end{equation}
in which the prime denotes differentiation with respect to $r$.

To satisfy the imposed mechanical boundary conditions (\ref{new_incremental_boundary_conditions})$_2$  and ensure that the determinantal condition for non-trivial solutions exists, we obtain the frequency equation of the prismatic vibrations as
\begin{equation} \label{determinant}
	\left| d_{ij} \right| = 0, \quad  \left( i, j=1 \thicksim 4 \right),
\end{equation}
where the first two rows $d_{ij}(i=1,2)$ of the determinant's elements that correspond to the boundary conditions on the inner surface $r=a$ are written as
%
%
%
\begin{equation} \label{dij}
	\begin{array}{l}
		d_{11} = \dfrac{m}{a^2}\left(c_{12} - c_{11}   \right) J_{m} \left( \alpha_{3} a \right) 
		+ 
		\dfrac{m \alpha_{3}}{2a} \left( c_{11} - c_{12} \right) \left[J_{m-1} \left( \alpha_{3} a \right) -  J_{m+1} \left( \alpha_{3} a \right)\right],
		\\\\
		d_{12} = \dfrac{m}{a^2}\left(c_{12} - c_{11}   \right) Y_{m} \left( \alpha_{3} a \right) 
		+ 
		\dfrac{m \alpha_{3}}{2a} \left( c_{11} - c_{12} \right) \left[Y_{m-1} \left( \alpha_{3} a \right) -  Y_{m+1} \left( \alpha_{3} a \right)\right],
		\\\\
		d_{13} = a^{m-2}\left[ \left( c_{12} - c_{11} \right) m \left( m-1 \right) + \rho \omega^2 a^2   \right],
		\quad 		
		d_{23} = m \left( m-1 \right) a^{m-2} \left( c_{66} + c_{69} \right),
		\\\\
		d_{14} = a^{-m-2}\left[ \left( c_{12} - c_{11} \right) m \left( m+1 \right) + \rho \omega^2 a^2   \right],
		\quad
		d_{24} = -m \left( m+1 \right) a^{-m-2} \left( c_{66} + c_{69} \right),
		\\\\
		d_{21} = -\dfrac{m^2 c_{69}}{a^2}J_{m} \left( \alpha_{3} a \right) 
		+
		\dfrac{\alpha_{3} c_{69}}{2a} \left[ J_{m-1} \left( \alpha_{3} a \right) - J_{m+1}\left( \alpha_{3} a \right)  \right] 
		\\\\ \qquad \quad
		- \dfrac{\alpha_{3}^2 c_{66}}{4} \left[ J_{m-2} \left( \alpha_{3} a \right) - 2 J_{m}\left( \alpha_{3} a \right) + J_{m+2} \left( \alpha_{3} a \right) \right],  
		\\\\
		d_{22} = -\dfrac{m^2 c_{69}}{a^2} Y_{m} \left( \alpha_{3} a \right) 
		+
		\dfrac{\alpha_{3} c_{69}}{2a} \left[ Y_{m-1} \left( \alpha_{3} a \right) - Y_{m+1}\left( \alpha_{3} a \right)  \right] 
		\\\\ \qquad \quad
		- \dfrac{\alpha_{3}^2 c_{66}}{4} \left[ Y_{m-2} \left( \alpha_{3} a \right) - 2 Y_{m}\left( \alpha_{3} a \right) + Y_{m+2} \left( \alpha_{3} a \right) \right]. 
	\end{array}
\end{equation}
For the mechanical boundary conditions on the outer surface $r=b$, we replace the inner radius $a$ with the outer radius $b$ in Eq.~(\ref{dij}) to obtain the elements $d_{ij}(i=3,4)$ of the final two rows of the determinant Eq.~(\ref{determinant}).

\newpage
\bibliographystyle{elsarticle-harv.bst}
\bibliography{ref.bib}







\end{document}